\documentclass[a4paper,11pt]{article}
\pdfoutput=1 % if your are submitting a pdflatex (i.e. if you have
\usepackage{jheppub_modified} % for details on the use of the package, please
\usepackage{amsthm} %for theorems and demonstrations 
\usepackage[T1]{fontenc} % if needed

\usepackage{amsmath,amsthm,verbatim,amssymb,amsfonts,amscd,graphicx,collectbox,mathtools,float}

\usepackage{tikz}
\usetikzlibrary{math}
\usetikzlibrary{arrows,shapes,positioning}
\usetikzlibrary{decorations.markings}
\tikzstyle arrowstyle=[scale=1]
\tikzstyle directed=[postaction={decorate,decoration={markings,
    mark=at position .65 with {\arrow[arrowstyle]{stealth}}}}]
\tikzstyle reverse directed=[postaction={decorate,decoration={markings,
    mark=at position .65 with {\arrowreversed[arrowstyle]{stealth};}}}]

\newlength{\mywidth}
\usepackage[colorlinks=true,urlcolor=blue,anchorcolor=blue,citecolor=blue,filecolor=blue,linkcolor=blue,menucolor=blue,pagecolor=blue,linktocpage=true,pdfproducer=medialab,pdfa=true]{hyperref}

\definecolor{bostonuniversityred}{rgb}{0.8, 0.0, 0.0} %1
\definecolor{bananayellow}{rgb}{1.0, 0.88, 0.21}
\definecolor{ao(english)}{rgb}{0.0, 0.5, 0.0} %4

\renewcommand{\vec}[1]{\mathbf{#1}}

\newcommand{\norm}[1]{\lVert#1\rVert}

\definecolor{blizzardblue}{rgb}{0.67, 0.9, 0.93}

\newtheorem{theorem}{Property}
\newtheorem*{theorem*}{Definition}

\newtheorem{mytheorem}{Property}
\numberwithin{mytheorem}{section}

\newcommand{\g}{g}

\title{\boldmath  Tree level integrability in 2d quantum field theories and affine Toda models}

\author[\clubsuit]{Patrick Dorey,} 
\author[\clubsuit]{Davide Polvara}

\affiliation[\clubsuit]{Department of Mathematical Sciences, Durham University, Durham DH1 3LE, United Kingdom }

\emailAdd{p.e.dorey@durham.ac.uk}
\emailAdd{davide.polvara@durham.ac.uk}

\abstract{
We investigate the perturbative integrability of  massive (1+1)-dimensional bosonic
quantum field theories, focusing on the conditions for them to have a purely elastic S-matrix, with no particle production and diagonal scattering, at tree level.
For theories satisfying what we call `simply-laced scattering conditions', by which we mean that
poles in inelastic $2$ to $2$ processes cancel in pairs, and 
poles in allowed processes are only due to one on-shell propagating
particle at a time,
the requirement that all inelastic amplitudes
must vanish is shown to imply the so-called
area rule, connecting the $3$-point couplings
$C^{(3)}_{abc}$ to the masses $m_a$, $m_b$, $m_c$ of the coupled
particles in a universal way.
We prove that the constraints we find are universally satisfied by all affine Toda theories, connecting pole cancellations in amplitudes to properties of the underlying root systems, and
develop a number of tools that we expect will be relevant for the study of loop amplitudes.
}

\begin{document} 
\maketitle
\flushbottom

%\section{Introduction}
%\label{sec:intro}

\section{Introduction}\label{sect_Intro}

% \section{A MODIFIED INTRODUCTION}
%\label{sect_Intro}

A key feature of integrable quantum field theories in 1+1 dimensions is the existence of an infinite tower of higher spin conserved charges. From these, momentum-dependent  translation operators can be constructed, and used to show that any non-vanishing scattering amplitude can be factorised into a product of 2 to 2 amplitudes \cite{a3,a4}. 
Combined with the constraints of crossing, unitarity, and either
the Yang Baxter or bootstrap equations, this has led to proposals for the exact S-matrices of many theories, which
can then be checked further using standard perturbation theory. From this perturbative
perspective, integrability should reveal itself in perhaps unexpected cancellation of sums of Feynman diagrams contributing to production
processes, as reviewed for some simple examples in \cite{Dorey:1996gd}. In the following we will use the phrase perturbative integrability to mean the vanishing of all such sums; this might be at tree level, or including all loop diagrams as well.
The general
mechanisms by which these cancellations are achieved are mathematically intricate, interesting to understand in their own right, and are the subject of the current paper. 

We start by considering a general Lagrangian for a quantum field theory of $r$ interacting massive scalar fields, possibly with different masses, in two dimensions
\begin{equation}\label{eq0_1}
\mathcal{L}=\sum_{a=1}^r \biggl( \frac{1}{2} \partial_\mu \phi_a  \partial^\mu \phi_{\bar{a}} - \frac{1}{2} m_a^2 \phi_a \phi_{\bar{a}}\biggr) - \sum_{n=3}^{+\infty} \sum_{a_1,\ldots,a_n=1}^r  \frac{1}{n!}C^{(n)}_{a_1\ldots a_n} \phi_{a_1} \ldots \phi_{a_n}.
\end{equation}
Here $a$ is a label for the possible types of particles in the model, which correspond to the possible asymptotic states of the theory. We adopt the convention $\phi_{\bar{a}}=\phi_a^*$. If the component $\phi_a$ is real, we assume $a=\bar{a}$, while if $\phi_a$ is a complex field $\bar{a}$ represents an index $\in \{1,\dots,r \}$ different from $a$.\footnote{For example if we have a theory of two real fields $\phi_1$ and $\phi_2$ we consider $\bar{1}=1$ and $\bar{2}=2$ so that the non-interacting part of the Lagrangian is given by $ \frac{1}{2} \partial_\mu \phi_1  \partial^\mu \phi_{1}+\frac{1}{2} \partial_\mu \phi_2  \partial^\mu \phi_{2} - \frac{1}{2} m_1^2 \phi_1 \phi_{1}-\frac{1}{2} m_2^2 \phi_2 \phi_{2}$. On the other hand if the fields are one the complex conjugate of the other we assume $\bar{1}=2$ and $\bar{2}=1$, so that the free Lagrangian is given by $ \partial_\mu \phi_1  \partial^\mu \phi_{\bar{1}} -  m_1^2 \phi_1 \phi_{\bar{1}}$.}
In this way we take into account both the case in which the fields in \eqref{eq0_1} are real and the one in which they are complex. 
We will label a generic $n$-point amplitude by $M^{(n)}$, by which we mean the sum over all relevant connected Feynman diagrams without inserting additional normalization factors. Contrarily we will refer to the S-matrix as the amplitude properly normalized and multiplied by the Dirac delta function of overall energy-momentum conservation.

We wish to find the possible sets of masses and couplings  for which the theory defined by \eqref{eq0_1} is perturbatively integrable at tree level, with a purely elastic S-matrix for $2$ to $2$ processes, meaning that the only $4$-point S-matrix elements different from zero are those in which the two incoming and outgoing particles are of the same type and carry the same set of momenta.
The standard way of computing Feynman diagrams is to consider all  propagators to come with an $i\epsilon$ factor in the denominator in such a way to avoid possible singularities. 
After summing over all diagrams, we take the limit $\epsilon \to 0$. 
In this limit, the connected part of any $n$-point amplitude decomposes into two pieces
$$
\lim_{\epsilon \to 0^+} M^{(n)}(\epsilon)=M_{fact}^{(n)}+M_{prod}^{(n)}
$$
of which the first corresponds to factorized scattering, and contains additional delta functions of the momenta that force the incoming and outgoing particles to carry the same set of momenta.
On the other hand $M_{prod}^{(n)}$ contributes to production processes. At the tree level it is not affected by $\epsilon$ and can be obtained by summing all the different Feynman diagrams, imposing $\epsilon=0$ in all the propagators from the beginning.

In a generic quantum field theory $M_{prod}^{(n)}$ can contain singularities arising from on-shell propagating particles in internal lines; 
however, in an integrable model, all such infinities  
must
cancel each other, since otherwise the production amplitude would not vanish\footnote{An exception happens in massless theories where an expansion around a trivial vacuum generally leads in two dimensions to an IR catastrophe. This  generates ambiguities in perturbation theory, and integrable Lagrangians can manifest production at tree-level \cite{Hoare:2018jim,Nappi:1979ig}.}.
This means that if a certain Feynman diagram is singular for particular values of the external momenta, we expect at least one other diagram to become singular for the same choice of the external momenta in such a way that the infinities cancel, making the total $M_{prod}^{(n)}$ free of singularities.

Early studies of the absence of production in integrable models include \cite{Arefeva:1974bk,Goebel:1986na}, which discuss the cancellation mechanism of Feynman diagrams contributing to production processes in the sine-Gordon theory at tree level and at one loop.
Starting from other simple models as reviewed in \cite{Dorey:1996gd} it is then possible to move to more complicated cases. For example the absence of production in the world-sheet scattering has been used to rule out the tree-level integrability for different string theories \cite{Kalousios:2009ey,Wulff:2017hzy,Wulff:2017vhv}. 

A number of recent works have studied the question of how to constrain the structure of higher-point couplings of general massive two-dimensional quantum field theories by imposing the
absence of particle production at tree level \cite{Khastgir:2003au,Gabai:2018tmm,Bercini:2018ysh}.
In particular, in \cite{Gabai:2018tmm} an explicit condition determining higher-point couplings in integrable theories of a single massive
bosonic field in terms of couplings at lower order was found and solved, leading to the `rediscovery' of the sinh-Gordon and Bullough-Dodd theories. The tree-level perturbative integrability of these models was then confirmed by proving the vanishing of off-diagonal scattering processes, again at tree level, for any number of external legs. The fact that, if this procedure works at all, there would only be these two options was already remarked in \cite{Dorey:1996gd}, so this completed the story of tree-level perturbative integrability for theories of a single massive scalar field.
The purpose of this paper is to make progress in extending these results to any two-dimensional bosonic quantum field theory of the form \eqref{eq0_1} by searching for the constraints on $3$- and $4$-point couplings are necessary to have the absence of off-diagonal processes at the tree level. We prove that these constraints are satisfied for the affine Toda models, making use of various properties of the associated root systems. This allows us to provide a general proof of the tree-level integrability of these quantum field theories. While we have not established that the affine Toda theories are the only solutions to these constraints, we also obtain partial results in that direction, including the establishing of the so-called area rule for a particular class of scalar theories.

The paper is structured as follows. In section \ref{sect0} we start
from a theory with a Lagrangian of type~\eqref{eq0_1}
and discuss the logic that needs to be followed to obtain the set of masses and couplings making the model perturbatively integrable at tree level.
In particular we review the
multi-Regge limit technique of \cite{Gabai:2018tmm} to find
a recursion relation on the couplings necessary for perturbative integrability.
We explain that this relation becomes also sufficient  for
integrability if the absence of off-diagonal
processes in $4$-, $5$- and $6$-point interactions can be proved. Then
section \ref{sect2} gives a detailed study of these processes, which
we call the `seeds of integrability', because they are the foundations
on which the induction procedure is built. Starting from $2$ to
$2$ non-diagonal collisions we recall the flipping rule
of \cite{Braden:1990wx} 
and build on it to show that a rule connecting the magnitudes of the $3$-point couplings with the
areas of the corresponding mass triangles needs to hold. In particular it is
shown how the proportionality constants between such couplings and
areas respect the same properties of the structure constants of some
Lie algebra. By the sole requirement of the absence of particle production
we rediscover also the factorisation properties and bootstrap relations connecting 
the different tree-level elements of the S-matrix. We then focus on the class of theories with what we called simply-laced scattering conditions, meaning that they possess
one possible on-shell propagator at a time for a given
choice of external momenta in allowed $4$-point processes, and have
singularities cancelling in pairs in non-allowed collisions. Imposing the absence of $5$-point amplitudes we prove that the proportionality constants connecting the $3$-point couplings in such theories to mass triangle areas, up to an overall interaction scale, are phases. In this way we obtain an area rule for $3$-point couplings that matches that  found in the past for simply-laced affine Toda theories. In section \ref{sect1} we present a general way to express all couplings of affine Toda theories in terms of roots and Lie algebra quantities,  similar in spirit to previous results in \cite{a24, Freeman:1991xw,Fring:1991me,Fring:1992tt}; in addition we show how such results contribute to the cancellation of all the  tree level particle production, concluding a tree-level perturbative proof of the integrability 
in the affine Toda models. 
Finally, the appendices contain some tools relevant to different parts of this work. 
In appendix \ref{App:2} we record a result related to the Cayley-Menger determinant, needed to find the constraints for the cancellation of $4$- and $5$-point processes; through such constraints we recover the area rule previously discussed.
Appendix \ref{App:0_2}
is devoted to the computation of the residues in $5$-point amplitudes, while appendix \ref{App:1} reviews some useful properties of Lie algebras needed for the treatment of the affine Toda models.

\section{Integrability by induction}\label{sect0}

This section starts with a review of
the inductive method introduced in \cite{Gabai:2018tmm} through which, by imposing no tree-level particle production in a particular high energy limit, higher point couplings in perturbatively integrable theories can be determined in terms of the particle masses and lower point couplings. 
In that paper the authors discussed how, if they were able to prove the absence of poles in $4$-, $5$-  and $6$-point processes,  the constraint 
\eqref{eq0_6}
on higher point couplings  found by inductively adopting their multi-Regge  
limit could be a sufficient condition to ensure the tree-level integrability of the theory. However their analysis was only performed completely for the sinh-Gordon and Bullough–Dodd theories, making use of the fact that the scattering amplitudes in theories with just a single type of particle, after imposing the momentum conservation constraint, can be shown to be rational functions of the light cone components of the momenta. 
In general models with an arbitrary number of different particles, each one  with its own mass, this property is no longer true. In addition to reviewing the inductive argument of \cite{Gabai:2018tmm}, we also fill this gap.

\subsection{The logic step by step}
\label{summarizing_the_logic_of_the_paper}
We work in light-cone components so that the on-shell momenta of the external particles in a scattering process can be written, depending on the situation, in the following different forms
\begin{equation}\label{eq0_2}
P_j=(p_j , \bar{p}_j)=m_j (a_j, \frac{1}{a_j})=m_j (e^{\theta_j} , e^{-\theta_j})
\end{equation}
where $j$ labels the external particles and $\theta_j$ are their associated rapidities. 

To find the set of masses and couplings making sums of Feynman diagrams contributing to inelastic processes equal to zero at tree level, we use the following logic: 

(i) We start with a generic $n$-point on-shell tree-level amplitude $M^{(n)}$ depending on $n-2$ momentum
parameters $a_1,\ldots, a_{n-2}$ ($a_{n-1}$ and $a_n$ are fixed by imposing the momentum conservation constraints). 
By exploiting some universal properties of the amplitude we prove that the absence of poles in $M^{(n)}$ (including poles at infinity) implies that such a function is a constant, not depending on the particular choice of $a_1, \dots, a_{n-2}$. This fact is rather easily-seen if the scattering involves particles with the same mass, since in this case $M^{(n)}$ is a rational function of $a_1,\ldots,a_{n-2}$ and by Liouville's theorem the fact that such a function is bounded
implies that it is a constant. This is the case considered in \cite{Gabai:2018tmm}. More work is required if $M^{(n)}$ involves particles with different masses, since after imposing momentum conservation 
we introduce square roots in $M^{(n)}$. This second situation, to our knowledge not covered in elsewhere, is
discussed in section~\ref{Meromorphicity_constant_implication_on_a_torus}.

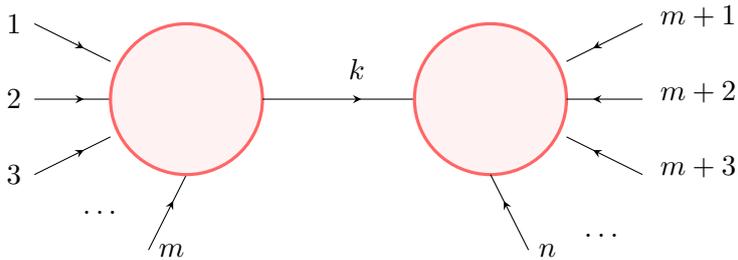
\begin{figure}
\begin{center}
\begin{tikzpicture}
\draw[directed] (6,1) -- (7,0.5);
\draw[directed] (6,0) -- (7,0);
\draw[directed] (6,-1) -- (7,-0.5);
\draw[directed] (7.5,-2) -- (8,-1);
\filldraw[color=red!60, fill=red!5, very thick](8,0) circle (1);

\draw[directed] (9,0) -- (11,0);
\filldraw[black] (10,0.4)  node[anchor=west] {$k$};

\filldraw[color=red!60, fill=red!5, very thick](12,0) circle (1);

\draw[directed] (14,1) -- (13,0.5);
\draw[directed] (14,0) -- (13,0);
\draw[directed] (14,-1) -- (13,-0.5);
\draw[directed] (12.5,-2) -- (12,-1);

\filldraw[black] (5.5,1)  node[anchor=west] {$1$};
\filldraw[black] (5.5,0)  node[anchor=west] {$2$};
\filldraw[black] (5.5,-1)  node[anchor=west] {$3$};
\filldraw[black] (6.5,-1.5)  node[anchor=west] {$\ldots$};
\filldraw[black] (7.5,-2)  node[anchor=west] {$m$};

\filldraw[black] (14.1,1.1)  node[anchor=west] {$m+1$};
\filldraw[black] (14.1,0.1)  node[anchor=west] {$m+2$};
\filldraw[black] (14.1,-0.9)  node[anchor=west] {$m+3$};
\filldraw[black] (13.1,-1.8)  node[anchor=west] {$\ldots$};
\filldraw[black] (12.5,-2)  node[anchor=west] {$n$};

\end{tikzpicture}
\caption{The residue of $M^{(n)}$ in the limit in which the
$k$-propagator is on-shell factorises into the product of two
amplitudes involving scattering processes of less particles. Since at
least one of them is equal to zero for the induction procedure the
full amplitude does not present singularities.}
\label{fig05_1}
\end{center}
\end{figure}

(ii) We then proceed inductively by supposing that $M^{(j)}$ is not just constant but zero for $j \in
\{5,6,7,\dots, n-1 \}$ with $n-1 \ge 6$;
in this case it is possible to prove that
$M^{(n)}$ is a constant. Indeed suppose that $M^{(n)}$ has a pole. This
corresponds to putting a propagator on-shell as shown in figure
\ref{fig05_1} and factorising the amplitude into two on-shell
sub-amplitudes $M^{(m+1)}$ and $M^{(n+1-m)}$. Since $n-1 \ge 6$,
at least one of $M^{(m+1)}$ and $M^{(n+1-m)}$ involves a
scattering process of five or more particles and therefore it is equal
to zero by the induction hypothesis. Since the residue at the pole is
proportional to the product of $M^{(m+1)}$ and $M^{(n+1-m)}$ in the
limit in which the $k$-particle in figure  \ref{fig05_1} goes on shell,
the residue goes to zero and we do not obtain a singularity. 
Due to the previous point the fact that $M^{(n)}$ is free of any singularities implies that such a amplitude is a constant.

(iii) The next step is to determine what constant such an $n$-point amplitude is equal to, and subsequently tune the next higher point coupling $C_{a_1,\dots,a_n}^{(n)}$ in \eqref{eq0_1} in such a way to cancel that constant. To achieve this we adopt the multi-Regge limit defined in \cite{Gabai:2018tmm}, which corresponds to a particular kinematical configuration in which most of the Feynman diagrams are suppressed, making the computation of the amplitude particularly simple. In this manner, by imposing that $M^{(n)}$ has to vanish, we can obtain the $n$-point coupling in terms of the masses and the lower point couplings. The method is described in  detail in section~\ref{Multi_Regge_limit_subsection} where we review the technique of \cite{Gabai:2018tmm}. 
For the sake of clarity we emphasise once again how the induction procedure explained in (ii) and (iii) works in a step by step way. Suppose that we have proved the absence of production in 5- and 6-point processes, which corresponds to have $M^{(5)}=M^{(6)}=0$. Then as explained above the amplitude $M^{(7)}$ cannot have any poles and, by point (i), is a constant, not depending on the particular choice of the external momenta. Since the choice of the kinematics does not affect the value of the 7-point amplitude it is not restrictive at this point to adopt a particularly simple kinematical configuration to tune the value of the 7-point couplings. In this manner we have added a new amplitude to our set of null processes and we get $M^{(5)}=M^{(6)}=M^{(7)}=0$. We can then proceed inductively to all the higher point amplitudes. The relation allowing all the $n$-point couplings to be found (with $n \ge 5$) in terms of the masses, $3$- and $4$-point couplings is given in equation \eqref{eq0_6}, and was first obtained in \cite{Gabai:2018tmm}.

(iv) Finally and most importantly we need to find sets of masses, 3- and 4-point couplings that ensure the
absence of particle production in $5$- and $6$-point processes so as to provide the basis for the induction procedure. 

In the rest of this section we first prove point (i) (in \ref{Meromorphicity_constant_implication_on_a_torus}), and then 
review the multi-Regge limit mentioned in (iii) to obtain the recursion relation for higher point couplings. This is done in \ref{Multi_Regge_limit_subsection}.
Sections \ref{sect2} and \ref{sect1} are mainly focused on point (iv) which provides the basis of the entire induction hypothesis. Defining the set of allowed masses, $3$- and $4$-point couplings making the induction possible corresponds of defining the space of tree-level integrable theories with a Lagrangian of type \eqref{eq0_1}. This opens the door to the possibility of classifying integrable models by imposing the absence of production.

\subsection{Constant amplitudes from absence of singularities} \label{Meromorphicity_constant_implication_on_a_torus}

Consider the scattering of $n$ particles that by convention we assume are all incoming,
with possibly-different masses
$$
P_1 + P_2 + \ldots + P_n \to 0.
$$
Written in the light cone components~\eqref{eq0_2}, the constraints of overall energy-momentum conservation become
\begin{subequations}
\label{energy_momentum_conservation_constraints}
\begin{align}
\label{first_energy_momentum_conservation_constraint}
\sum_{i=1}^n m_i a_i=0, \\
\label{second_energy_momentum_conservation_constraint}
\sum_{i=1}^n \frac{m_i}{a_i}=0.
\end{align}
\end{subequations}
We keep $a_1,\ldots, a_{n-2}$ as independent variables and fix $a_{n-1}$ and $a_n$ in terms of them.
Using \eqref{first_energy_momentum_conservation_constraint} to write $a_{n-1}$ as a negative linear combinations of the other light cone components and substituting the result into \eqref{second_energy_momentum_conservation_constraint} we end up with a quadratic equation for $a_n$. By solving this equation we obtain two different solutions for $a_n$ containing the same rational term in $a_1, \dots, a_{n-2}$ and differing by a square root quantity coming with opposite sign in the two different solutions. It is possible to check, due to the structure of the constraints \eqref{energy_momentum_conservation_constraints}, that the argument of the square root is a homogeneous polynomial, that we call $S^{(2n-4)}$, of order $2n-4$ in the variables $a_1,\dots, a_{n-2}$, where $n$ is the number of scattered particles. Moreover $S^{(2n-4)}$ is a polynomial of order four in each one of the $a_j$, with $j \in \{1,\ldots, n-2\}$. For example in the scattering of $5$ particles $S^{(2n-4)}=S^{(6)}$ is a homogeneous polynomial of order six in $a_1, a_2, a_3$; possible terms of order six admitted in $S^{(6)}$ are $a_1^2 a_2^2 a_3^2$ or $a_1^4 a_2 a_3$ while $a_1^5 a_2$, though it is of order six, is not an admitted term since it is of order $5$ in $a_1$; indeed, for any $n$, $S^{(2n-4)}$ is a polynomial of order four in each one of its variable, so that power of the $a_j$s bigger than four are not admitted.

Without imposing the overall energy-momentum conservation there are many different ways to write an $n$-point amplitude, corresponding to different rational functions in $a_1, \ldots, a_n$. 
However, no matter the initial rational function in $a_1, \ldots, a_n$ we choose, on the kinematical surface conserving the total energy and momentum the amplitude becomes a uniquely defined function of the form
\begin{equation}
\label{amplitude_after_momentum_conservation_with_sqrt}
M^{(n)}(a_1, \dots, a_{n-2})= \frac{Q_1^{(N)} + Q_2^{(N+2-n)} \sqrt{S^{(2n-4)}}}{Q_3^{(N)} + Q_4^{(N+2-n)} \sqrt{S^{(2n-4)}}}.
\end{equation}
The different quantities in the numerator and denominator on the RHS of \eqref{amplitude_after_momentum_conservation_with_sqrt} are homogeneous polynomials in the variables $a_1, \dots, a_{n-2}$ of degree indicated in their superscripts. For example we have 
$S^{(2n-4)}(\lambda a_1, \dots, \lambda a_{n-2})= \lambda^{2n-4} S^{(2n-4)}( a_1, \ldots, a_{n-2})$. The lowercase letter $n$ indicates the number of particles involved in the scattering while the capital letter $N$ depends on $n$ and
the number of Feynman diagrams contributing to the process and it is a generic positive integer. 

It is worth noting that the total amplitude, due to the Lorentz invariance of the Lagrangian \eqref{eq0_1}, does not scale under a global transformation $a_j \to \lambda a_j$. This fact is guaranteed by the matching between the degrees of the different polynomials in \eqref{amplitude_after_momentum_conservation_with_sqrt}.
We note also that in principle the amplitude is not a single valued function for a particular choice of $a_1, \dots, a_{n-2}$, but instead has two  branches of solutions that correspond to taking the positive or negative sign in front of the square root term $\sqrt{S^{(2n-4)}}$. The two signs correspond to the two possible kinematical configurations obtained solving the constraints \eqref{energy_momentum_conservation_constraints} in terms of $a_{n-1}$ and $a_n$.

In the special situation in which the $n$- and $(n-1)$-particle are the same, we have a symmetry in the two solutions obtained by imposing energy-momentum conservation; they can be mapped one into the other by exchanging $a_n$ and $a_{n-1}$. Since the amplitude is also symmetric under this transformation, it has to be invariant by mapping one branch into the other. The effect of this is that the two polynomials $Q_2^{(N+2-n)}$ and $Q_4^{(N+2-n)}$ in front of the square root terms in \eqref{amplitude_after_momentum_conservation_with_sqrt} have to be zero.  Such situations, in which the amplitude continues to be rational also on the kinematical region satisfying the conservation constraints, was already analysed in \cite{Gabai:2018tmm}. In that paper the authors discussed how proving the absence of poles in $M^{(n)}$ is equivalent to proving that it is a constant. Indeed the only rational function without any poles is a polynomial; moreover the only polynomial in $a_1, \ldots, a_{n-2}$ invariant under a scaling $a_j \to \lambda a_j$ is a constant. 

A bit more tricky is the case in which the $n$- and $(n-1)$-particle have different masses. In this case the square root could actually be present in \eqref{amplitude_after_momentum_conservation_with_sqrt} and 
Liouville's theorem cannot be directly applied to show
that $M^{(n)}$ is a constant. This  was discussed in \cite{Braden:1991vz} in the context of the cancellation of inelastic $2$ to $2$ events in simply-laced affine Toda models. First the authors checked, in spite of the potential
square root, that the amplitude is in fact meromorphic in the light cone components surviving  momentum conservation. This required a set of constraints on the lower point couplings to hold, that were verified for the different affine Toda theories. They then proved that all the poles due to on-shell propagating bound states cancel against each other, making the $4$-point amplitude a constant. However such an approach, that requires a check of the absence of branch points in the amplitude before verifying the absence of poles, becomes more and more complicated as the number of scattered particles increases. Fortunately we will show in a moment how this first step in the logic used in \cite{Braden:1991vz} is actually unnecessary since the absence of poles in the amplitude automatically implies that $M^{(n)}$ is a constant. This means that the constraints obtained by imposing that $M^{(n)}$ is bounded in both the two kinematical configurations obtained by solving \eqref{energy_momentum_conservation_constraints} imply the constraints for the branch points cancellations.
To prove this fact we work with one free variable at a time.

We define $a_1=z$ to be our independent free parameter and we keep $a_2,\ldots, a_{n-2}$ fixed in \eqref{amplitude_after_momentum_conservation_with_sqrt}.
The term $S^{(2n-4)}(z,a_2,\ldots, a_{n-2})$ appearing below the square root, for fixed values of $a_2,\ldots,a_{n-2}$, as previously mentioned, is a polynomial of order four in $z$; 
it is therefore proportional to
\begin{equation}
\label{S_written_in_terms_of_z_with_branch_points_explicit}
S(z)=(z-z_1)(z-z_2)(z-z_3)(z-z_4),
\end{equation}
where $z_1$, $z_2$, $z_3$ and $z_4$ are four branch points depending on the particular values of $a_2,\ldots, a_{n-2}$.
This makes $M^{(n)}$ a rational function of the two arguments $z$ and $\sqrt{S(z)}$. 
We can choose the branch cuts of $\sqrt{S(z)}$ to be any copy of non-intersecting segments connecting the branch points in pairs. For example we can choose one branch cut to connect $z_1$ with $z_2$ and the other one to connect $z_3$ with $z_4$. By circling around the branch points we move from one cover of the complex plane, corresponding to one kinematical solution of the energy-momentum conservation \eqref{energy_momentum_conservation_constraints}, to a second on which the sign of $\sqrt{S(z)}$ is flipped. The domain of the amplitude is therefore a two-sheeted covering of $\mathbb{C}$, $\Sigma_1$ and $\Sigma_2$, each one corresponding to one of the two kinematical configurations obtained by solving \eqref{energy_momentum_conservation_constraints}. If we now open the cuts and we add the point at infinity we can see that the double cover of the complex plane is homeomorphic to a torus, with $\Sigma_1$ and $\Sigma_2$ corresponding to the two halves of the doughnut. All this comes from standard considerations on Riemann surfaces as described for example in~\cite{Complex_Functions}. To be more specific let us explain how the map between the torus and the double cover of $\mathbb{C}$ is realized. 
If we have a generic fourth-order polynomial $S(z)$ with non-equal roots it is possible, by applying a conformal mapping
% \DDD{Is it right calling it conformal transformation?}
\begin{equation}
S'(z') \equiv \frac{(z'+\delta)^4}{(\beta-\alpha \delta)^2} S(z),
\end{equation}
with
\begin{equation}
z=\frac{\alpha z' + \beta}{z' + \delta},
\end{equation}
to map one of the branch points to infinity; in this way, by properly tuning $\alpha$, $\beta$ and $\gamma$, the new polynomial $S'(z')$ becomes of the form
\begin{equation}
\label{Weierstrass_form_of_the_polynomial}
S'(z')=4z'^3 - g_2 z'-g_3,
\end{equation}
for some values of $g_2$ and $g_3$. The explicit values of $\alpha$, $\beta$ and $\gamma$ mapping $S(z)$ into \eqref{Weierstrass_form_of_the_polynomial} can be easily derived  (see, for example, \cite{Mittag_Leffler_Complex} for a full derivation).

From this transformation it follows that any rational function in the variables $z$ and $\sqrt{S(z)}$, where $S$ is a generic polynomial of order four presenting different roots, can be written as a rational function $f$ in $z'$ and $\sqrt{S'(z')}$. In this case we have
\begin{equation}
\label{conversion_of_rational_amplitude_Mn_to_a_function_f_appendix}
M^{(n)}(z,\sqrt{S(z)})=M^{(n)}\Bigl( \frac{\alpha z' + \beta}{z' + \delta},\frac{(\beta-\alpha \delta)}{(z'+\delta)^2} \sqrt{S'(z')} \Bigr)\equiv f(z', \sqrt{S'(z')}).
\end{equation}
The initial amplitude $M^{(n)}$ is not single valued on $\mathbb{C}$, indeed for each point $z$ there are two possible values of $\sqrt{S(z)}$ defined over two different Riemann sheets. However after the conversion \eqref{conversion_of_rational_amplitude_Mn_to_a_function_f_appendix} it is easy to map such a double cover of the complex plane to a torus. This is achieved using the parametrization $z'=\wp(x)$, where the symbol $\wp$ represents a Weierstrass elliptic function identified by  the parameters $g_2$ and $g_3$ in \eqref{Weierstrass_form_of_the_polynomial}. To each value $z'$ on the complex plane there are two different points $x_1$ and $x_2$ on a torus such that $\wp(x_1)=\wp(x_2)=z'$. Moreover, the lattice $\Omega$ associated to $\wp$ is completely defined by a pair of numbers $g_2$ and $g_3$ in terms of which $\wp$ satisfies the following differential equation
$$
(\wp(x)')^2= 4 \wp(x)^3 - g_2 \wp(x) -g_3.
$$
Then the values of the derivative of $\wp$ at the points $x_1$ and $x_2$ are one the opposite of the other and correspond to the two branches of solutions of $\sqrt{S'(z')}$. The starting double valued function $M^{(n)}(z,\sqrt{S(z)})$ can therefore be mapped into a single valued periodic meromorphic function
$$
f(\wp(x), \wp(x)')
$$
defined on a torus. This implies that if $M^{(n)}(z,\sqrt{S(z)})$ does not present any pole in both its Riemann sheets, then $f(\wp(x), \wp(x)')$ has to be bounded on the torus and by Liouville's theorem it needs to be a constant not depending on $x$.

Therefore if the amplitude does not present any singularity in $a_1$ in both the Riemann sheets, no matter the choice of $a_2, \dots,a_{n-2}$, then it has to be a constant in $a_1$. This means
$$
\frac{\partial}{\partial a_1} M^{(n)}(a_1,\ldots, a_{n-2})=0.
$$
Repeating the previous analysis one variable at a time we conclude that if $M^{(n)}$ is completely free of singularities, both at finite values of the momenta and at infinity, we obtain
$$
\frac{\partial}{\partial a_1} M^{(n)}(a_1,\ldots, a_{n-2})=\frac{\partial}{\partial a_2} M^{(n)}(a_1,\ldots, a_{n-2})=\ldots=\frac{\partial}{\partial a_{n-2}} M^{(n)}(a_1,\ldots, a_{n-2})=0.
$$
We conclude that a necessary and sufficient condition an amplitude has to satisfy in order to be constant is to have a bounded absolute value. This concludes the proof of point (i). 

The fact that the Riemann surface $\Lambda$ over which the amplitude is defined has the topology of a torus ensures that there exists a map between such a torus to $\Lambda$ making the amplitude single valued on the doughnut. Previously we explicitly wrote a parametrization on the torus using the Weierstrass $\wp$ elliptic function anyway it is possible applying this map by using other elliptic functions. In \ref{Elastic_scattering_from_degenerate_torus} we explain how this map can be realised in a simple example using the Jacobi function $sn$. From it we derive the elastic scattering as the limit case in which the torus becomes degenerate.

\subsection{Elastic scattering from degenerate doughnuts} 
\label{Elastic_scattering_from_degenerate_torus}
A useful exercise to check that things go as expected is parameterizing  a four-point inelastic scattering on a torus, and recovering the case in which the initial and final masses are equal in a second time. We discover that in the limit in which the initial and final particles are equal, the torus over which the amplitude is a single valued meromorphic function becomes degenerate. In this limit the torus splits into two separated regions, over which the amplitude is still meromorphic, but cannot anymore be analytically continued from one region to the other by circling around the branch points. The two values of the amplitude over these two separate regions correspond to two distinct functions that represent respectively a transmission and a reflection event.

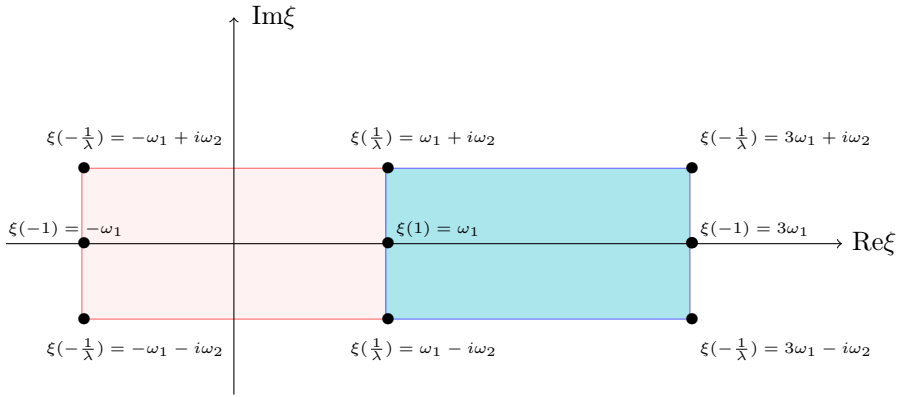
\begin{figure}
\begin{center}
    
\begin{tikzpicture}
\filldraw[color=red!60, fill=red!5](5 ,-1) (-2,-1) -- (+2,-1) -- (+2,1) --(-2,1) -- (-2,-1);
\filldraw[color=blue!60, fill=blizzardblue](5 ,-1) (2,-1) -- (+6,-1) -- (+6,1) --(2,1) -- (2,-1);
\draw[->] (-3,0) -- (8,0);
\draw[->] (0,-2) -- (0,3);
\filldraw[black] (8,0)  node[anchor=west] {\small{Re$\xi$}};
\filldraw[black] (0.1,3)  node[anchor=west] {\small{Im$\xi$}};

\filldraw[black] (-2.2,-1)  node[anchor=west] {$\bullet$};
\filldraw[black] (+1.8,-1)  node[anchor=west] {$\bullet$};
\filldraw[black] (+1.8,1)  node[anchor=west] {$\bullet$};
\filldraw[black] (-2.2,1)  node[anchor=west] {$\bullet$};
\filldraw[black] (-2.2,0)  node[anchor=west] {$\bullet$};
\filldraw[black] (1.8,0)  node[anchor=west] {$\bullet$};
\filldraw[black] (5.8,0)  node[anchor=west] {$\bullet$};
\filldraw[black] (5.8,1)  node[anchor=west] {$\bullet$};
\filldraw[black] (5.8,-1)  node[anchor=west] {$\bullet$};
\filldraw[black] (5.8,0)  node[anchor=west] {$\bullet$};

\filldraw[black] (-2.6,-1.4)  node[anchor=west] {\tiny{$\xi(-\frac{1}{\lambda})=-\omega_1-i\omega_2$}};
\filldraw[black] (-2.6,1.4)  node[anchor=west] {\tiny{$\xi(-\frac{1}{\lambda})=-\omega_1+i\omega_2$}};
\filldraw[black] (-3.1,0.2)  node[anchor=west] {\tiny{$\xi(-1)=-\omega_1$}};
\filldraw[black] (2,0.2)  node[anchor=west] {\tiny{$\xi(1)=\omega_1$}};
\filldraw[black] (1.4,-1.4)  node[anchor=west] {\tiny{$\xi(\frac{1}{\lambda})=\omega_1-i \omega_2$}};
\filldraw[black] (1.4,1.4)  node[anchor=west] {\tiny{$\xi(\frac{1}{\lambda})=\omega_1+i \omega_2$}};
\filldraw[black] (6,-1.4)  node[anchor=west] {\tiny{$\xi(-\frac{1}{\lambda})=3 \omega_1-i \omega_2$}};
\filldraw[black] (6,1.4)  node[anchor=west] {\tiny{$\xi(-\frac{1}{\lambda})=3 \omega_1+i \omega_2$}};
\filldraw[black] (6,0.2)  node[anchor=west] {\tiny{$\xi(-1)=3 \omega_1$}};

\end{tikzpicture}
\caption{Torus surface over which the 4-point amplitude takes values. The red and blue regions correspond to the two different kinematical configurations obtained by solving the momentum conservation constraints and are the two halves of the torus.}
\label{From_C_to_Torus_JacobiSN_map}
\end{center}
\end{figure}

Let us consider the scattering of two initial states, $(a, b)$, carrying respectively momenta $(P_a, P_b)$; suppose that after the scattering the particles change their types into $(c,d)$, with respective momenta $(P_c,P_d)$
$$
P_a+P_b \to P_c+P_d. 
$$
We define the Mandelstam variables as follows
$$ s=(P_a+P_b)^2 \hspace{3mm},\hspace{3mm}  t=(P_a-P_c)^2 \hspace{3mm},\hspace{3mm} u=(P_a-P_d)^2.
$$
Considering the external particles on-shell, in two dimensions the values of $t$ and $u$ can be completely fixed in terms of $s$
\begin{equation}
\begin{split}
t&=\frac{m_a^2+m_b^2+m_c^2+m_d^2-s}{2}+\frac{(m^2_a-m^2_b)(m^2_d-m^2_c)-\Sigma(s)}{2s}\\
u&=\frac{m_a^2+m_b^2+m_c^2+m_d^2-s}{2}+\frac{(m^2_a-m^2_b)(m^2_c-m^2_d)+\Sigma(s)}{2s},
\end{split}
\end{equation}
where $\Sigma$ is a double-valued function of $s$ 
\begin{equation}
\label{Sigma_si_branch_points}
\Sigma(s)=\sqrt{(s-s_1) (s-s_2) (s-s_3) (s-s_4)}
\end{equation}
presenting two Riemann sheets. The branch point positions are given by
$$
s_1=(m_a+m_b)^2 \hspace{2mm}, \hspace{2mm} s_2=(m_a-m_b)^2 \hspace{2mm}, \hspace{2mm} s_3=(m_c+m_d)^2 \hspace{2mm} , \hspace{2mm} s_4=(m_c-m_d)^2.
$$
We can fix the branch cuts along the two segments on the real axis of the $s$ plane connecting $s_1$ with $s_3$ and $s_2$ with $s_4$.
Then $\Sigma$ is defined over a double cover of the complex plane and we can move from one to the other sheet over which $\Sigma$ takes values by rotating of a $2 \pi$  angle around the branch points. In order to study an elastic process it is our intention to take the limit in which the length of the branch cuts shrink to zero, which is $m_d \to m_a$ and $m_c \to m_b$, in such a way to close the tunnels between the two Riemann sheets. 
To this end we choose particular values of the masses given by
$$
m_a=\mu \cos \alpha \hspace{3mm} , \hspace{3mm} m_b=\mu \sin \alpha
$$
and
$$
m_c=\mu \sin \beta \hspace{3mm} , \hspace{3mm} m_d=\mu \cos \beta. 
$$
In the limit $\beta \to \alpha$ we obtain the same values for the initial and final masses, moreover with these special values of the masses we have $m_a^2+m_b^2=m_c^2+m_d^2=\mu^2$. Therefore if we apply the change of variable 
$$
s=m_a^2+m_b^2+2 m_a m_b y,
$$
by defining $\lambda=\frac{m_a m_b}{m_c m_d}$, we can express the relation \eqref{Sigma_si_branch_points} in terms of $y$
\begin{equation}
\label{Sigma_as_function_of_y_elliptic_functions}
\Sigma(y)= 4 m_a m_b m_c m_d\sqrt{(1-y^2) (1-\lambda^2 y^2)}.
\end{equation}
If we look at \eqref{Sigma_as_function_of_y_elliptic_functions} we note that the argument under square root is a known expression appearing in Jacobi elliptic functions. Using a Schwarz–Christoffel transformation defined by
\begin{equation}
\label{Schwarz_Christoffel_transformation_inverse_Jacobi}
\xi(y)=\int_0^y \frac{dt}{\sqrt{(1-t^2) (1-\lambda^2 t^2)}}.
\end{equation}
we can map the entire complex plane, over which $y$ takes values, into a rectangle composed by points $\xi$. 
For each $y \in \mathbb{C}$ there are two possible values for $\xi(y)$ depending on the Riemann sheet over which the function $w(y)=\frac{1}{\sqrt{(1-y^2) (1-\lambda^2 y^2)}}$ is integrated over. Integrating over the sheet where $w(0)=1$ we map the entire complex plane into the red rectangle in figure \ref{From_C_to_Torus_JacobiSN_map}. The branch points, corresponding to the values of $y$ at which $w$ is singular, are mapped into the bullets located on the frame of the red rectangle.
Similarly by performing the integration \eqref{Schwarz_Christoffel_transformation_inverse_Jacobi} over the surface on which $w(0)=-1$ and translating by a $\lambda$ dependent parameter $2\omega_1$ we map the second cover of $\mathbb{C}$ into the blue rectangle in figure \ref{From_C_to_Torus_JacobiSN_map}. The union of the red and blue rectangles correspond to a torus having periodicity along the real and imaginary axis given respectively by $4 \omega_1$ and $2 \omega_2$, where the quantities $\omega_1$, $\omega_2$ depend on the ratio $\lambda$ between the masses. The function $\Sigma$ which presents branch cuts on the complex plane and it is not single-valued on $\mathbb{C}$, is a meromorphic function on the torus \ref{From_C_to_Torus_JacobiSN_map}. If we want to map back the double cover of the complex plane from the torus we need to invert the integral expression \eqref{Schwarz_Christoffel_transformation_inverse_Jacobi}. This generates a known function in complex analysis known as Jacobi elliptic sine $sn(x,\lambda^2)$. We can parametrize $y=sn(x,\lambda^2)$, then for each $y \in \mathbb{C}$ there are two values of $x$, $x_r$ taking  value on the red rectangle, and $x_b$ lying on on the blue rectangle, such that $sn(x_r,\lambda^2)=sn(x_b,\lambda^2)=y$.
On the other hand the Jacobi sine satisfies the following differential equation
$$
(y')^2=(1-y^2) (1-\lambda^2 y^2).
$$
so that at the point $x_r$ and $x_b$ at which $y$ takes the same value we get $y'(x_r)=-y'(x_b)$. It is clear now why $\Sigma$ is meromorphic on the torus; up to the multiplicative factor $4 m_a m_b m_c m_d$ it is exactly the derivative of the Jacobi sine function, so that the Mandelstam variables can be entirely parametrized in terms of $sn(x,\lambda^2)$ and $sn'(x,\lambda^2)$, with $x$ taking values on the two halved of the torus in figure \ref{From_C_to_Torus_JacobiSN_map}, each one corresponding to a cover of $\mathbb{C}$.

To obtain the degenerate limit in which the initial and final masses are the same, we approach the parameter $\lambda$ to one from the left, $\lambda\to 1^-$. In this limit the quantity $\omega_1$ tends to infinity and the red and blue regions in figure \ref{From_C_to_Torus_JacobiSN_map} become infinite far away from one another. This is the case in which the branch points collide in pairs $s_3=s_1$, $s_4=s_2$ and $\Sigma$ reduces to
\begin{equation}
\label{degenerate_situation_for_sigma_s1s2}
\Sigma(s)=\pm (s-s_1)(s-s_2).
\end{equation}
\begin{figure}[htp]
    \centering
    \includegraphics[width=\linewidth]{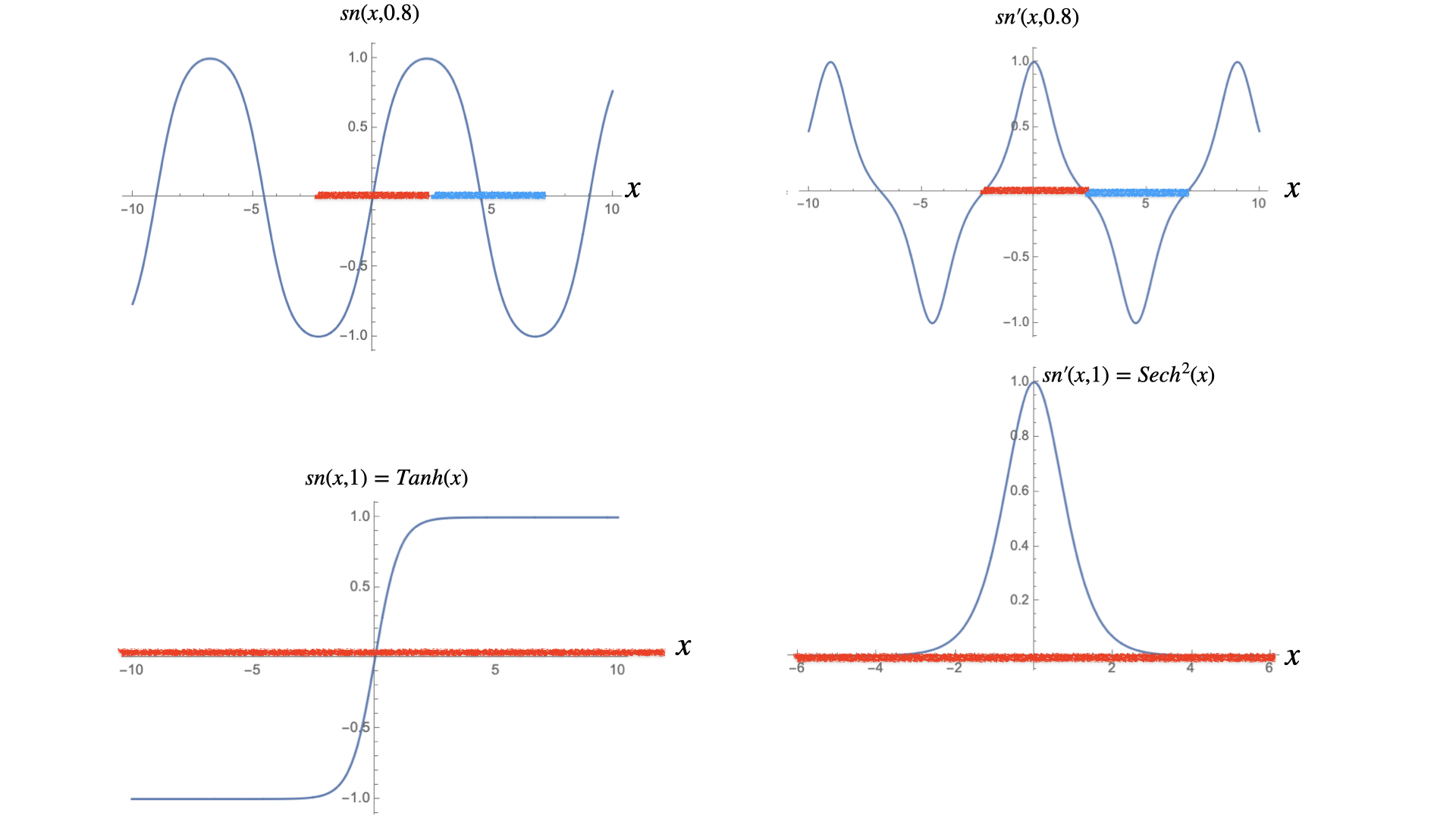}
    \caption{Example of Jacobi elliptic function $sn(x,\lambda^2)$ and its derivative, plotted for real values of $x$. In the first row, the red and blue segments correspond to the intersection of the red and blue regions of the torus in figure \ref{From_C_to_Torus_JacobiSN_map} with the real axis. Approaching $\lambda^2$ to one, as shown in the second row, we push the blue part infinitely far away, and the Jacobi elliptic functions become hyperbolic functions defined over half of the torus.}
    \label{Example_of_degenerate_situation_in_which_real_periodicity_tends_to_infinity}
\end{figure}
The two disjoint red and blue regions on the torus correspond to the different choices of sign in \eqref{degenerate_situation_for_sigma_s1s2};
once we pick up a sign, we cannot move to the region with opposite sign since the branch cuts have collapsed to points and the torus has become  degenerate. The limit $\lambda \to 1$ corresponds to the special case in which Jacobi elliptic functions reduce to hyperbolic functions. In this situation one of the two doughnut periodicities (that one along the imaginary axis) is preserved while the other one becomes infinite. In figure \ref{Example_of_degenerate_situation_in_which_real_periodicity_tends_to_infinity} we show an example of degeneracy by plotting the functions $sn(x,\lambda^2)$ and $sn'(x,\lambda^2)$ for real values of $x$. We see that in the particular situation $\lambda=1$ the red half part of the torus occupies the entire $x$ axis not leaving possibility to pass in a continuous way to the blue part of the doughnut. To do that we would have to flip the sign in \eqref{degenerate_situation_for_sigma_s1s2} which corresponds of sending $sn'(x,1)=Sech^2(x)\to -Sech^2(x)$.

The red and the blue regions correspond to the two kinematical configurations in which transmission and reflection occur. In the first situation the scattering is elastic; it is the case in which the initial and final sets, $(a,b)$ and $(c,d)$, are the same and both carry the same momenta $(P_d=P_a,P_c=P_b) $. In this configuration the Mandelstam variable $u$ is equal to zero while $t=2m_a^2+2m_b^2-s$. If the theory is integrable, the amplitude in this region does not have to be zero since the scattering conserves all the quantum numbers. Contrarily the blue half part of the degenerate torus corresponds to a reflection process in which the incoming and outgoing particles have different momenta. This is the region in which we expect the amplitude (that is not the same function defined on the red cover, since we cannot analytically continue from one domain to the other) to be zero and where all the singularities coming from different Feynman diagrams should cancel.

Having proved that the absence of poles implies a constant amplitude,
we can now move on to find the constraints on the masses and couplings leading to perturbative integrability at tree level. Point (ii) in section~\ref{summarizing_the_logic_of_the_paper} was already discussed and does not need further study. It is based on the consideration that once we have a certain number of zero amplitudes $M^{(5)}=M^{(6)}=\dots=M^{(n-1)}=0$ by having properly tuned the masses and the couplings ($C^{(3)}, C^{(4)},\dots, C^{(n-1)})$ in \eqref{eq0_1}, we can fix the next $n$-point coupling uniquely by requiring the corresponding $n$-point scattering process to be zero. This mechanism and its first steps were discussed in \cite{Dorey:1996gd} but it was only in \cite{Gabai:2018tmm} that it was explained how to handle the problem to all orders, making use of a particular multi-Regge limit of the amplitude. We proceed therefore to review the analysis carried out in \cite{Gabai:2018tmm} to find one by one the higher point couplings; this is what we called point (iii) in the logic summarized in section~\ref{summarizing_the_logic_of_the_paper}.

\subsection{The multi-Regge limit}
\label{Multi_Regge_limit_subsection}
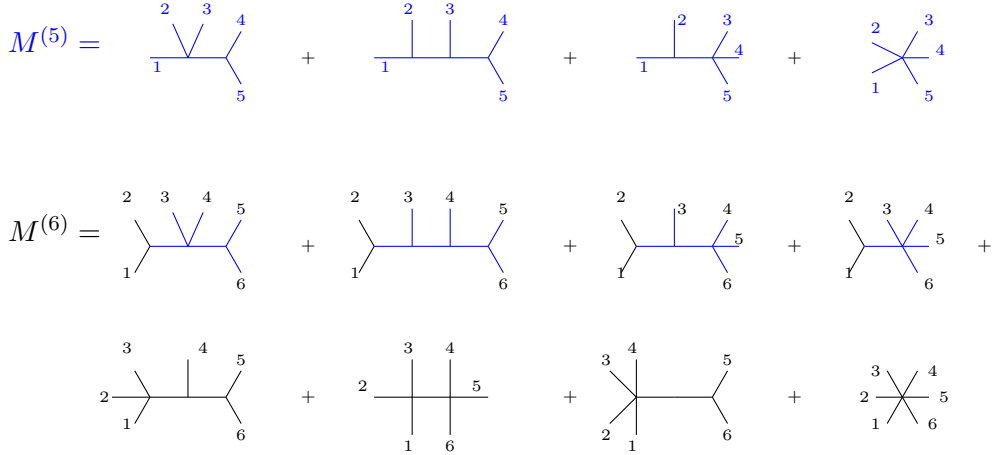
\begin{figure}
\begin{tikzpicture}
\tikzmath{\y=0.5;}

%5-POINT PROCESS
\filldraw[blue] (-5*\y,0.5*\y)  node[anchor=west] {$M^{(5)}=$};

\draw[blue] (-1*\y,0) -- (0,0);
\draw[blue] (-0.4*\y,0.9*\y) -- (0,0);
\draw[blue] (0.4*\y,0.9*\y) -- (0,0);
\draw[blue] (1*\y,0*\y) -- (0,0);
\draw[blue] (1*\y,0*\y) -- (1.4*\y,0.7*\y);
\draw[blue] (1*\y,0*\y) -- (1.4*\y,-0.7*\y);

\filldraw[blue] (-1.2*\y,-0.25*\y)  node[anchor=west] {\tiny{$1$}};
\filldraw[blue] (-1*\y,1.3*\y)  node[anchor=west] {\tiny{$2$}};
\filldraw[blue] (0.1*\y,1.3*\y)  node[anchor=west] {\tiny{$3$}};
\filldraw[blue] (1*\y,1*\y)  node[anchor=west] {\tiny{$4$}};
\filldraw[blue] (1*\y,-1*\y)  node[anchor=west] {\tiny{$5$}};

\filldraw[black] (2.7*\y,0*\y)  node[anchor=west] {\tiny{$+$}};

\draw[blue] (4.9*\y,0) -- (5.9*\y,0);
\draw[blue] (5.9*\y,1*\y) -- (5.9*\y,0);
\draw[blue] (7.9*\y,0*\y) -- (5.9*\y,0);
\draw[blue] (6.9*\y,1*\y) -- (6.9*\y,0);
\draw[blue] (7.9*\y,0*\y) -- (8.3*\y,0.7*\y);
\draw[blue] (7.9*\y,0*\y) -- (8.3*\y,-0.7*\y);

\filldraw[blue] (4.8*\y,-0.25*\y)  node[anchor=west] {\tiny{$1$}};
\filldraw[blue] (5.4*\y,1.3*\y)  node[anchor=west] {\tiny{$2$}};
\filldraw[blue] (6.5*\y,1.3*\y)  node[anchor=west] {\tiny{$3$}};
\filldraw[blue] (7.9*\y,1*\y)  node[anchor=west] {\tiny{$4$}};
\filldraw[blue] (7.9*\y,-1*\y)  node[anchor=west] {\tiny{$5$}};

\filldraw[black] (9.6*\y,0*\y)  node[anchor=west] {\tiny{$+$}};

\draw[blue] (11.8*\y,0) -- (12.8*\y,0);
\draw[blue] (12.8*\y,1*\y) -- (12.8*\y,0);
\draw[blue] (12.8*\y,0) -- (13.8*\y,0);
\draw[blue] (14.5*\y,0) -- (13.8*\y,0);
\draw[blue] (13.8*\y,0*\y) -- (14.2*\y,0.7*\y);
\draw[blue] (13.8*\y,0*\y) -- (14.2*\y,-0.7*\y);

\filldraw[blue] (11.6*\y,-0.25*\y)  node[anchor=west] {\tiny{$1$}};
\filldraw[blue] (12.6*\y,1*\y)  node[anchor=west] {\tiny{$2$}};
\filldraw[blue] (13.8*\y,1*\y)  node[anchor=west] {\tiny{$3$}};
\filldraw[blue] (14.1*\y,0.2*\y)  node[anchor=west] {\tiny{$4$}};
\filldraw[blue] (13.8*\y,-1*\y)  node[anchor=west] {\tiny{$5$}};

\filldraw[black] (15.5*\y,0*\y)  node[anchor=west] {\tiny{$+$}};

\draw[blue] (18*\y,0.4*\y) -- (18.8*\y,0);
\draw[blue] (18*\y,-0.4*\y) -- (18.8*\y,0);
\draw[blue] (19.5*\y,0) -- (18.8*\y,0);
\draw[blue] (18.8*\y,0*\y) -- (19.2*\y,0.7*\y);
\draw[blue] (18.8*\y,0*\y) -- (19.2*\y,-0.7*\y);

\filldraw[blue] (17.7*\y,-0.8*\y)  node[anchor=west] {\tiny{$1$}};
\filldraw[blue] (17.7*\y,0.8*\y)  node[anchor=west] {\tiny{$2$}};
\filldraw[blue] (19.1*\y,1*\y)  node[anchor=west] {\tiny{$3$}};
\filldraw[blue] (19.4*\y,0.2*\y)  node[anchor=west] {\tiny{$4$}};
\filldraw[blue] (19.1*\y,-1*\y)  node[anchor=west] {\tiny{$5$}};

%6-POINT PROCESS
\filldraw[black] (-5*\y,-4.5*\y)  node[anchor=west] {$M^{(6)}=$};

\draw[] (-1*\y,-5*\y) -- (-1.4*\y,-5.7*\y);
\draw[] (-1*\y,-5*\y) -- (-1.4*\y,-4.3*\y);
\draw[blue] (-1*\y,-5*\y) -- (0,-5*\y);
\draw[blue] (-0.4*\y,-4.1*\y) -- (0,-5*\y);
\draw[blue] (0.4*\y,-4.1*\y) -- (0,-5*\y);
\draw[blue] (1*\y,-5*\y) -- (0,-5*\y);
\draw[blue] (1*\y,-5*\y) -- (1.4*\y,-4.3*\y);
\draw[blue] (1*\y,-5*\y) -- (1.4*\y,-5.7*\y);

\filldraw[] (-2*\y,-5.7*\y)  node[anchor=west] {\tiny{$1$}};
\filldraw[] (-2*\y,-3.7*\y)  node[anchor=west] {\tiny{$2$}};
\filldraw[] (-1*\y,-3.7*\y)  node[anchor=west] {\tiny{$3$}};
\filldraw[] (0.1*\y,-3.7*\y)  node[anchor=west] {\tiny{$4$}};
\filldraw[] (1*\y,-4*\y)  node[anchor=west] {\tiny{$5$}};
\filldraw[] (1*\y,-6*\y)  node[anchor=west] {\tiny{$6$}};

\filldraw[black] (2.7*\y,-5*\y)  node[anchor=west] {\tiny{$+$}};

\draw[] (4.9*\y,-5*\y) -- (4.5*\y,-5.7*\y);
\draw[] (4.9*\y,-5*\y) -- (4.5*\y,-4.3*\y);
\draw[blue] (4.9*\y,-5*\y) -- (5.9*\y,-5*\y);
\draw[blue] (5.9*\y,-4*\y) -- (5.9*\y,-5*\y);
\draw[blue] (7.9*\y,-5*\y) -- (5.9*\y,-5*\y);
\draw[blue] (6.9*\y,-4*\y) -- (6.9*\y,-5*\y);
\draw[blue] (7.9*\y,-5*\y) -- (8.3*\y,-4.3*\y);
\draw[blue] (7.9*\y,-5*\y) -- (8.3*\y,-5.7*\y);

\filldraw[] (4*\y,-5.7*\y)  node[anchor=west] {\tiny{$1$}};
\filldraw[] (4*\y,-3.7*\y)  node[anchor=west] {\tiny{$2$}};
\filldraw[] (5.4*\y,-3.7*\y)  node[anchor=west] {\tiny{$3$}};
\filldraw[] (6.5*\y,-3.7*\y)  node[anchor=west] {\tiny{$4$}};
\filldraw[] (7.9*\y,-4*\y)  node[anchor=west] {\tiny{$5$}};
\filldraw[] (7.9*\y,-6*\y)  node[anchor=west] {\tiny{$6$}};

\filldraw[black] (9.6*\y,-5*\y)  node[anchor=west] {\tiny{$+$}};

\draw[] (11.8*\y,-5*\y) -- (11.4*\y,-5.7*\y);
\draw[] (11.8*\y,-5*\y) -- (11.4*\y,-4.3*\y);
\draw[blue] (11.8*\y,-5*\y) -- (12.8*\y,-5*\y);
\draw[blue] (12.8*\y,-4*\y) -- (12.8*\y,-5*\y);
\draw[blue] (12.8*\y,-5*\y) -- (13.8*\y,-5*\y);
\draw[blue] (14.5*\y,-5*\y) -- (13.8*\y,-5*\y);
\draw[blue] (13.8*\y,-5*\y) -- (14.2*\y,-4.3*\y);
\draw[blue] (13.8*\y,-5*\y) -- (14.2*\y,-5.7*\y);

\filldraw[] (11*\y,-5.7*\y)  node[anchor=west] {\tiny{$1$}};
\filldraw[] (11*\y,-3.7*\y)  node[anchor=west] {\tiny{$2$}};
\filldraw[] (12.6*\y,-4*\y)  node[anchor=west] {\tiny{$3$}};
\filldraw[] (13.8*\y,-4*\y)  node[anchor=west] {\tiny{$4$}};
\filldraw[] (14.1*\y,-4.8*\y)  node[anchor=west] {\tiny{$5$}};
\filldraw[] (13.8*\y,-6*\y)  node[anchor=west] {\tiny{$6$}};

\filldraw[black] (15.5*\y,-5*\y)  node[anchor=west] {\tiny{$+$}};

\draw[] (17.8*\y,-5*\y) -- (17.4*\y,-5.7*\y);
\draw[] (17.8*\y,-5*\y) -- (17.4*\y,-4.3*\y);
\draw[blue] (17.8*\y,-5*\y) -- (18.8*\y,-5*\y);
\draw[blue] (18.8*\y,-5*\y) -- (18.4*\y,-4.3*\y);
\draw[blue] (19.5*\y,-5*\y) -- (18.8*\y,-5*\y);
\draw[blue] (18.8*\y,-5*\y) -- (19.2*\y,-4.3*\y);
\draw[blue] (18.8*\y,-5*\y) -- (19.2*\y,-5.7*\y);

\filldraw[] (17*\y,-5.7*\y)  node[anchor=west] {\tiny{$1$}};
\filldraw[] (17*\y,-3.7*\y)  node[anchor=west] {\tiny{$2$}};
\filldraw[] (18*\y,-4*\y)  node[anchor=west] {\tiny{$3$}};
\filldraw[] (19.1*\y,-4*\y)  node[anchor=west] {\tiny{$4$}};
\filldraw[] (19.4*\y,-4.8*\y)  node[anchor=west] {\tiny{$5$}};
\filldraw[] (19.1*\y,-6*\y)  node[anchor=west] {\tiny{$6$}};

\filldraw[black] (20.5*\y,-5*\y)  node[anchor=west] {\tiny{$+$}};

\draw[] (-1*\y,-9*\y) -- (-2*\y,-9*\y);
\draw[] (-1*\y,-9*\y) -- (-1.4*\y,-9.7*\y);
\draw[] (-1*\y,-9*\y) -- (-1.4*\y,-8.3*\y);
\draw[] (-1*\y,-9*\y) -- (0,-9*\y);
\draw[] (-0*\y,-8*\y) -- (0,-9*\y);
\draw[] (1*\y,-9*\y) -- (0,-9*\y);
\draw[] (1*\y,-9*\y) -- (1.4*\y,-8.3*\y);
\draw[] (1*\y,-9*\y) -- (1.4*\y,-9.7*\y);

\filldraw[] (-2*\y,-9.7*\y)  node[anchor=west] {\tiny{$1$}};
\filldraw[] (-2.6*\y,-9*\y)  node[anchor=west] {\tiny{$2$}};
\filldraw[] (-2*\y,-7.7*\y)  node[anchor=west] {\tiny{$3$}};
\filldraw[] (0*\y,-7.7*\y)  node[anchor=west] {\tiny{$4$}};
\filldraw[] (1*\y,-8*\y)  node[anchor=west] {\tiny{$5$}};
\filldraw[] (1*\y,-10*\y)  node[anchor=west] {\tiny{$6$}};

\filldraw[black] (2.7*\y,-9*\y)  node[anchor=west] {\tiny{$+$}};

\draw[] (5.9*\y,-8*\y) -- (5.9*\y,-10*\y);
\draw[] (7.9*\y,-9*\y) -- (4.9*\y,-9*\y);
\draw[] (6.9*\y,-8*\y) -- (6.9*\y,-10*\y);

\filldraw[] (5.4*\y,-10.3*\y)  node[anchor=west] {\tiny{$1$}};
\filldraw[] (4.2*\y,-8.7*\y)  node[anchor=west] {\tiny{$2$}};
\filldraw[] (5.4*\y,-7.7*\y)  node[anchor=west] {\tiny{$3$}};
\filldraw[] (6.5*\y,-7.7*\y)  node[anchor=west] {\tiny{$4$}};
\filldraw[] (7.2*\y,-8.7*\y)  node[anchor=west] {\tiny{$5$}};
\filldraw[] (6.5*\y,-10.3*\y)  node[anchor=west] {\tiny{$6$}};

\filldraw[black] (9.6*\y,-9*\y)  node[anchor=west] {\tiny{$+$}};

\draw[] (11.8*\y,-9*\y) -- (11.8*\y,-10*\y);
\draw[] (11.8*\y,-9*\y) -- (11.8*\y,-8*\y);

\draw[] (11.8*\y,-9*\y) -- (11.1*\y,-9.7*\y);
\draw[] (11.8*\y,-9*\y) -- (11.1*\y,-8.3*\y);
\draw[] (11.8*\y,-9*\y) -- (12.8*\y,-9*\y);
\draw[] (12.8*\y,-9*\y) -- (13.8*\y,-9*\y);
\draw[] (13.8*\y,-9*\y) -- (14.2*\y,-8.3*\y);
\draw[] (13.8*\y,-9*\y) -- (14.2*\y,-9.7*\y);

\filldraw[] (11.3*\y,-10.3*\y)  node[anchor=west] {\tiny{$1$}};
\filldraw[] (10.6*\y,-10*\y)  node[anchor=west] {\tiny{$2$}};
\filldraw[] (10.6*\y,-8*\y)  node[anchor=west] {\tiny{$3$}};
\filldraw[] (11.3*\y,-7.7*\y)  node[anchor=west] {\tiny{$4$}};
\filldraw[] (13.8*\y,-8*\y)  node[anchor=west] {\tiny{$5$}};
\filldraw[] (13.8*\y,-10*\y)  node[anchor=west] {\tiny{$6$}};

\filldraw[black] (15.5*\y,-9*\y)  node[anchor=west] {\tiny{$+$}};

\draw[] (18.8*\y,-9*\y) -- (18.1*\y,-9*\y);
\draw[] (18.8*\y,-9*\y) -- (19.5*\y,-9*\y);
\draw[] (18.8*\y,-9*\y) -- (19.2*\y,-8.3*\y);
\draw[] (18.8*\y,-9*\y) -- (18.4*\y,-8.3*\y);
\draw[] (18.8*\y,-9*\y) -- (19.2*\y,-9.7*\y);
\draw[] (18.8*\y,-9*\y) -- (18.4*\y,-9.7*\y);

\filldraw[] (17.7*\y,-9.7*\y)  node[anchor=west] {\tiny{$1$}};
\filldraw[] (17.4*\y,-9*\y)  node[anchor=west] {\tiny{$2$}};
\filldraw[] (17.7*\y,-8.3*\y)  node[anchor=west] {\tiny{$3$}};
\filldraw[] (19.2*\y,-8.3*\y)  node[anchor=west] {\tiny{$4$}};
\filldraw[] (19.5*\y,-9*\y)  node[anchor=west] {\tiny{$5$}};
\filldraw[] (19.2*\y,-9.7*\y)  node[anchor=west] {\tiny{$6$}};

\end{tikzpicture}
\caption{Nonzero diagrams contributing to a $5$- and a $6$-point process in the multi-Regge limit. In the result for $M^{(6)}$ it is contained the amplitude $M^{(5)}$; such amplitude is depicted in blue.}
\label{figuresect1_1}
\end{figure}

To find a condition on the $n$-point coupling we work in a particular multi-Regge limit in which one particle is at rest while $n-3$ particles are extremely energetic 
\begin{equation}\label{eq0_3}
a_1=1 \ \ \text{and} \ \ a_j=-x^{j-2} \ \ \text{for} \ \ j=3,\ldots , n-1
\end{equation}
where $x \gg 1$. We will assume by convention that all the momenta are incoming. Imposing momentum conservation, the remaining two momenta, on one of the two branches of solutions, are
\begin{equation}\label{eq0_4}
a_2=-\frac{m_2}{m_1} + o(x^{-1}) \ \  \text{and} \ \ a_n= \frac{m_{n-1}}{m_n} x^{n-3} + \frac{m_{n-2}}{m_n}x^{n-4}+\ldots + \frac{m_{3}}{m_n} x + o(x^0) .
\end{equation}
In a tree-level scattering process any internal propagator inside a Feynman diagram splits the external particles into two subsets. On one side we have a subset $\alpha \subset \{1,\dots,n \}$ while on the other side we have its complement. 
If the $a_j$ are chosen as in (\ref{eq0_3}), (\ref{eq0_4}),
the only nonzero propagators
in the limit $x \to \infty$ 
are those that divide the diagram into subsets $\{1, 2,\ldots, k\}$ and $\{ k + 1,\ldots , n-1, n\}$,
since in all  other cases the momentum transferred diverges. 
In particular if we consider the propagator $G_a(\alpha)$ of a particle of type $a$ splitting a diagram into a subset $\alpha$ and its complement, we have
\begin{align}\label{eq0_5}
G_a(\alpha) &= \frac{i}{(\sum_{j \in \alpha} m_j a_j)(\sum_{j \in \alpha} \frac{m_j}{a_j})-m_a^2+i \epsilon} \nonumber \\
& \to
\begin{cases}
-\frac{i}{m_a^2} \ \ \text{if} \ \  
\alpha= \{k+1, \dots ,n-1,n \} \quad (\mbox{or equivalently~}
\alpha= \{1,2, \dots ,k \} ),\\
0 \ \  \text{otherwise} .\\
\end{cases}
\end{align}
Therefore in this limit the only surviving tree-level diagrams are chains with all the particles ordered from left to right as shown in figure \ref{figuresect1_1} . 
Any nonzero diagram contributes with a factor $(-i)^{V+P}$ where $V$ is the number of vertices and $P$ is the number of propagators and any time we have a propagator between two vertices we need to sum over all the possible propagating particles.

Let us consider the scattering of $5$ particles in this limit. We label 
the types of the particles with the letters $b_1,\ldots, b_5$,
and the parameters of the light cone components of their momenta as defined in~\eqref{eq0_2}  with the letters $a_1, \ldots, a_5$. Proving that such an 
amplitude is a constant is not trivial, and finding the conditions on the masses, $3$- and $4$-point couplings making this possible will be the concern of the next section. But if we assume that they have been tuned in such a way that the conditions hold, and therefore $M^{(5)}_{b_1 b_2 b_3 b_4 b_5}$ is a constant not depending on the external momenta, 
we can use the multi-Regge limit \eqref{eq0_3}, \eqref{eq0_4} to find the value of this constant. As already explained, in this limit only Feynman diagrams corresponding to ordered chains survive, and we can read the value of the amplitude from the first (blue) row in figure~\ref{figuresect1_1}
\begin{equation}
    \label{5_point_amplitude_in_Multi_Regge_limit}
    \begingroup\color{blue} 
    M^{(5)}_{b_1 b_2 b_3 b_4 b_5}  \simeq  \sum_i \frac{C^{(4)}_{b_1 b_2 b_3 \bar{i}} C^{(3)}_{i b_4 b_5} }{m_i^2} - \sum_{i,j} \frac{C^{(3)}_{b_1 b_2\bar{i}} C^{(3)}_{i b_3 \bar{j}} C^{(3)}_{j b_4 b_5}}{m_i^2 m_j^2}+\sum_i \frac{C^{(3)}_{b_1 b_2 \bar{i}} C^{(4)}_{i b_3 b_4 b_5} }{m_i^2} - C^{(5)}_{b_1 b_2 b_3 b_4 b_5}.
    \endgroup
\end{equation}
The equality \eqref{5_point_amplitude_in_Multi_Regge_limit} is valid up to an overall multiplicative factor containing a power of the imaginary unit coming from vertices and propagators. By requiring that the $5$-point process~\eqref{5_point_amplitude_in_Multi_Regge_limit} is zero, for any choice of types $\{b_1, b_2, b_3, b_4, b_5 \}$, we fix the values of the $5$-point couplings
$C^{(5)}_{b_1 b_2 b_3 b_4 b_5}$ in terms of the masses and the $3$- and $4$-point couplings. Once we prove that all $5$-point processes are null we move to $M^{(6)}$. Once again the six-point amplitudes could in principle depend on the external kinematics adopted and it will be a matter of the next sections to check the conditions for a six-point amplitude to be a constant. However if we assume that $M^{(6)}$ does not depend on the external momenta, then the values of $6$-point couplings can be fixed by making these amplitudes equal to zero. Let define certain type labels  $b_1, \ldots, b_6$ for the external particles and assume that these particles have momentum parameters $a_1, \ldots, a_6$, defined as in \eqref{eq0_3}, \eqref{eq0_4}.
Then the value of $M^{(6)}_{b_1 b_2 b_3 b_4 b_5 b_6}$ is given by summing over all the Feynman diagrams in which the external legs are ordered, as recorded in the second and third rows in figure~\ref{figuresect1_1}. The algebraic expression for this sum is
\begin{equation}
    \label{6_point_amplitude_in_Multi_Regge_limit}
    \begin{split}
    M^{(6)}_{b_1 b_2 b_3 b_4 b_5 b_6}  &\simeq  \sum_a \frac{C^{(3)}_{b_1 b_2 \bar{a}}}{m_a^2} \begingroup\color{blue} 
    \Bigl[ \sum_i \frac{C^{(4)}_{a b_3 b_4 \bar{i}} C^{(3)}_{i b_5 b_6} }{m_i^2} - \sum_{i,j} \frac{C^{(3)}_{a b_3\bar{i}} C^{(3)}_{i b_4 \bar{j}} C^{(3)}_{j b_5 b_6}}{m_i^2 m_j^2}+\sum_i \frac{C^{(3)}_{a b_3 \bar{i}} C^{(4)}_{i b_4 b_5 b_6} }{m_i^2} - C^{(5)}_{a b_3 b_4 b_5 b_6} \Bigr]
    \endgroup\\
&+\sum_{i,j} \frac{C^{(4)}_{b_1 b_2 b_3 \bar{i}} C^{(3)}_{i b_4 \bar{j}} C^{(3)}_{j b_5 b_6} }{m_i^2 m_j^2}- \sum_{i} \frac{C^{(4)}_{b_1 b_2 b_3\bar{i}} C^{(4)}_{i b_4 b_5 b_6}}{m_i^2}-\sum_{i} \frac{C^{(5)}_{b_1 b_2 b_3 b_4 \bar{i}} C^{(3)}_{i b_5 b_6}}{m_i^2}+ C^{(6)}_{b_1 b_2 b_3 b_4 b_5 b_6}.
    \end{split}
\end{equation}
We note that the expression we found for the six-point amplitude is not completely new. Indeed the blue part in \eqref{6_point_amplitude_in_Multi_Regge_limit}, that matches the blue pictures in the second row of figure~\ref{figuresect1_1}, is exactly the value of a $5$-point amplitude in the multi-Regge limit. Since we already tuned the $5$-point couplings in such a way to make such part null, the blue terms in~\eqref{6_point_amplitude_in_Multi_Regge_limit} can be ignored and the constraint on the six-point couplings to have a theory with null processes with six external legs can be obtained by imposing that the second row in~\eqref{6_point_amplitude_in_Multi_Regge_limit} is equal to zero. 

We point out that the values of momenta entering into the $5$-point process in the first row in figure~\ref{figuresect1_1} are all on-shell since all the momenta are associated to external particles. On the other hand the amplitude $M^{(5)}$ that can be read as the blue expression in the second row of figure~\ref{figuresect1_1} is off-shell since one of the momenta (let us call it $P$) is flowing in an internal propagator and satisfies $P^2=0$ due to the multi-Regge limit with six external legs. In spite of this fact, the value of the off-shell $5$-point amplitude appearing inside $M^{(6)}$, where the external parameters have been fixed to be in the multi-Regge limit according with~\eqref{eq0_3} with $n=6$,  is exactly the same as the value of the on-shell amplitude $M^{(5)}$ verifying the multi-Regge limit with $n=5$. This consideration, deriving from the fact that for $x\gg1$ the Minkowski norm of momenta flowing inside propagators can only be zero or infinity, allows us to identify in a generic $n$-point amplitude all the $k$-point amplitudes, with $k<n$,  that have already been derived in the previous steps.
In principle these amplitudes are off-shell, but as an effect of this multi-Regge limit, their values are exactly the same as if they are on-shell in a multi-Regge limit with fewer external particles.
For example if now we want the value of a $7$-point amplitude in this high energy limit we obtain, up to an overall factor, 
\begin{equation}
    \label{7_point_amplitude_in_Multi_Regge_limit}
    \begin{split}
     M^{(7)}_{b_1\ldots b_7} &\simeq \sum_a \frac{C^{(3)}_{b_1 b_2 \bar{a}}}{m_a^2} M^{(6)}_{a b_3 \ldots b_7} + \sum_a \frac{C^{(4)}_{b_1 b_2 b_3 \bar{a}}}{m_a^2} M^{(5)}_{a b_4 \ldots b_7}\\
     &+i\sum_{i,j} \frac{C^{(5)}_{b_1 b_2 b_3 b_4 \bar{i}} C^{(3)}_{i b_5 \bar{j}} C^{(3)}_{j b_6 b_7} }{m_i^2 m_j^2}- i \sum_{i} \frac{C^{(5)}_{b_1 b_2 b_3 b_4 \bar{i}} C^{(4)}_{i b_5 b_6 b_7}}{m_i^2}- i \sum_{i} \frac{C^{(6)}_{b_1 b_2 b_3 b_4 b_5 \bar{i}} C^{(3)}_{i b_6 b_7}}{m_i^2}+ i C^{(7)}_{b_1 \ldots b_7}.
     \end{split}
\end{equation}
Once again the first two contributions on the RHS of \eqref{7_point_amplitude_in_Multi_Regge_limit} contain respectively a $6$- and $5$-point process in their multi-Regge limit and have to be zero by the previous analysis. The constraint on the $7$-point coupling is therefore derived by setting to zero the second row of~\eqref{7_point_amplitude_in_Multi_Regge_limit}. 

\begin{figure}
\begin{tikzpicture}
\draw[] (-1,-1) -- (1,1);
\draw[] (1,-1) -- (-1,1);
\draw[] (-1,0) -- (1,0);
\draw[] (0,-1) -- (0,1);

\filldraw[black] (1,-1)  node[anchor=west] {\footnotesize{$1$}};
\filldraw[black] (0,-1)  node[anchor=west] {\footnotesize{$2$}};
\filldraw[black] (-0.7,-1)  node[anchor=west] {\footnotesize{$\ldots$}};
\filldraw[black] (0.6,1.2)  node[anchor=west] {\footnotesize{$n-1$}};
\filldraw[black] (1,0)  node[anchor=west] {\footnotesize{$n$}};

\draw[] (2,-1) -- (3,0);
\draw[] (4,-1) -- (2,1);
\draw[] (2,0) -- (5,0);
\draw[] (3,-1) -- (3,1);
\draw[] (4,0) -- (4,1);

\filldraw[black] (4,-1)  node[anchor=west] {\footnotesize{$1$}};
\filldraw[black] (3,-1)  node[anchor=west] {\footnotesize{$2$}};
\filldraw[black] (2.2,-1)  node[anchor=west] {\footnotesize{$\ldots$}};
\filldraw[black] (4,1)  node[anchor=west] {\footnotesize{$n-1$}};
\filldraw[black] (5,0)  node[anchor=west] {\footnotesize{$n$}};

\draw[] (6,-1) -- (7,0);
\draw[] (7,0) -- (6,1);
\draw[] (6,0) -- (9,0);
\draw[] (7,-1) -- (7,1);
\draw[] (8,-1) -- (8,1);

\filldraw[black] (7,-1)  node[anchor=west] {\footnotesize{$1$}};
\filldraw[black] (6,-1)  node[anchor=west] {\footnotesize{$2$}};
\filldraw[black] (6,-0.3)  node[anchor=west] {\footnotesize{$\vdots$}};

\filldraw[black] (8,1)  node[anchor=west] {\footnotesize{$n-2$}};
\filldraw[black] (8.6,0.3)  node[anchor=west] {\footnotesize{$n-1$}};
\filldraw[black] (8,-1)  node[anchor=west] {\footnotesize{$n$}};

\draw[] (10,-1) -- (11,0);
\draw[] (11,0) -- (10,1);
\draw[] (10,0) -- (14,0);
\draw[] (11,-1) -- (11,1);
\draw[] (12,0) -- (12,1);
\draw[] (13,0) -- (13,1);

\filldraw[black] (11,-1)  node[anchor=west] {\footnotesize{$1$}};
\filldraw[black] (10,-1)  node[anchor=west] {\footnotesize{$2$}};
\filldraw[black] (10,-0.3)  node[anchor=west] {\footnotesize{$\vdots$}};

\filldraw[black] (11.5,1.2)  node[anchor=west] {\footnotesize{$n-2$}};
\filldraw[black] (12.5,1.2)  node[anchor=west] {\footnotesize{$n-1$}};
\filldraw[black] (14,0)  node[anchor=west] {\footnotesize{$n$}};

\end{tikzpicture}
\caption{Diagrams surviving in an $n$-point scattering process in the multi-Regge limit having fixed $M^{(4)}$, $M^{(5)}$, \dots , $M^{(n-1)}$ to be zero by tuning the vertices up to $C_{a_1 \dots a_{n-1}}^{(n-1)}$}
\label{figuresect1_2}
\end{figure}
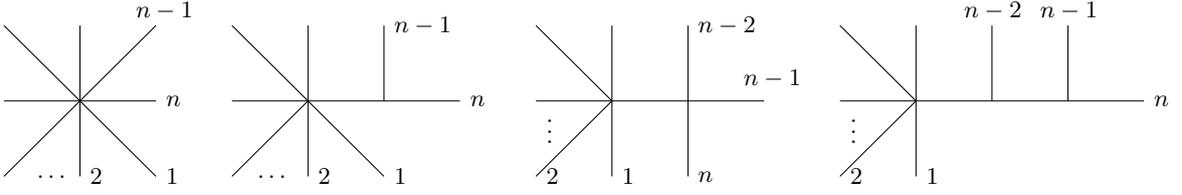

With these preliminaries over, we can construct the general Lagrangian in \eqref{eq0_1} by induction. We assume  that the couplings up to $C_{b_1 \dots b_{n-1}}^{(n-1)}$ have been tuned so as
to set the amplitudes $M^{(5)}=M^{(6)}=\ldots=M^{(n-1)}=0$. If we now consider a scattering process involving $n$ external legs, the only Feynman diagrams surviving  are those shown in figure \ref{figuresect1_2}. All the other diagrams involve processes contained in the amplitudes that have been fixed to zero by the induction hypothesis. Imposing that the $n$-point scattering is also null, from figure \ref{figuresect1_2} we read the following equation for the $n$-point coupling:
 \begin{equation}\label{eq0_6}
\begin{split}
C^{(n)}_{b_1 \ldots b_n}&- \sum_l C^{(n-1)}_{b_1 \ldots b_{n-2} \bar{l}} \frac{1}{m^2_l}C^{(3)}_{l  b_{n-1}b_{n}}- \sum_s C^{(n-2)}_{b_1 \ldots b_{n-3} \bar{s}} \frac{1}{m^2_s} C_{s b_{n-2} b_{n-1} b_n}^{(4)}\\
&+ \sum_l C^{(n-2)}_{b_1 \ldots b_{n-3} \bar{s}} \frac{1}{m^2_s} C_{s b_{n-2} \bar{l}}^{(3)} \frac{1}{m_l^2} C_{l b_{n-1} b_n}^{(3)}=0 .
\end{split}\end{equation}
This equation was found in  \cite{Gabai:2018tmm} and allows the value of $C^{(n)}$ to be found given the values of the masses and the $3$-, $4$-, $(n-2)$- and $(n-1)$-point couplings.

What we have just found is a necessary condition for the absence of particle production at the tree level. Of course the relation in \eqref{eq0_6} is not a sufficient condition to ensure the tree-level integrability of the theory, and indeed it leaves the choice of $3$- and $4$-point couplings completely free, as well as the masses of the particles. What we do know is that once we have fixed these parameters, all the higher-point couplings are uniquely fixed by equation \eqref{eq0_6}.

With equation~\eqref{eq0_6} we conclude the proof of points (i), (ii), (iii) from section~\ref{summarizing_the_logic_of_the_paper}. What is still missing is point (iv), corresponding to the finding of possible sets of masses, $3$- and $4$-point couplings that make the induction possible. 
To find these values we need to impose the cancellation of poles in 
$4$-point inelastic processes\footnote{In the present paper we construct integrable Lagrangians of theories presenting purely elastic scattering, i.e. with a diagonal S-matrix in $2$ to $2$ interactions.} and in events involving production with $5$ and $6$ external legs.
If we are able to do this we have provided the basis for the induction procedure and, by the analysis carried out in (i), (ii) and (iii), the relation in \eqref{eq0_6} becomes a sufficient condition to prove the absence of particle
production in the theory. 
Since these lower point processes are the basis of the entire induction we call the masses and the 3/4-point couplings
the `seeds of integrability'.

\section{Seeds of integrability}\label{sect2} 

In order to find all the couplings iteratively using \eqref{eq0_6} we need to figure out what `seed' $3$- and $4$-point couplings allow only elastic tree-level scattering in $4$-, $5$- and $6$-point interactions. We start by studying $2$ to $2$ non-diagonal scattering amplitudes, which we require to be null since the theories we want to construct 
are purely elastic,  and then we will move to higher point processes.

\subsection{Simplification processes in 4-point non-diagonal scattering} \label{sect2_2to2notallowed} 
We examine the case in which two incoming particles $a$ and $b$ evolve to two outgoing particles $c$ and $d$ of different types
\begin{equation}\label{eq4_0}
a(p_1)+b(p_2) \to c(p_3)+d(p_4) ,
\end{equation}
considering for the moment only $3$-point couplings. If the theory is integrable such
processes should be forbidden, a fact that should be visible perturbatively.  In particular, this requires that a so-called flipping rule on the masses and couplings
of the particles should exist,
as introduced in context of the perturbative study of higher poles in Toda theories
in \cite{Braden:1990wx}.
The idea is that any time a Feynman diagram contributing to a
non-diagonal 2 to 2 process has a pole at a particular value of the external momenta, there must be a pole in at least one other diagram for the same value of the momenta, so as to obtain a finite (and therefore constant) overall 
result which can be cancelled by a suitably-chosen $4$-point 
coupling, as explained in the last section. Note that this includes the possibility to have diagrams with on-shell bound state particles propagating in all three different $s$-, $t$- and $u$-channels, whose sum of residues is equal to zero. 
We adopt the convention for the Mandelstam variables defining
\begin{equation}
s=(P_1+P_2)^2 \ \ \ , \ \ \  t=(P_1-P_3)^2   \ \ \ , \ \ \ u=(P_1-P_4)^2 .
\end{equation}

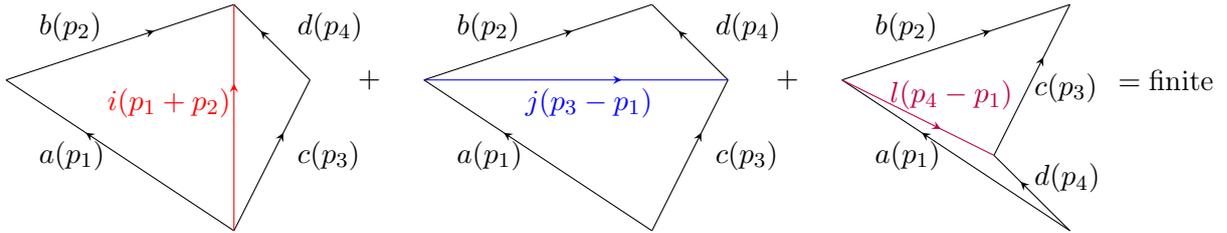
\begin{figure}

\begin{tikzpicture}

\draw[directed] (0,-2) -- (-3,0);
\draw[directed] (-3,0) -- (0,1);
\draw[directed] (0,-2) -- (1,0);
\draw[directed] (1,0) -- (0,1);
\draw[directed][red] (0,-2) -- (0,1);

\filldraw[black] (-2.7,-1)  node[anchor=west] {$a(p_1)$};
\filldraw[black] (-2.7,0.7)  node[anchor=west] {$b(p_2)$};
\filldraw[black] (0.7,-1)  node[anchor=west] {$c(p_3)$};
\filldraw[black] (0.7,0.7)  node[anchor=west] {$d(p_4)$};
\filldraw[black][red] (-1.8,-0.3)  node[anchor=west] {$i(p_1+p_2)$};

\filldraw[black] (1.5,0)  node[anchor=west] {$+$};

\draw[directed] (0+5.5,-2) -- (-3+5.5,0);
\draw[directed] (-3+5.5,0) -- (0+5.5,1);
\draw[directed] (0+5.5,-2) -- (1+5.5,0);
\draw[directed] (1+5.5,0) -- (0+5.5,1);
\draw[directed][blue] (-3+5.5,0) -- (1+5.5,0);

\filldraw[black] (-2.7+5.5,-1)  node[anchor=west] {$a(p_1)$};
\filldraw[black] (-2.7+5.5,0.7)  node[anchor=west] {$b(p_2)$};
\filldraw[black] (0.7+5.5,-1)  node[anchor=west] {$c(p_3)$};
\filldraw[black] (0.7+5.5,0.7)  node[anchor=west] {$d(p_4)$};
\filldraw[black][blue] (-1.8+5.5,-0.3)  node[anchor=west] {$j(p_3-p_1)$};

\filldraw[black] (1.5+5.5,0)  node[anchor=west] {$+$};

\draw[directed] (0+11,-2) -- (-3+11,0);
\draw[directed] (-3+11,0) -- (0+11,1);
\draw[directed] (1+10,0-2) -- (+10,1-2);
\draw[directed] (+10,1-2) -- (0+11,1);
\draw[directed][purple] (-3+11,0) -- (+10,1-2);

\filldraw[black] (-2.7+11,-1)  node[anchor=west] {$a(p_1)$};
\filldraw[black] (-2.7+11,0.7)  node[anchor=west] {$b(p_2)$};
\filldraw[black] (-0.6+11,0.8-0.9)  node[anchor=west] {$c(p_3)$};
\filldraw[black] (-0.6+11,0.7-2)  node[anchor=west] {$d(p_4)$};
\filldraw[purple] (8.5,-0.2)  node[anchor=west] {$l(p_4-p_1)$};

\filldraw[black] (11.5,0)  node[anchor=west] {$= \text{finite}$};

\end{tikzpicture}
\caption{Poles in the $s$-, $t$- and $u$-channel in a $2$ to $2$
off-diagonal process. The poles cancel so that the total contribution is finite.}
\label{figure_t/u_channel_poles}
\end{figure}

By assumption we consider all the particles contained in \eqref{eq0_1} to be possible asymptotic states of the theory; this implies, in order to prevent
decay processes, that any time
a coupling $C^{(3)}_{abc}$ is nonzero the mass
of each particle appearing in the vertex should be smaller than the
sum of the other two. 
As a consequence of this the masses of three
particles admitting a nonzero $3$-point coupling can be
drawn in Euclidean space as the sides of a  triangle.
To find the poles we need to consider the analytic continuation of the amplitude and study the region in which the external momenta are complex and the $a_j$ factors in \eqref{eq0_2} are phases (i.e. the rapidities $\theta_j$ are imaginary). From now on we will work considering always the first component of the momenta in \eqref{eq0_2}, that will be a complex number with absolute value given by the particle mass. We can assert that if there exists a coupling $C^{(3)}_{ijk}$ different from zero mediating the  interaction between  three asymptotic states $i$, $j$ and $k$ then there are three complex numbers $p_i$, $p_j$ and $p_k$ corresponding to the particles $i$, $j$ and $k$ such that $p_i+p_j+p_k=m_i e^{iU_i} + m_j e^{iU_j}+ m_k e^{iU_k}= 0$ and the absolute values of such momenta are respectively $m_i$, $m_j$ and $m_k$. We use such triangular relations to construct Feynman diagrams with internal propagators on-shell as shown in figure \ref{figure_t/u_channel_poles}. We have called $i U_j$ the imaginary value of the rapidity $\theta_j$ defined on the RHS of \eqref{eq0_2}.

Figure \ref{figure_t/u_channel_poles} shows the flipping rule in a case for which poles appear simultaneously in three Feynman diagrams with three possibly-different particles $i$, $j$ and $l$ propagating on-shell  in the $s$-, $t$- and $u$-channels respectively. Since the particle momenta are complex numbers with absolute values given by the particle masses we can draw dual versions of the Feynman diagrams on the complex plane as tiled polygons. As shown in the figure, while the diagrams having particles $i$ and $j$ propagating  in the $s$- and $t$-channels are represented by convex quadrilaterals, the diagram with the $l$ particle propagating in the $u$ channel is concave. In such a situation on the pole position we have $s=m_i^2$, $t=m_j^2$ and $u=m_l^2$ for the same value of the external momenta. Remembering that in two dimensions only one Mandelstam variable is independent we can Taylor expand $t$ and $u$ respect to $s$ as
\begin{equation}\label{eq4_1}
\begin{split}
t(s)&=t(m_i^2)+ \frac{dt}{ds}\Bigr|_{m_i^2} \ (s-m_i^2)+\ldots=m_j^2+ \frac{dt}{ds}\Bigr|_{m_i^2} \ (s-m_i^2)+\ldots\\
u(s)&=u(m_i^2)+ \frac{du}{ds}\Bigr|_{m_i^2} \ (s-m_i^2)+\ldots=m_l^2+ \frac{du}{ds}\Bigr|_{m_i^2} \ (s-m_i^2)+\ldots
\end{split}\end{equation}
Summing the three diagrams in figure \ref{figure_t/u_channel_poles} the amplitude near the pole is
\begin{equation}\label{eq4_2}
M^{(4)} \sim C^{(3)}_{a\bar{i}b} \  \frac{1}{s-m_i^2} \ C^{(3)}_{\bar{c} i \bar{d}}  + C^{(3)}_{a j \bar{c}}  \ \frac{1}{\frac{dt}{ds}\Bigr|_{m_i^2} (s-m_i^2)} \ C^{(3)}_{b\bar{j}\bar{d}} + C^{(3)}_{a l \bar{d}} \ \frac{1}{\frac{du}{ds}\Bigr|_{m_i^2} (s-m_i^2)} \ C^{(3)}_{b\bar{l}\bar{c}} .
\end{equation}
From a general property of the diagonals of quadrilaterals (in appendix \ref{App:2} we give an explicit derivation) we know that for the second diagram in figure \ref{figure_t/u_channel_poles},
\begin{equation}\label{New_eq4_3convex}
\frac{dt}{ds}\Bigr|_{m_i^2}= - \frac{ \Delta_{a j c}  \Delta_{b j d} }{\Delta_{a i b} \Delta_{ c i d} }
\end{equation}
where $\Delta_{ABC}$ is the area of the triangle having for sides the masses $m_A$, $m_B$ and $m_C$.
The minus sign in \eqref{New_eq4_3convex} reflects the 
fact that the  diagram is convex, so stretching the $i$ diagonal  keeping the lengths of
the external sides fixed causes the $j$ diagonal to get shorter. 
On the other hand for the concave quadrilateral (the last diagram in figure \ref{figure_t/u_channel_poles}) increasing the $i$ diagonal also increases the $l$ diagonal too, and we have
\begin{equation}\label{New_eq4_3concave}
\frac{du}{ds}\Bigr|_{m_i^2}= \frac{ \Delta_{a l d} \Delta_{b l c}} {\Delta_{a i b} \Delta_{c i d }}.
\end{equation}
Substituting \eqref{New_eq4_3convex} and \eqref{New_eq4_3concave} into \eqref{eq4_2} we have
\begin{equation}\label{New_eq4_5}
M^{(4)} \sim \frac{\Delta_{a i b} \Delta_{c i \bar{d}}}{s-m_i^2} \biggl[ \ \frac{C^{(3)}_{a \bar{i} b} C^{(3)}_{\bar{c} i \bar{d}}}{\Delta_{a i b} \Delta_{c i d}} - \frac{C^{(3)}_{a j \bar{c}} C^{(3)}_{b \bar{j}  \bar{d} }}{ \Delta_{a j c} \Delta_{b j  d}} + \frac{ C^{(3)}_{a l \bar{d}} C^{(3)}_{b \bar{l}  \bar{c}}}{ \Delta_{a l d} \Delta_{b l c}} \  \biggr].
\end{equation}
We have written the singular contribution to the amplitude in the neighbourhood of the pole $s \sim m_i^2$,  corresponding choosing a value of the red diagonal in figure \ref{figure_t/u_channel_poles} close to $m_i$ with the external sides kept fixed at their mass-shell values.
This formula makes it natural to incorporate the area of the corresponding mass triangles into the parametrisation of the $3$-point couplings by setting
\begin{equation}\label{New_eq4_6}
C^{(3)}_{i j k}= \ \Delta_{ijk} \ f_{ijk}.
\end{equation}
While the area of triangle does 
not distinguish particles from antiparticles, since their masses are equal, $f_{ijk}$ needs to differentiate indices of particles from those of antiparticles that as usual we indicate
respectively with $i$ and $\bar{i}$. A first 
feature of these parameters, coming from the 
way in which we wrote the Lagrangian \eqref{eq0_1}, is that $f_{ijk}$ has to be symmetric under exchange of 
any pair of indices. Moreover the reality of the Lagrangian in \eqref{eq0_1} requires
\begin{equation}\label{New_eq4_11}
f_{\bar{a} \bar{b} \bar{c}}=f^*_{abc}\,.
\end{equation} 
Substituting \eqref{New_eq4_6} into \eqref{New_eq4_5}, the residue for a $2$ to $2$ inelastic amplitude is proportional to 
\begin{equation}\label{New_eq4_9}
Res(M^{(4)})\sim f_{a\bar{i} b} f_{\bar{c} i \bar{d}} - f_{a j \bar{c}} f_{b \bar{j} \bar{d} } + f_{a l \bar{d}} f_{b \bar{l} \bar{c}}  
\end{equation}
and the requirement that it be equal to zero implies the following constraint 
\begin{equation}\label{New_eq4_12}
f_{a \bar{i} b} f_{\bar{c} i \bar{d}  } - f_{a j \bar{c}} f_{b \bar{j} \bar{d} } + f_{a l \bar{d} } f_{b \bar{l} \bar{c}}=0 .
\end{equation}
The situation can be generalized to the degenerate case in which more than a single particle propagates on-shell in each one of the channels.  If for example there are different intermediate states, all with mass $m_i$, propagating on-shell in the $s$-channel we need to sum over all the possible particles $i$ in \eqref{New_eq4_12} with that mass. The more general situation is therefore obtained by introducing in the relation \eqref{New_eq4_12} three different sums over all the possible particles $i$, $j$ and $l$ with respective masses $m_i$, $m_j$ and $m_l$. 
However, such degenerate situations never occur in affine Toda theories, where the cancellations of singularities always happen between pairs or triplets of Feynman diagrams with at most one particle propagating on-shell in each one of the possible channels. 
Cases for which the cancellation of poles happens between pairs of Feynman diagrams with opposite residues, 
corresponding to particles propagating in just two different channels, are also contained in relation~\eqref{New_eq4_12} 
by simply setting one of the terms  to zero. 

\begin{figure}
\medskip

\begin{center}
\begin{tikzpicture}
\tikzmath{\y=0.8;}

%1st diagram 
\draw[] (5.8*\y,0*\y+5*\y) -- (7.3*\y,1.5*\y+5*\y);
\draw[] (7.3*\y,-1.8*\y+5*\y) -- (5.8*\y,0*\y+5*\y);
\draw[] (7.3*\y,-1.8*\y+5*\y) -- (7.6*\y,0.1*\y+5*\y);
\draw[] (7.6*\y,0.1*\y+5*\y) -- (7.3*\y,1.5*\y+5*\y);
\draw[] (7.3*\y,-1.8*\y+5*\y) -- (7.3*\y,1.5*\y+5*\y);

\filldraw[black] (6.85*\y,0.1*\y+5*\y)  node[anchor=west] {\tiny{$i$}};
\filldraw[black] (6.1*\y,1*\y+5*\y)  node[anchor=west] {\tiny{$b$}};
\filldraw[black] (6.1*\y,-1*\y+5*\y)  node[anchor=west] {\tiny{$a$}};
\filldraw[black] (7.4*\y,0.7*\y+5*\y)  node[anchor=west] {\tiny{$d$}};
\filldraw[black] (7.4*\y,-1*\y+5*\y)  node[anchor=west] {\tiny{$c$}};

\filldraw[black] (9.8*\y,1.1*\y+5*\y)  node[anchor=west] {\tiny{type III}};
\filldraw[black] (9.9*\y,0.7*\y+5*\y)  node[anchor=west] {\tiny{$+/-$}};
\draw[] (9.5*\y,0.4*\y+5*\y) -- (11.3*\y,0.4*\y+5*\y);

\filldraw[black] (6.4*\y,0.7*\y+2*\y)  node[anchor=west] {\tiny{$s$-channel}};

%2nd diagram 
\draw[] (5.8*\y+3.5*\y,0*\y+1.5*\y) -- (7.3*\y+3.5*\y,1.5*\y+1.5*\y);
\draw[] (7.3*\y+3.5*\y,-1.8*\y+1.5*\y) -- (5.8*\y+3.5*\y,0*\y+1.5*\y);
\draw[] (7.3*\y+3.5*\y,-1.8*\y+1.5*\y) -- (7.6*\y+3.5*\y,0.1*\y+1.5*\y);
\draw[] (7.6*\y+3.5*\y,0.1*\y+1.5*\y) -- (7.3*\y+3.5*\y,1.5*\y+1.5*\y);
\draw[] (5.8*\y+3.5*\y,0*\y+1.5*\y) -- (7.6*\y+3.5*\y,0.1*\y+1.5*\y);

\filldraw[black] (6.85*\y+3.5*\y,0.3*\y+1.5*\y)  node[anchor=west] {\tiny{$j$}};
\filldraw[black] (6.1*\y+3.5*\y,1*\y+1.5*\y)  node[anchor=west] {\tiny{$b$}};
\filldraw[black] (6.1*\y+3.5*\y,-1*\y+1.5*\y)  node[anchor=west] {\tiny{$a$}};
\filldraw[black] (7.4*\y+3.5*\y,0.7*\y+1.5*\y)  node[anchor=west] {\tiny{$d$}};
\filldraw[black] (7.4*\y+3.5*\y,-1*\y+1.5*\y)  node[anchor=west] {\tiny{$c$}};

\filldraw[black] (8.4*\y,0.2*\y+3*\y)  node[anchor=west] {\tiny{$+/+$}};
\filldraw[black] (8.3*\y,0.6*\y+3*\y)  node[anchor=west] {\tiny{type I}};
\draw[] (8*\y,3.4*\y) -- (9.1*\y,2.4*\y);
\filldraw[black] (9.8*\y,1.2*\y-2*\y)  node[anchor=west] {\tiny{$t$-channel}};

%3rd diagram 
\draw[] (5.8*\y+7*\y,0*\y+5*\y) -- (7.3*\y+7*\y,1.5*\y+5*\y);
\draw[] (7.3*\y+7*\y,-1.8*\y+5*\y) -- (5.8*\y+7*\y,0*\y+5*\y);
\draw[] (7*\y+7*\y,-0.4*\y+5*\y) -- (7.3*\y+7*\y,1.5*\y+5*\y);
\draw[] (7.3*\y+7*\y,-1.8*\y+5*\y) -- (7*\y+7*\y,-0.4*\y+5*\y);
\draw[] (5.8*\y+7*\y,0*\y+5*\y) -- (7*\y+7*\y,-0.4*\y+5*\y);

\filldraw[black] (6.3*\y+7*\y,0*\y+5*\y)  node[anchor=west] {\tiny{$l$}};
\filldraw[black] (6.1*\y+7*\y,1*\y+5*\y)  node[anchor=west] {\tiny{$b$}};
\filldraw[black] (6.1*\y+7*\y,-1*\y+5*\y)  node[anchor=west] {\tiny{$a$}};
\filldraw[black] (7.4*\y+7*\y,0.7*\y+5*\y)  node[anchor=west] {\tiny{$c$}};
\filldraw[black] (7.4*\y+7*\y,-1*\y+5*\y)  node[anchor=west] {\tiny{$d$}};

\filldraw[black] (11.7*\y,0.6*\y+3*\y)  node[anchor=west] {\tiny{type II}};
\filldraw[black] (11.7*\y,0.2*\y+3*\y)  node[anchor=west] {\tiny{$+/+$}};
\draw[] (13*\y,3.4*\y) -- (11.9*\y,2.4*\y);
\filldraw[black] (13.4*\y,0.7*\y+2*\y)  node[anchor=west] {\tiny{$u$-channel}};

\end{tikzpicture}
\end{center}
\caption{Possible simultaneous poles in a non-allowed four-point process. In theories satisfying simply-laced scattering conditions, only two among the three diagrams diverge simultaneously for a particular choice of the external kinematics. In this case the product of the three-point couplings changes sign flipping from the $s$- to the $u$-channel whereas it does not change sign flipping from the $t$- to the $s$-channel and flipping from the $t$- to the $u$-channel.}
\label{2_to_2_scattering_tree_level_simultaneous_poles_changing_sign}
\end{figure}
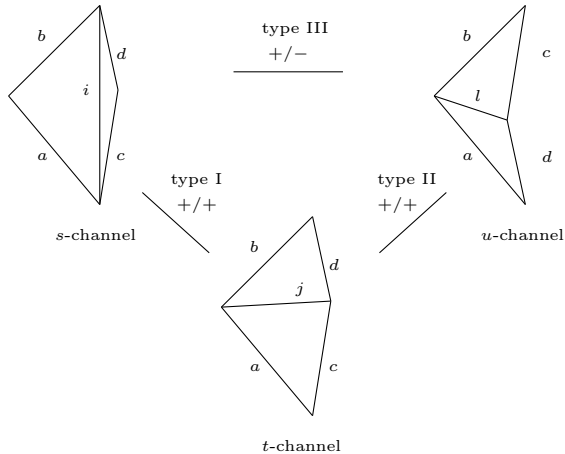

It is worth noting that the pole cancellation condition in inelastic $4$-point processes not only relates
 the values of different $3$-point couplings, but also gives strong constraints on the possible sets of 
masses. The requirement that poles always have to appear at least in pairs in order to cancel, as in
figure~\ref{figure_t/u_channel_poles}, is highly non-trivial, and leaves very little 
freedom on the possible masses of the theory. 
 Moreover, the fact that the amplitude should not 
present any singularities, no matter the branch of the kinematics we are considering in the solution 
of the energy-momentum constraints, together with the flipping move, allows us 
to construct networks of Feynman 
diagrams related each other. We will show below
how this works in an example of the $e_8^{(1)}$ 
affine Toda theory. This model, as all the other affine Toda field theories constructed from simply-laced Dynkin 
diagrams, is characterised by satisfying the following `simply-laced scattering conditions'
\begin{theorem}
\label{Simply_laced_scattering_conditions}
A theory respects `simply-laced scattering conditions' if in $2$ to $2$ non-diagonal scattering the poles cancel in pairs (flip $s/t$, $s/u$ or $t/u$) and in  $2$ to $2$ diagonal interactions it presents only one on-shell propagating particle at a time.
\end{theorem}
We will study these conditions further in the analysis of $5$-point interactions where they will play a crucial role
in constraining the values of the couplings. If they are satisfied, the cancellation mechanism of the singularities
happens between flipped copies of Feynman diagrams with particles propagating in a pair of channels: $s/t$, $s/u$ or $t/u$. It never happens that three poles appear simultaneously in Feynman diagrams with on-shell bound states propagating in the three different channels. The two flipped diagonals correspond to the masses of two particles propagating in different channels cancelling each other. We distinguish three types of flip depending on the way in which we replace a diagonal with its flipped version. Flips of type I are characterised by maintaining the external convex shape of the polygon during the replacement procedure (in figure \ref{2_to_2_scattering_tree_level_simultaneous_poles_changing_sign} it is the flip connecting the $s$-  to the $t$-channel). Contrarily flips of type II and type III change the order of the external particles in passing from one channel to its flip, obtaining in one case a concave polygon. These two kinds of flip are distinguished by the fact that in the former, 
one of the two points remains the same (in figure \ref{2_to_2_scattering_tree_level_simultaneous_poles_changing_sign} we see indeed that both the $j$ and $l$ vectors starts from the meeting point of the sides $a$ and $b$)
while in latter
both the starting and the ending point of the diagonal change.
Depending on the type of flip connecting the cancelling diagrams, the product of the $f$-functions entering in the three-point vertices may or may not change sign.
Assume for example to have a type I flip, connecting in our case the $s$- and the $t$-channels (which means the pair of cancelling singularities corresponds to a copy of diagrams with particles propagating in the $s$- and $t$-channels). In such a case only the first two terms in \eqref{New_eq4_12} are different from zero implying that the product of the corresponding $f$-functions, in order to avoid the singularity, does not change sign
$$
f_{ab\bar{i}} f_{i \bar{c} \bar{d}} = f_{a\bar{c}\bar{j}} f_{j c \bar{d}} .
$$
Using the same argument we see that the sign does not change in type II flips, while it does change with a flip of type III. We summarise these different situations in figure \ref{2_to_2_scattering_tree_level_simultaneous_poles_changing_sign}. This sign rule is useful also in the study of loop diagrams to understand the cancellation mechanism at loop level and in the computation of the total sum of Landau singularities \cite{Braden:1990wx}. The original paper \cite{Braden:1990wx} differentiated between two different kinds of flips that were the two different types they encountered in the construction of loop networks of singular Feynman diagrams. We distinguish a third type of flip here in order to understand the sign rule connecting products of different $3$-point couplings, as also shown in figure~\ref{2_to_2_scattering_tree_level_simultaneous_poles_changing_sign}.

Let show how to generate a network of Feynman diagrams entering into a certain non-allowed processes in the $e_8^{(1)}$ affine model. In the present case we use different colours to indicate the different masses of the theory, that we label in increasing order $m_1<\ldots<m_8$. Consider the following inelastic process
\begin{equation}
\label{A_Particular_colourful_scattering_process}
P_{\textcolor{blue}{\bullet}}+P_{\textcolor{black}{\bullet}} \to P_{\textcolor{bostonuniversityred}{\bullet}} +P_{\textcolor{orange}{\bullet}},
\end{equation}
in which a `blue'- and a `black'-particle (with masses $m_5$ and $m_3$ respectively in $e_8^{(1)}$) evolve into a `red'- and an `orange'-particle (with masses $m_1$ and $m_2$). If we know the on-shell shape of a single Feynman diagram contributing to this scattering process, corresponding to a certain kinematical configuration of the external states at which an internal diagonal is on-shell, we can figure out from it all the remaining diagrams. Let explain how the generation of all the on-shell graphs can be realized. 

\begin{figure}
\begin{center}
\begin{tikzpicture}
\tikzmath{\y=0.69;}

\filldraw[] (-14.5,8)  node[anchor=west] {\small{$m_1:$}};
\fill[bostonuniversityred] (-13.6,8.1) -- (-13.6+0.2,8.1) -- (-13.6+0.2,8.1-0.2) -- (-13.6,8.1-0.2);
\filldraw[] (-11.5,8)  node[anchor=west] {\small{$m_2:$}};
\fill[orange] (-10.6,8.1) -- (-10.6+0.2,8.1) -- (-10.6+0.2,8.1-0.2) -- (-10.6,8.1-0.2);
\filldraw[] (-8.5,8)  node[anchor=west] {\small{$m_3:$}};
\fill[black] (-7.6,8.1) -- (-7.6+0.2,8.1) -- (-7.6+0.2,8.1-0.2) -- (-7.6,8.1-0.2);

\filldraw[] (-5.5,8)  node[anchor=west] {\small{$m_4:$}};
\fill[ao(english)] (-4.6,8.1) -- (-4.6+0.2,8.1) -- (-4.6+0.2,8.1-0.2) -- (-4.6,8.1-0.2);

\filldraw[] (-2.5,8)  node[anchor=west] {\small{$m_5:$}};
\fill[blue] (-1.6,8.1) -- (-1.6+0.2,8.1) -- (-1.6+0.2,8.1-0.2) -- (-1.6,8.1-0.2);

\filldraw[] (-14.5,6.5)  node[anchor=west] {\small{$m_6:$}};
\fill[brown] (-13.6,6.6) -- (-13.6+0.2,6.6) -- (-13.6+0.2,6.6-0.2) -- (-13.6,6.6-0.2);

\filldraw[] (-11.5,6.5)  node[anchor=west] {\small{$m_7$ and $m_8$: not present in the picture}};

%Diagram 1
\draw[blue][directed] (-12,3) -- (-3.090169941-12, 3);
\draw[][directed] (-3.090169941-12, 3) -- (-1.868095685-12, 1.682040911+3);
\draw[orange][directed] (-1.022442654-12,0.217326896+3) -- (-1.868095685-12,1.682040911+3);
\draw[bostonuniversityred][directed] (-12,3) -- (-1.022442654-12,0.217326896+3);
\draw[][] (-3.090169941-12, 3) -- (-1.022442654-12,0.217326896+3);
\filldraw[black] (-3.090169941-11, 5.2)  node[anchor=west] {\tiny{(12)}};

\draw[][] (-3.090169941-9, 4) -- (-3.090169941-8, 4);
\filldraw[black] (-3.090169941-9.2, 4.5)  node[anchor=west] {\tiny{type III flip}};

%Diagram 2
\draw[blue][directed] (0-8,3) -- (-3.090169941-8, 3);
\draw[][directed] (-3.090169941-8, 3) -- (-1.868095685-8, 1.682040911+3);
\draw[orange][directed] (-1.022442654-8,0.217326896+3) -- (-8-1.868095685,1.682040911+3);
\draw[bostonuniversityred][directed] (0-8,3) -- (-1.022442654-8,0.217326896+3);
\draw[ao(english)][] (-8, 3) -- (-1.868095685-8, 1.682040911+3);
\filldraw[black] (-3.090169941-7, 5.2)  node[anchor=west] {\tiny{(1)}};

\draw[][] (-3.090169941-5, 4) -- (-3.090169941-4, 4);
\filldraw[black] (-3.090169941-4.8, 4.5)  node[anchor=west] {\tiny{jump}};

%Diagram 3
\draw[blue][directed] (0-4,3) -- (-3.090169941-4, 3);
\draw[][directed] (-3.090169941-4, 3) -- (-1.868095685-4, 1.682040911+3);
\draw[orange][directed] (-4, 3) -- (-1.545084971-4, 0.6879161499+3);
\draw[bostonuniversityred][directed] (-1.545084971-4, 0.6879161499+3) -- (-1.868095685-4, 1.682040911+3);
\draw[ao(english)][] (-4, 3) -- (-1.868095685-4, 1.682040911+3);
\filldraw[black] (-3.090169941-3, 5.2)  node[anchor=west] {\tiny{(2)}};

\draw[][] (-3.090169941-1, 4) -- (-3.090169941, 4);
\filldraw[black] (-3.090169941-1.2, 4.5)  node[anchor=west] {\tiny{type III flip}};

%Diagram 4
\draw[blue][directed] (0,3) -- (-3.090169941, 3);
\draw[][directed] (-3.090169941, 3) -- (-1.868095685, 1.682040911+3);
\draw[orange][directed] (0, 3) -- (-1.545084971, 0.6879161499+3);
\draw[bostonuniversityred][directed] (-1.545084971, 0.6879161499+3) -- (-1.868095685, 1.682040911+3);
\draw[orange][] (-1.545084971, 0.6879161499+3) -- (-3.090169941, 3);
\filldraw[black] (-3.090169941+1, 5.2)  node[anchor=west] {\tiny{(3)}};

\draw[][] (-3.090169941+2, 2) -- (-3.090169941+2, 1);
\filldraw[black] (-3.090169941+2.2, 1.5)  node[anchor=west] {\tiny{jump}};

%Diagram 5
\draw[blue][directed] (0,0) -- (-3.090169941, 0);
\draw[][directed] (-3.090169941, 0) -- (-1.022442653, -0.2173268950);
\draw[orange][directed] (0, 0) -- (-1.545084971, 0.6879161499);
\draw[bostonuniversityred][directed] (-1.545084971, 0.6879161499) -- (-1.022442653, -0.2173268950);
\draw[orange][] (-1.545084971, 0.6879161499) -- (-3.090169941, 0);
\filldraw[black] (-3.090169941+3, 0.5)  node[anchor=west] {\tiny{(4)}};

\draw[][] (-3.090169941+2, -0.5) -- (-3.090169941+2, -1.5);
\filldraw[black] (-3.090169941+2.2, -1)  node[anchor=west] {\tiny{type II flip}};

%Diagram 6
\draw[blue][directed] (0,-3) -- (-3.090169941, -3);
\draw[][directed] (-3.090169941, -3) -- (-1.022442653, -0.2173268950-3);
\draw[orange][directed] (0.522642318, -0.905243045-3) -- (-1.022442653, -0.2173268950-3);
\draw[bostonuniversityred][directed] (0,-3) -- (0.522642318, -0.905243045-3);
\draw[bostonuniversityred][] (-1.022442653, -0.2173268950-3) -- (0,-3);
\filldraw[black] (-3.090169941+3, 0.5-3)  node[anchor=west] {\tiny{(5)}};

\draw[][] (-3.090169941-0.5, 4-6.5) -- (-3.090169941+0.5, 4-6.5);
\filldraw[black] (-3.090169941-0.5, -2)  node[anchor=west] {\tiny{jump}};

%Diagram 7
\draw[blue][directed] (0-4,-3) -- (-3.090169941-4, -3);
\draw[][directed] (-3.090169941-4, -3) -- (-1.022442653-4, -0.2173268950-3);
\draw[bostonuniversityred][] (-1.022442653-4, -0.2173268950-3) -- (0-4,-3);
\draw[bostonuniversityred][directed] (0-4,-3) -- (0.1092619958-4, 1.039558453-3);
\draw[orange][directed] (0.1092619958-4, 1.039558453-3) -- (-1.022442653-4, -0.2173268950-3);
\filldraw[black] (-3.090169941-2.5, -3-1)  node[anchor=west] {\tiny{(6)}};

\draw[][] (-3.090169941-2, -0.5) -- (-3.090169941-2, -1.5);
\filldraw[black] (-3.090169941-1.8, -1)  node[anchor=west] {\tiny{type II flip}};

%Diagram 8
\draw[][directed] (-3.090169941-4, 0) -- (-1.022442653-4, -0.2173268950);
\draw[orange][directed] (0.1092619958-4, 1.039558453) -- (-1.022442653-4, -0.2173268950);
\draw[brown][] (0.1092619958-4, 1.039558453) -- (-3.090169941-4, 0);
\draw[blue][directed] (0-4,0) -- (-3.090169941-4, 0);
\draw[bostonuniversityred][directed] (0-4,0) -- (0.1092619958-4, 1.039558453);
\filldraw[black] (-3.090169941-2.5, 1.5)  node[anchor=west] {\tiny{(7)}};

\draw[][] (-3.090169941-0.5-4, 4-3.5) -- (-3.090169941+0.5-4, 4-3.5);
\filldraw[black] (-3.090169941-0.5-4, 1)  node[anchor=west] {\tiny{jump}};

%Diagram 9
\draw[brown][] (0.1092619958-8, 1.039558453) -- (-3.090169941-8,0);
\draw[blue][directed] (0-8,0) -- (-3.090169941-8, 0);
\draw[bostonuniversityred][directed] (0-8,0) -- (0.1092619958-8, 1.039558453);
\draw[][directed] (-3.090169941-8,0) -- (1.545084971-3.090169941-8, 1.391200757);
\draw[orange][directed] (0.1092619958-8, 1.039558453) -- (1.545084971-3.090169941-8, 1.391200757);
\filldraw[black] (-3.090169941-6.2, 1.5)  node[anchor=west] {\tiny{(8)}};

\draw[][] (-3.090169941-6, -0.5) -- (-3.090169941-6, -1.5);
\filldraw[black] (-3.090169941-1.8-4, -1)  node[anchor=west] {\tiny{type I flip}};

%Diagram 10
\draw[blue][directed] (0-8,-3) -- (-3.090169941-8, -3);
\draw[bostonuniversityred][directed] (0-8,-3) -- (0.1092619958-8, 1.039558453-3);
\draw[][directed] (-3.090169941-8,-3) -- (-1.545084970-8, 1.391200757-3);
\draw[orange][directed] (0.1092619958-8, 1.039558453-3) -- (1.545084971-3.090169941-8, 1.391200757-3);
\draw[][] (0-8,-3) -- (-1.545084970-8, 1.391200757-3);
\filldraw[black] (-3.090169941-6.5, -3-1)  node[anchor=west] {\tiny{(9)}};

%Diagram 11
\draw[blue][directed] (0-12,-3) -- (-3.090169941-12, -3);
\draw[][directed] (-3.090169941-12,-3) -- (-1.545084970-12, 1.391200757-3);
\draw[bostonuniversityred][directed] (0-12,-3) -- (-1.022442653-12, -0.2173268970-3);
\draw[orange][directed] (-1.022442653-12, -0.2173268970-3) -- (-1.545084970-12, 1.391200757-3);
\draw[][] (0-12,-3) -- (-1.545084970-12, 1.391200757-3);
\filldraw[black] (-3.090169941-10.5, -3-1)  node[anchor=west] {\tiny{(10)}};

\draw[][] (-3.090169941-9, 4-6.5) -- (-3.090169941-8, 4-6.5);
\filldraw[black] (-3.090169941-9, -2)  node[anchor=west] {\tiny{jump}};

%Diagram 12
\draw[blue][directed] (0-12,0) -- (-3.090169941-12, 0);
\draw[][directed] (-3.090169941-12,0) -- (-1.545084970-12, 1.391200757);
\draw[bostonuniversityred][directed] (0-12,0) -- (-1.022442653-12, -0.2173268970);
\draw[orange][directed] (-1.022442653-12, -0.2173268970) -- (-1.545084970-12, 1.391200757);
\draw[][] (-1.022442653-12, -0.2173268970) -- (-3.090169941-12, 0);
\filldraw[black] (-3.090169941-12.5, 1)  node[anchor=west] {\tiny{(11)}};

\draw[][] (-3.090169941-10, -0.5) -- (-3.090169941-10, -1.5);
\filldraw[black] (-3.090169941-1.8-8, -1)  node[anchor=west] {\tiny{type II flip}};

\draw[][] (-3.090169941-10, -0.5+3) -- (-3.090169941-10, -1.5+3);
\filldraw[black] (-3.090169941-1.8-8, 2)  node[anchor=west] {\tiny{jump}};

\end{tikzpicture}
\end{center}
\caption{Network of Feynman diagrams with different on-shell configurations of external momenta contributing to an inelastic process in the $e_8^{(1)}$ affine Toda field theory. Segments depicted with different colours are equal in length to the different masses of the theory. By properly ordering the external lines it is possible, for each pair of diagrams connected by one flip, to recognize the corresponding flip shown in figure~\ref{2_to_2_scattering_tree_level_simultaneous_poles_changing_sign}.}
\label{Network_non_elastic_two_to_two_E8_affine_Toda_model}
\end{figure}
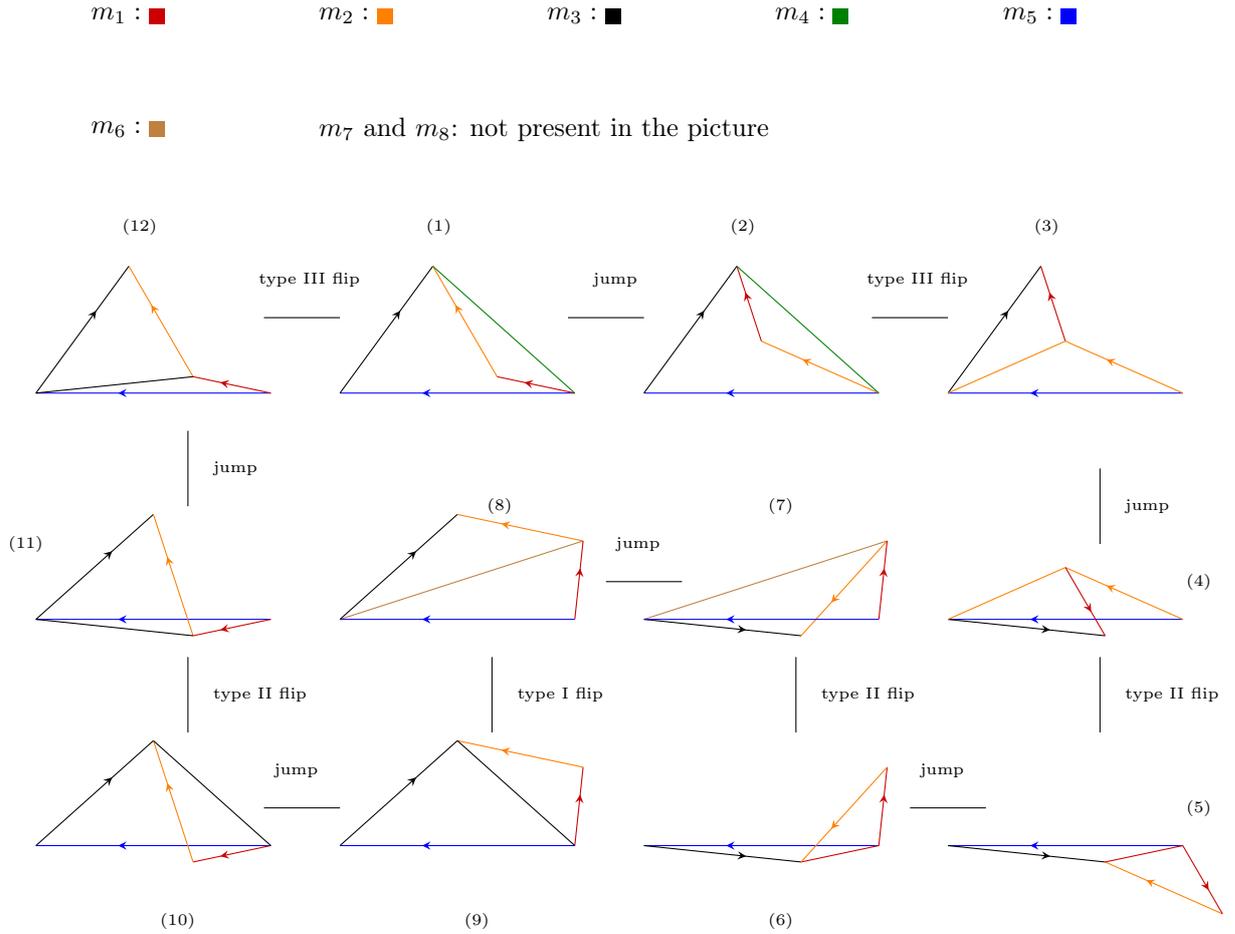

We start by taking the quadrilateral number (1) in figure~\ref{Network_non_elastic_two_to_two_E8_affine_Toda_model}, with an on-shell green diagonal corresponding to a particle with mass $m_4$ propagating in the $s$ channel. Starting from this configuration we can move to the other two diagrams by applying two different moves. We can flip the green propagating particle finding what other diagonal is equal in length to one of the possible eight different values of masses present in the theory. We find in this manner that a type III flip can be applied moving the diagram to the configuration (12), in which a `black'-particle propagates in the crossed-channel. The two diagrams (1) and (12) are one the flipped of the other, and the associated poles, that appear for the same external kinematical configuration, cancel in the sum. On the other hand we also know that there is another choice of the external kinematics for which the `green'-particle of diagram (1) is on-shell. Such a second choice corresponds of reflecting the outgoing `red'- and `orange'-particle with respect to this green segment, and represents the other solution of $\Sigma$ in \eqref{Sigma_si_branch_points} satisfying  energy-momentum conservation. This second move, that we call a `jump' in figure~\ref{Network_non_elastic_two_to_two_E8_affine_Toda_model}, corresponds of keeping  the diagonal associated to the propagating bound state fixed
and reflecting two external sides with
respect to it. This move preserves the singular propagator while changing the 
external kinematical configuration at which this propagator diverges.
Starting from diagram (1), in figure~\ref{Network_non_elastic_two_to_two_E8_affine_Toda_model}, we apply the two moves, alternating `jumps' and `flips'. While the `flip' changes the type of propagator entering into the diagram, connecting therefore two different Feynman diagrams contributing to the process, the `jump' does not change the particles and the vertices of the diagram but only the values of the external momenta. In figure~\ref{Network_non_elastic_two_to_two_E8_affine_Toda_model}, (1) and (2) correspond to the same diagram with different choices of external kinematics, similarly (3) and (4) and so on. After a finite number of jumps and flips we return to the graph we started from. This generates in total six different Feynman diagrams contributing to the process, whose on-shell configurations are contained in the copies of pictures [(1),(2)], [(3),(4)], [(5),(6)], [(7),(8)], [(9),(10)] and [(11),(12)] (each pair of graphs contains two different kinematical configurations of the same diagram).

This discussion can be extended to models (such as the non simply-laced affine Toda
theories) in which it is also possible to find  $3$ simultaneously
on-shell propagating particles for a single kinematical configuration.
Although figure \ref{Network_non_elastic_two_to_two_E8_affine_Toda_model} shows a particular
scattering from the $e^{(1)}_8$ theory for which the flipping rule happens between pairs of diagrams, similar networks need to be present in any integrable model constructed from a Lagrangian of the form in \eqref{eq0_1}.

To any flip in a theory with simply-laced scattering conditions we can
associate a  constraint of the form \eqref{New_eq4_12}, where we set to zero the contribution that is not present.
Since the network is closed in a
circle this leads to a finite number
of constraints on the functions $f_{ijk}$, that in the present case become
\begin{equation}
\label{constraints_on_f_two_to_two_non_allowed_flip_and_jumps}
 f_{\textcolor{blue}{\bullet} \bullet \textcolor{ao(english)}{\bullet}} f_{\textcolor{ao(english)}{\bullet} \textcolor{bostonuniversityred}{\bullet} \textcolor{orange}{\bullet}}=-f_{ \textcolor{bostonuniversityred}{\bullet} \textcolor{black}{\bullet}\textcolor{orange}{\bullet}} f_{ \textcolor{orange}{\bullet} \textcolor{orange}{\bullet} \textcolor{blue}{\bullet}}=-f_{ \textcolor{blue}{\bullet} \textcolor{black}{\bullet}\textcolor{bostonuniversityred}{\bullet}} f_{ \textcolor{bostonuniversityred}{\bullet} \textcolor{bostonuniversityred}{\bullet} \textcolor{orange}{\bullet}}=-f_{ \textcolor{black}{\bullet} \textcolor{orange}{\bullet}\textcolor{brown}{\bullet}} f_{ \textcolor{brown}{\bullet} \textcolor{bostonuniversityred}{\bullet} \textcolor{blue}{\bullet}}=-f_{ \textcolor{black}{\bullet} \textcolor{blue}{\bullet}\textcolor{black}{\bullet}} f_{ \textcolor{black}{\bullet} \textcolor{bostonuniversityred}{\bullet} \textcolor{orange}{\bullet}}=-f_{ \textcolor{black}{\bullet} \textcolor{orange}{\bullet}\textcolor{black}{\bullet}} f_{ \textcolor{black}{\bullet} \textcolor{bostonuniversityred}{\bullet} \textcolor{blue}{\bullet}},   
\end{equation}
where we used the different colours to label the different particles. The signs connecting the products of the different $f$-functions come from the sign rule (summarized in figure~\ref{2_to_2_scattering_tree_level_simultaneous_poles_changing_sign}) for the different types of flip.
By simplifying some of the equal terms in \eqref{constraints_on_f_two_to_two_non_allowed_flip_and_jumps} it follows from these constraints that $f_{ \textcolor{black}{\bullet} \textcolor{blue}{\bullet}\textcolor{black}{\bullet}}=f_{ \textcolor{orange}{\bullet} \textcolor{orange}{\bullet} \textcolor{blue}{\bullet}}$ and $f_{ \textcolor{black}{\bullet} \textcolor{orange}{\bullet}\textcolor{black}{\bullet}}=f_{ \textcolor{bostonuniversityred}{\bullet} \textcolor{bostonuniversityred}{\bullet} \textcolor{orange}{\bullet}}
$.
It is interesting to note  that the network of graphs in figure~\ref{Network_non_elastic_two_to_two_E8_affine_Toda_model} includes all of the
Feynman diagrams involving $3$-point couplings contributing to the scattering \eqref{A_Particular_colourful_scattering_process}, so, perhaps surprisingly, all these different graphs could be found starting from a single Feynman diagram.
However the story is not quite over; indeed at this point we know a set of nonzero couplings 
and from them we can start studying further processes. For example we observe that both 
$f_{ \textcolor{blue}{\bullet} \textcolor{black}{\bullet}\textcolor{bostonuniversityred}{\bullet}} \ne 0$ and 
$f_{ \textcolor{bostonuniversityred}{\bullet} \textcolor{orange}{\bullet}\textcolor{ao(english)}{\bullet}} \ne 0$;
this implies that we can draw a diagram in which a `blue' and a `black'-particle fuse into a `red'-propagator 
that then decays into a `orange'- and a `green'-state. Starting from this diagram we can therefore starting 
applying `jumps' and `flips' to obtain the different graphs contributing to the process
$$
P_{\textcolor{blue}{\bullet}}+P_{\textcolor{black}{\bullet}} \to P_{\textcolor{ao(english)}{\bullet}} +P_{\textcolor{orange}{\bullet}}.
$$
This would relate the $3$-point couplings of this second process to the $3$-point couplings entering into \eqref{A_Particular_colourful_scattering_process}. 

Only a few special sets of masses allow  closed networks of diagrams to be obtained
for all the different processes. If we instead start with a generic diagram, drawn as a
quadrilateral having sides of random length, and start applying `jumps' and `flips' to it, we will go on adding more and more particles to the theory in order to make the singularity cancellations possible. In the end the network will never close.
In order to obtain a closed loop of graphs, we need to start with a special set of masses, corresponding an integrable theory.

So far we have found the conditions under which the $2$ to $2$ non-diagonal scattering amplitude constructed using only $3$-point vertices has no poles and is therefore a constant not depending on the choice of the momenta. Now we need to set the $4$-point coupling so as to cancel this constant and obtain at the end a null process.

\subsection{Setting the 4-point couplings} \label{4pointsby3points}

Focusing on the non-diagonal process in equation \eqref{eq4_0}, the amplitude obtained using only $3$-point vertices is given by summing over all the possible particles propagating in the $s$-, $t$- and $u$-channels
\begin{equation}\label{New_eq4_13_multiple}
M^{(4)}=-i \sum_s C_{ab\bar{s}}^{(3)} \ \frac{1}{s-m_s^2} \ C_{s\bar{c}\bar{d}}^{(3)}-i \sum_j C_{a \bar{c} \bar{j}}^{(3)} \ \frac{1}{t-m_j^2} \ C_{j b \bar{d}}^{(3)}-i \sum_l C_{a \bar{d} \bar{l}}^{(3)} \ \frac{1}{u-m_l^2} \ C_{l b\bar{c}}^{(3)} .
\end{equation}
Since we have already proved that after having properly tuned the masses and $3$-point couplings such an amplitude is a constant (for $\{c,d\} \ne \{a,b\}$) we can take a particular choice of momenta that simplifies the computation. Two choices of multi-Regge limit can be adopted; we can set $s=+\infty$, $t=-\infty$ and $u=0$ or $s=+\infty$, $t=0$ and $u=-\infty$. This corresponds of solving the limit $s\to +\infty$ in the two different Riemann sheets over which $\Sigma$ in \eqref{Sigma_si_branch_points} takes values. In both the cases the result needs to be the same, since in the absence of poles the amplitude is a constant, and the $4$-point coupling cancelling the process is 
\begin{equation}\label{New_eq4_13_fixed}
C^{(4)}_{ab\bar{c}\bar{d}}= \sum_j C_{a \bar{c} \bar{j}}^{(3)} \ \frac{1}{m_j^2} \ C_{j b \bar{d}}^{(3)} = \sum_l C_{a \bar{d} \bar{l}}^{(3)} \ \frac{1}{m_l^2} \ C_{l b\bar{c}}^{(3)} .
\end{equation}
A similar result can be obtained in the case in which the particles in the initial and final state are equal, so that $\{c,d\}=\{a,b\}$. In this case we need to set the $4$-point coupling so as to cancel the reflection process, corresponding to the function defined over the blue half of the degenerate torus described in section~\ref{Elastic_scattering_from_degenerate_torus}, and allow only transmission. The only situation in which the $4$-point coupling cannot be obtained from the masses and $3$-point vertices is when it involves the scattering of $4$ real equal particles. This is a situation in which we cannot distinguish the reflection from the transmission and the event is always allowed. In this case to tune the $4$-point coupling correctly
we need to analyse the interaction of five external states.

\subsection{No-particle production in 5-point processes and tree level bootstrap} \label{No particles production in 5-point processes}

The cancellation of $4$-point non-diagonal processes is made possible by the flipping rule, cancelling all the possible poles appearing in the sums of Feynman diagrams. We now study  how the same rule permits the cancellation of all singularities in the $5$-point scattering amplitudes, provided we impose extra constraints on the values of the $f_{ijk}$ and the $4$-point couplings.

\begin{figure}
\begin{tikzpicture}
\draw[directed] (-1,0) -- (0.2,0);
\draw[directed] (-2,-1) -- (-1,0);
\draw[directed] (-2,1) -- (-1,0);
\filldraw[color=red!60, fill=red!5, very thick](0.6 ,0) circle (0.4);

\draw[directed] (1,0) -- (2,0);
\draw[directed] (0.6,-1.2) -- (0.6,-0.4);
\draw[directed] (0.6,0.4) -- (0.6,1.2);

\filldraw[black] (-2,1)  node[anchor=west] {\scriptsize{$b(p_2)$}};
\filldraw[black] (-2,-1)  node[anchor=west] {\scriptsize{$a(p_1)$}};
\filldraw[black] (-0.5,-0.2)  node[anchor=west] {\scriptsize{$c$}};
\filldraw[black] (1.5,-0.2)  node[anchor=west] {\scriptsize{$c(p_4)$}};
\filldraw[black] (0.5,-1.1)  node[anchor=west] {\scriptsize{$d(p_3)$}};
\filldraw[black] (0.5,1.1)  node[anchor=west] {\scriptsize{$d(p_5)$}};

\filldraw[black] (3,0)  node[anchor=west] {\scriptsize{$+$}};

\draw[directed] (6,0) -- (7.2,0);
\draw[directed] (5-1*0.8,-1*0.8-1) -- (-1*0.2+5,-1-1*0.2);
\draw[directed] (5-1*0.2,-1-1*0.2) -- (6,0);
\draw[directed] (5,1) -- (6,0);
\filldraw[color=red!60, fill=red!5, very thick](5 ,-1) circle (0.4);

\draw[directed] (5,-2.2) -- (5,-1.4);
\draw[directed] (5,-0.6) -- (5,0.2);

\filldraw[black] (5,1)  node[anchor=west] {\scriptsize{$b(p_2)$}};
\filldraw[black] (3.5,-1.5)  node[anchor=west] {\scriptsize{$a(p_1)$}};
\filldraw[black] (5.6,-0.4)  node[anchor=west] {\scriptsize{$a$}};
\filldraw[black] (6.5,-0.2)  node[anchor=west] {\scriptsize{$c(p_4)$}};
\filldraw[black] (4.9,-2.2)  node[anchor=west] {\scriptsize{$d(p_3)$}};
\filldraw[black] (4.1,0.1)  node[anchor=west] {\scriptsize{$d(p_5)$}};

\filldraw[black] (8,0)  node[anchor=west] {\scriptsize{$+$}};

\draw[directed] (11,0) -- (12.2,0);
\draw[directed] (10,-1) -- (11,0);
\draw[directed] (10.5,0.5) -- (11,0);
\draw[directed] (9.5,1.5) -- (10,1);
\filldraw[color=red!60, fill=red!5, very thick](10.2 ,0.8) circle (0.4);
\draw[directed] (10.2,-0.3) -- (10.2,0.4);
\draw[directed] (10.2,1.2) -- (10.2,1.8);

\filldraw[black] (8.8,1.6)  node[anchor=west] {\scriptsize{$b(p_2)$}};
\filldraw[black] (10.6,0.5)  node[anchor=west] {\scriptsize{$b$}};
\filldraw[black] (9.5,-1.3)  node[anchor=west] {\scriptsize{$a(p_1)$}};
\filldraw[black] (11.5,-0.2)  node[anchor=west] {\scriptsize{$c(p_4)$}};
\filldraw[black] (10.1,1.8)  node[anchor=west] {\scriptsize{$d(p_5)$}};
\filldraw[black] (9.3,-0.3)  node[anchor=west] {\scriptsize{$d(p_3)$}};

\filldraw[black] (13,0)  node[anchor=west] {\scriptsize{$= \ \ $ finite}};

\end{tikzpicture}
\caption{Divergent contributions to the process $a + b + d \to c + d $. We list the poles that have nonzero residues when the propagators connecting the $3$-point vertices and the $4$-point on-shell amplitudes are mass-shell. Despite the three contributions are separately divergent we prove that their sum is finite if additional constraints are imposed on the $3$- and $4$-point couplings.}
\label{fig5amplit_blobdiagr}
\end{figure}
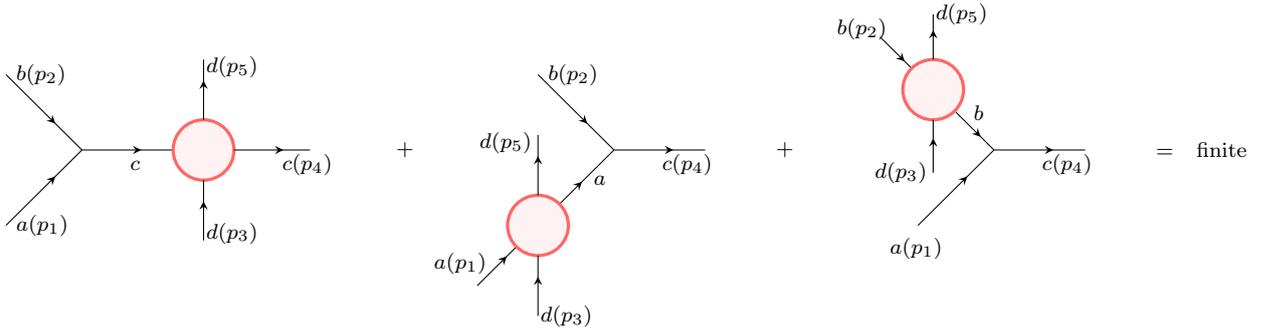

We start with the case in which all the interacting particles are
different. This is somewhat trivial, since whenever an
internal propagator goes on-shell it splits the amplitude into a $3$-point vertex and an on-shell $4$-point inelastic process. Since the inelastic $4$-point processes are null, entering into the analysis previously performed, the residues of such scattering processes are
all zero and no singularity appears. 
We therefore move directly to a less trivial case, in which the scattering involves two equal particles as in the following $3$ to $2$ event:
\begin{equation}\label{eq4_6}
a(p_1)+b(p_2)+d(p_3) \rightarrow c(p_4)+d(p_5) .
\end{equation}
We choose here to study a process with $3$ incoming and $2$ outgoing particles, but of course it is possible to bring particles on the left hand side or on the right hand side of the arrow by changing the sign of the momenta and particles to antiparticles. This process represents the most general case of a $5$-point scattering amplitude
that could in principle
contain poles. 
Such poles can  appear when, as shown in figure~\ref{fig5amplit_blobdiagr}, the propagator connecting the blobs\footnote{The blobs represent the sum of all tree-level Feynman diagrams having as external legs the types of particles entering into the blob.} and the $3$-point vertices is a particle of the same type as those present as external legs. 
This is indeed the only way to have a nonzero residue when the propagator diverges because the $2$ to $2$ scattering processes represented by the three blobs in figure~\ref{fig5amplit_blobdiagr} are diagonal.
We sum the three different combinations of Feynman diagrams in figure \ref{fig5amplit_blobdiagr}, choosing to write the momenta in light-cone components as in \eqref{eq0_2}. 
By using Lorentz invariance and the conservation of the overall energy and momentum we can remove three parameters from the amplitude, writing it as a function depending only on $a_3$ and $a_5$. Then we study its dependence on one parameter at a time, starting with $a_5$.
Taking the limit $a_5\to a_3$ (i.e. choosing the same momenta for the incoming and outgoing $d$ particle in \eqref{eq4_6}) we can isolate the residue of the amplitude at the pole
\begin{equation}\label{eq4_7}
M^{(5)}(a_3,a_5) \simeq \frac{1}{a_5 - a_3} \ \underset{a_5=a_3}{\mathrm{Res}}M^{(5)}(a_3,a_5).
\end{equation}
The residue in the expression above is a function of a single parameter $a_3$ and can be written as
\begin{equation}\label{Equation_Added_eq4_7}
\begin{split}
&\underset{a_5=a_3}{\mathrm{Res}}M^{(5)}(a_3,a_5)=\\
&-\frac{C^{(3)}_{ab\bar{c}}}{m_d} a_3 \Bigl[ \frac{1}{m_c} \frac{a_4 a_3}{a_4^2-a_3^2}M_{cd}^{(4)}(a_4,a_3)- \frac{1}{m_a} \frac{a_1 a_3}{a_1^2-a_3^2}M_{ad}^{(4)} (a_1,a_3)-\frac{1}{m_b} \frac{a_2 a_3}{a_2^2-a_3^2}M_{bd}^{(4)} (a_2,a_3)\Bigr].
\end{split}
\end{equation}
The parameters $a_1$, $a_2$ and $a_4$ in~\eqref{Equation_Added_eq4_7} are fixed and on the pole satisfy the fusing relation depicted in figure~\ref{Triangular_relation_bootstrap_angles_between_particles_a_b_and_c_on_the_pole_of_5_point_process_bootstrap}, so that the residue depends only on $a_3$.
$M_{cd}^{(4)}$, $M_{ad}^{(4)}$ and  $M_{bd}^{(4)}$ are the
on-shell values of the $2$ to $2$ amplitudes in figure
\ref{fig5amplit_blobdiagr}. This corresponds to taking the limit in
which the  blobs in figure \ref{fig5amplit_blobdiagr} go on-shell and
the intermediate propagators diverge. Since these blobs are allowed
processes, the three terms in the square brackets in \eqref{Equation_Added_eq4_7} do not vanish individually, but must cancel between themselves.
To prove this,
it is sufficient to prove that the sum of these terms has no poles
as a function of $a_3$.
This is
enough to show that it is a constant as a function of $a_3$,
which can be seen to be zero by taking the 
limit $a_3 \to \infty$.

Before continuing the study of the singularities connected to $5$-point interactions we show how the requirement that the expression in \eqref{eq4_7} has zero residue is equivalent to imposing tree-level bootstrap relations connecting the different S-matrix elements. 

\subsubsection{The tree level bootstrap}

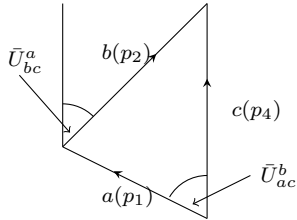
\begin{figure}
\begin{center}
\begin{tikzpicture}
\tikzmath{\y=1.9;}

\draw[][] (-1*\y,0.8*\y) arc(90:45:0.3*\y);
\draw[][] (0*\y,0.3*\y) arc(90:150:0.3*\y);

\draw[directed] (0*\y,0*\y) -- (0*\y,1.5*\y);
\draw[directed] (0*\y,0*\y) -- (-1*\y,0.5*\y);
\draw[directed] (-1*\y,0.5*\y) -- (0*\y,1.5*\y);
\draw[] (-1*\y,0.5*\y) -- (-1*\y,1.5*\y);
\draw[->] (-1.3*\y,1*\y) -- (-0.95*\y,0.6*\y);
\draw[->] (0.3*\y,0.3*\y) -- (-0.1*\y,0.1*\y);

\filldraw[black] (0.3*\y,0.3*\y)  node[anchor=west] {\scriptsize{$\bar{U}_{ac}^b$}};
\filldraw[black] (-1.45*\y,1.1*\y)  node[anchor=west] {\scriptsize{$\bar{U}_{bc}^a$}};
\filldraw[black] (0.1*\y,0.75*\y)  node[anchor=west] {\scriptsize{$c(p_4)$}};
\filldraw[black] (-0.8*\y,0.15*\y)  node[anchor=west] {\scriptsize{$a(p_1)$}};
\filldraw[black] (-0.8*\y,1.15*\y)  node[anchor=west] {\scriptsize{$b(p_2)$}};

\end{tikzpicture}
\caption{Mass triangle corresponding to the $3$-point vertex $C_{ab \bar{c}}$. Assuming the orientation of $p_4$ to be the axis respect to which we measure the angles following the counterclockwise direction we have $p_1=m_a e^{i \bar{U}_{ac}^b}$ , $p_2=m_b e^{-i \bar{U}_{bc}^a}$ , $p_4=m_c$.}
\label{Triangular_relation_bootstrap_angles_between_particles_a_b_and_c_on_the_pole_of_5_point_process_bootstrap}
\end{center}
\end{figure}

The $2$ to $2$ S-matrix element $S_{ij}$ is given in terms of $M^{(4)}_{ij}$ by
\begin{equation}
\label{Tree_level_bootstrap_conversion_between_S_matrix_in_theta_momentum_space}
S_{i j} (\theta_{i j})= \frac{1}{4m_i m_j \sinh \theta_{i j}} M_{ij}^{(4)} (a_i , a_j),
\end{equation}
where $\theta_{ij}$ is the difference between the rapidities of particles $i$ and $j$, and the conversion factor comes from writing the Dirac delta function associated to the total momentum conservation in terms of the rapidities and from the additional normalisation terms carried by the external particles.
Using this fact and writing the $a$-variables as $a_i=e^{\theta_i}$ the requirement that the residue of \eqref{eq4_7} is equal to zero is equivalent to the following constraint on the tree-level S-matrix
\begin{equation}
\label{Tree_level_bootstrap_first_relation}
S^{tree}_{cd} (\theta_{43})=S^{tree}_{ad}(\theta_{13})+S^{tree}_{bd}(\theta_{23}).
\end{equation}
In this equality the only free parameter is the rapidity $\theta_3$ of the $d$-particle, since all the other rapidities are frozen on their on-shell values making the diagrams in figure \ref{fig5amplit_blobdiagr} singular on the pole.
Defining the difference between the $c$- and $d$-particle rapidities to be $\theta_{43}=\theta$, the relation in \eqref{Tree_level_bootstrap_first_relation} can be written as
\begin{equation}
\label{Tree_level_bootstrap_second_relation}
S^{tree}_{cd} (\theta)=S^{tree}_{ad}(\theta+\theta_{14})+S^{tree}_{bd}(\theta+\theta_{24})
\end{equation}
The quantities $\theta_{14}$ and $\theta_{24}$ are  the differences between the rapidities of the $a$- and the $c$-particles, and between the rapidities of the $b$- and the $c$-particles interacting in the vertex $C^{(3)}_{ab\bar{c}}$. These are imaginary angles that are frozen on the pole position, where the on-shell particles fuse in the $3$-point vertex $C^{(3)}_{ab\bar{c}}$ satisfying the triangular relation in figure~\ref{Triangular_relation_bootstrap_angles_between_particles_a_b_and_c_on_the_pole_of_5_point_process_bootstrap}.
Referring to figure \ref{Triangular_relation_bootstrap_angles_between_particles_a_b_and_c_on_the_pole_of_5_point_process_bootstrap} we label 
$$
\theta_{41}=-i \bar{U}^b_{ac} \hspace{4mm}, \hspace{4mm} \theta_{42}=i \bar{U}^a_{bc}.
$$
Therefore we can write the constraint for the cancellation of the $5$-point process as
\begin{equation}
\label{tree_level_bootstrap_constraint_on_the_S_matrix_coming_from_5_point_pole_cancellation_equation}
 C_{ab \bar{c}} \ne 0 \implies S^{tree}_{dc} (\theta)= S^{tree}_{da} (\theta-i \bar{U}^b_{ac})+S^{tree}_{db} (\theta+i \bar{U}^a_{bc})
\end{equation}
where $\bar{U}^b_{ac}$ is the angle between sides $m_a$ and $m_c$ and $\bar{U}^a_{bc}$ is the angle between $m_b$ and $m_c$  in the mass triangle $\Delta_{abc}$. 

The relation in \eqref{tree_level_bootstrap_constraint_on_the_S_matrix_coming_from_5_point_pole_cancellation_equation} connects all the $2$ to $2$ tree level S-matrix elements of the theory together and represents the first order in perturbation theory of the bootstrap relation
\begin{equation}
\label{Quantum_exact_bootstrap_relation}
 C_{ab \bar{c}} \ne 0 \implies S_{dc} (\theta)= S_{da} (\theta-i \bar{U}^b_{ac}) S_{db} (\theta+i \bar{U}^a_{bc})
\end{equation}
that, if the integrability is preserved at quantum level, is valid order by order in the loop expansion
$$
S_{ij}(\theta)=1+S^{tree}_{ij}(\theta) + \ldots
$$
It is interesting to note how the bootstrap equations at the tree level completely emerges from the sole requirement that all the $5$-point processes are equal to zero and their pictorial representation can be recognised in the sum of the singular diagrams in figure \ref{fig5amplit_blobdiagr}.
Verifying the cancellation of the $5$-point processes for a general theory requires to prove that the tree level bootstrap relation \eqref{tree_level_bootstrap_constraint_on_the_S_matrix_coming_from_5_point_pole_cancellation_equation} is satisfied between all the different S-matrix elements. Such a relation is converted into particular constraints on the $3$- and $4$-point couplings of the theory.

\subsubsection{Pole cancellation and `Simply-laced scattering conditions'}

We now search  for the additional constraints on the couplings necessary for the  the tree-level bootstrap relations to be satisfied, and 
therefore for the vanishing of all the $5$-point processes, for models satisfying the `simply-laced scattering conditions' defined in property~\ref{Simply_laced_scattering_conditions}. 
We show how in these theories a necessary condition for the absence of particle production in $5$-point interactions is that the absolute values of the $f$-functions do not depend on the particular $3$-point vertex we are considering and from the relation in \eqref{New_eq4_6} we can then deduce the following area rule
\begin{equation}\label{Area_rule_abs_f_value}
|C^{(3)}_{i j k}|=  |f| \ \Delta_{ijk} ,
\end{equation}
where $|f|$ is the absolute value common to each $f$-function. We now proceed to
prove this assertion, adopting the  method already used for the $4$-point off-diagonal processes and imposing the absence of singularities. The requirement that the residue of \eqref{eq4_7} does not have any singularities in the variable $a_3$
is equivalent to requiring that the LHS and the RHS terms of the relation \eqref{tree_level_bootstrap_constraint_on_the_S_matrix_coming_from_5_point_pole_cancellation_equation} have the same pole structure.

In order to verify the cancellation
between the different poles appearing in  \eqref{Equation_Added_eq4_7} 
we split the possible singularities of  the residue
into two kinds. The
first type are singularities due to the possibility that some
propagators go on-shell inside the $4$-point amplitudes
$M_{cd}^{(4)}$, $M_{ad}^{(4)}$ and  $M_{bd}^{(4)}$ while the second
type are collinear singularities; this last kind happens when one of the denominators in
\eqref{Equation_Added_eq4_7} diverges, that is the situation in which $a_3^2 \to a_4^2$, $a_3^2 \to a_1^2$ or $a_3^2 \to a_2^2$. We study these two different situations separately and we show how, with a few additional constraints, the poles of \eqref{Equation_Added_eq4_7} cancel in both cases.
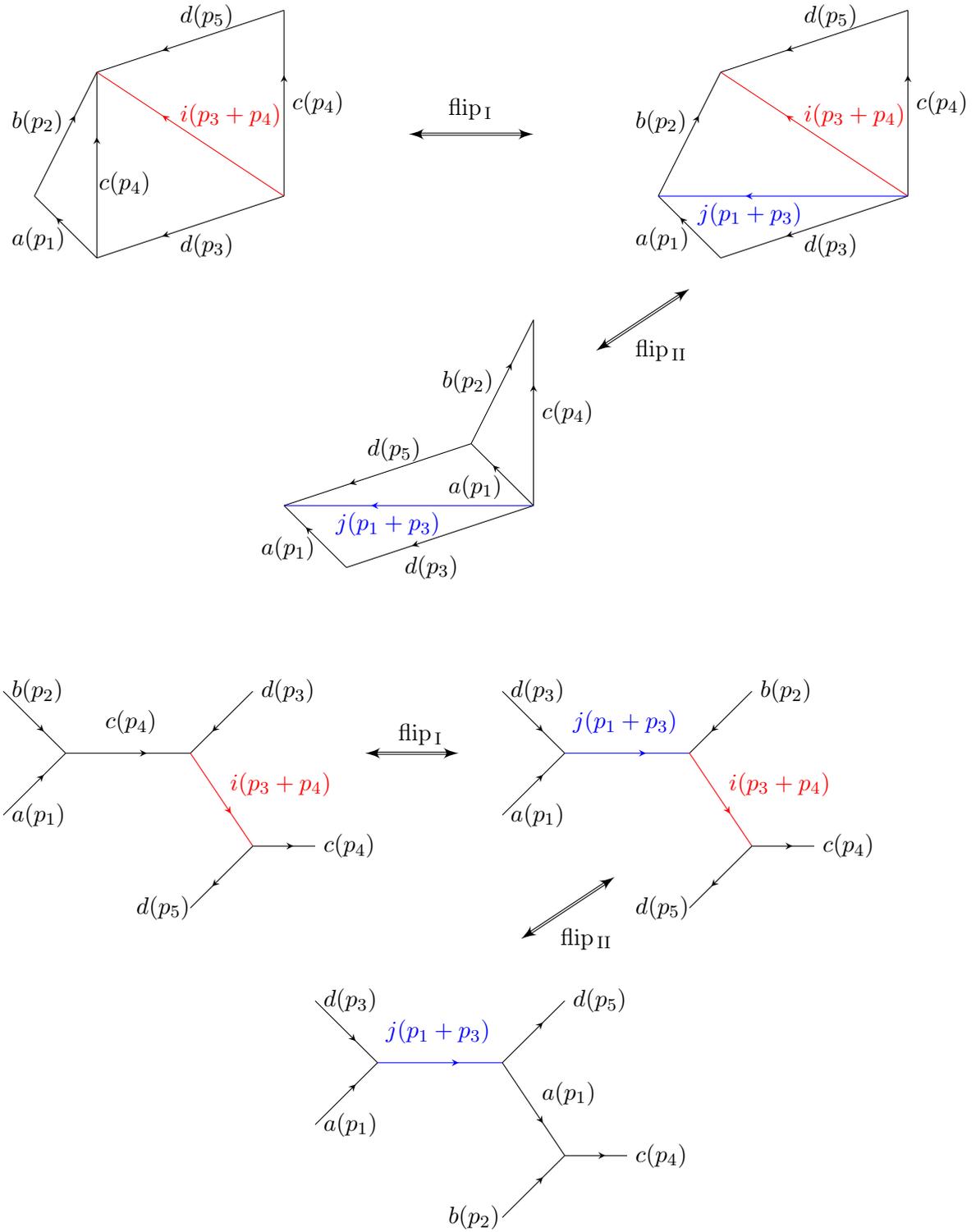
\begin{figure}
\begin{tikzpicture}
\draw[directed] (3.5,0) -- (4.5,2);
\draw[directed] (4.5,-1) -- (3.5,0);
\draw[directed] (4.5,-1) -- (4.5,2);
\draw[directed] (7.5,0) -- (4.5,-1);
\draw[directed] (7.5,3) -- (4.5,2);
\draw[directed] (7.5,0) -- (7.5,3);
\draw[color=red][directed] (7.5,0) -- (4.5,2);

\filldraw[black] (3,1.2)  node[anchor=west] {$b(p_2)$};
\filldraw[black] (3,-.7)  node[anchor=west] {$a(p_1)$};
\filldraw[black] (4.4,0.2)  node[anchor=west] {$c(p_4)$};
\filldraw[black] (5.7,-0.8)  node[anchor=west] {$d(p_3)$};
\filldraw[black] (5.7,2.9)  node[anchor=west] {$d(p_5)$};
\filldraw[black] (7.5,1.5)  node[anchor=west] {$c(p_4)$};
\filldraw[red] (5.7,1.3)  node[anchor=west] {$i(p_3+p_4)$};

\draw[latex'-latex',double] (9.5,1) -- (11.5,1);
\filldraw[black] (10,1.4)  node[anchor=west] {flip$_{\,\rm I}$};

\draw[directed] (3.5+10,0) -- (4.5+10,2);
\draw[directed] (4.5+10,-1) -- (3.5+10,0);
\draw[directed] (7.5+10,0) -- (4.5+10,-1);
\draw[directed] (7.5+10,3) -- (4.5+10,2);
\draw[directed] (7.5+10,0) -- (7.5+10,3);
\draw[color=red][directed] (7.5+10,0) -- (4.5+10,2);
\draw[directed][blue] (3.5+14,0) -- (3.5+10,0);

\filldraw[black] (3+10,1.2)  node[anchor=west] {$b(p_2)$};
\filldraw[black] (3+10,-.7)  node[anchor=west] {$a(p_1)$};
\filldraw[black] (5.7+10,-0.8)  node[anchor=west] {$d(p_3)$};
\filldraw[black] (5.7+10,2.9)  node[anchor=west] {$d(p_5)$};
\filldraw[black] (7.5+10,1.5)  node[anchor=west] {$c(p_4)$};
\filldraw[red] (5.7+10,1.3)  node[anchor=west] {$i(p_3+p_4)$};
\filldraw[blue] (14,-0.3)  node[anchor=west] {$j(p_1+p_3)$};

\draw[latex'-latex',double] (14,-1.5) -- (12.5,-2.5);
\filldraw[black] (13,-2.5)  node[anchor=west] {flip$_{\,\rm II}$};

\draw[directed] (6.5+4,-4) -- (7.5+4,-2);
\draw[directed] (7.5+4,-5) -- (6.5+4,-4);
\draw[directed] (7.5+4,-5) -- (7.5+4,-2);
\draw[directed] (7.5+4,-5) -- (4.5+4,-6);
\draw[directed] (6.5+4,-4) -- (3.5+4,-5);
\draw[directed] (4.5+4,-6) -- (3.5+4,-5);
\draw[directed][blue] (7.5+4,-5) -- (3.5+4,-5);

\filldraw[black] (3+4,-5.7)  node[anchor=west] {$a(p_1)$};
\filldraw[black] (6+4,-4.7)  node[anchor=west] {$a(p_1)$};
\filldraw[black] (5.9+4,-3)  node[anchor=west] {$b(p_2)$};
\filldraw[black] (7.5+4,-3.5)  node[anchor=west] {$c(p_4)$};
\filldraw[black] (5.3+4,-6)  node[anchor=west] {$d(p_3)$};
\filldraw[black] (4.7+4,-4.1)  node[anchor=west] {$d(p_5)$};
\filldraw[blue] (4.2+4,-5.3)  node[anchor=west] {$j(p_1+p_3)$};

\draw[directed] (11-7,1-10) -- (13-7,1-10);
\draw[directed] (10-7,0-10) -- (11-7,1-10);
\draw[directed] (10-7,2-10) -- (11-7,1-10);
\draw[directed] (14-7,2-10) -- (13-7,1-10);
\draw[directed][red] (13-7,1-10) -- (14-7,-0.5-10);
\draw[directed] (14-7,-0.5-10) -- (15-7,-0.5-10);
\draw[directed] (14-7,-0.5-10) -- (13-7,-1.5-10);

\filldraw[black] (10-7,2-10)  node[anchor=west] {$b(p_2)$};
\filldraw[black] (10-7,0-10)  node[anchor=west] {$a(p_1)$};
\filldraw[black] (11.5-7,1.5-10)  node[anchor=west] {$c(p_4)$};
\filldraw[black] (15-7,-0.5-10)  node[anchor=west] {$c(p_4)$};
\filldraw[black] (14-7,2-10)  node[anchor=west] {$d(p_3)$};
\filldraw[black] (12-7,-1.5-10)  node[anchor=west] {$d(p_5)$};
\filldraw[red] (13.5-7,0.5-10)  node[anchor=west] {$i(p_3+p_4)$};

\draw[latex'-latex',double] (8.8,-9) -- (10.3,-9);
\filldraw[black] (9.2,-8.7)  node[anchor=west] {flip$_{\,\rm I}$};

\draw[directed][blue] (12,1-10) -- (14,1-10);
\draw[directed] (11,0-10) -- (12,1-10);
\draw[directed] (11,2-10) -- (12,1-10);
\draw[directed] (15,2-10) -- (14,1-10);
\draw[directed][red] (14,1-10) -- (15,-0.5-10);
\draw[directed] (15,-0.5-10) -- (16,-0.5-10);
\draw[directed] (15,-0.5-10) -- (14,-1.5-10);

\filldraw[black] (11,2-10)  node[anchor=west] {$d(p_3)$};
\filldraw[black] (11,0-10)  node[anchor=west] {$a(p_1)$};
\filldraw[blue] (12,1.5-10)  node[anchor=west] {$j(p_1+p_3)$};
\filldraw[black] (16,-0.5-10)  node[anchor=west] {$c(p_4)$};
\filldraw[black] (15,2-10)  node[anchor=west] {$b(p_2)$};
\filldraw[black] (13,-1.5-10)  node[anchor=west] {$d(p_5)$};
\filldraw[red] (14.5,0.5-10)  node[anchor=west] {$i(p_3+p_4)$};

\draw[latex'-latex',double] (14.8-2,-13+2) -- (13.3-2,-14+2);
\filldraw[black] (11.8,-12)  node[anchor=west] {flip$_{\,\rm II}$};

\draw[directed][blue] (9,-4-10) -- (11,-4-10);
\draw[directed] (8,-5-10) -- (9,-4-10);
\draw[directed](8,-3-10) -- (9,-4-10);
\draw[directed]  (11,-4-10) -- (12,-3-10);
\draw[directed] (11,-4-10) -- (12,-5.5-10);
\draw[directed] (11,-6.5-10) -- (12,-5.5-10);
\draw[directed] (12,-5.5-10) -- (13,-5.5-10) ;

\filldraw[black] (8,-3-10)  node[anchor=west] {$d(p_3)$};
\filldraw[black] (8,-5-10)  node[anchor=west] {$a(p_1)$};
\filldraw[black][blue] (9,-3.5-10)  node[anchor=west] {$j(p_1+p_3)$};
\filldraw[black] (12,-3-10)  node[anchor=west] {$d(p_5)$};
\filldraw[black] (10,-6.5-10)  node[anchor=west] {$b(p_2)$};
\filldraw[black] (13,-5.5-10)  node[anchor=west] {$c(p_4)$};
\filldraw (11.5,-4.5-10)  node[anchor=west] {$a(p_1)$};

\end{tikzpicture}
\caption{Poles structure in the process $a + b + d \to c + d$. On the bottom and on the top are listed respectively the Feynman diagrams and their on-shell description.}
\label{fig5amplit_flipping}
\end{figure}

We start describing the first kind of poles, those due to simultaneous
singularities in $M_{cd}^{(4)}$, $M_{ad}^{(4)}$ and  $M_{bd}^{(4)}$. We represent an example of this situation in figure \ref{fig5amplit_flipping} where we have drawn both the Feynman diagrams (on the bottom) and their dual description in terms of vectors in the complex plane (on the top). We suppose there is a propagating $i$-particle going on-shell in $M_{cd}^{(4)}$ (it is represented by a red line in the first picture of figure \ref{fig5amplit_flipping}). Looking at the quadrilateral defined by vectors $a(p_1)$, $b(p_2)$, $i(p_3+p_4)$ and $d(p_3)$ we note that the flipping rule can be applied on the propagating particle $c(p_4)$. As explained previously there are two possibilities for the flip, one case in which $a(p_1)+d(p_3)=i(p_3+p_4)-b(p_2)=j(p_1+p_3)$ where $j$ is an on-shell propagating particle, and one in which the on-shell propagating particle is given by $b(p_2)+d(p_4)=i(p_1+p_3)-a(p_1)$. We are assuming that we cannot have both at the same time. In figure \ref{fig5amplit_flipping} the first situation is shown; it is evident how, after having applied two flips, the same values of momenta which contribute to a pole in $M_{cd}^{(4)}$ generate another pole in the amplitude $M_{ad}^{(4)}$. If the particles were all different we could continue to flip internal propagating particles and we would obtain a closed network with a finite number of diagrams connected by simple flips. In this situation since two particles are identical after two flips the network is completed and we have obtained all the divergent diagrams.
Focusing on the situation shown in figure \ref{fig5amplit_blobdiagr} in which $M_{cd}^{(4)}$ and $M_{ad}^{(4)}$ diverge simultaneously we compute the residue of the RHS of~\eqref{Equation_Added_eq4_7} with respect to the variable $a_3$.
Close to the pole position, at which the $i$- and $j$-particle are on-shell, \eqref{Equation_Added_eq4_7} becomes proportional to
\begin{equation}\label{NewNeweq4_8}
\frac{1}{p_i^2-m_i^2}  \biggl[  \frac{\mid C^{(3)}_{cd\bar{i}} \mid^2}{\Delta_{c d i}^2} - \frac{\mid C^{(3)}_{ad\bar{j}} \mid^2}{\Delta^2_{a d j}} \biggr].
\end{equation}
A detailed derivation of this expression is given in appendix \ref{App:0_2}.  Substituting \eqref{New_eq4_6} into~\eqref{NewNeweq4_8} and requiring the quantity in square 
brackets to be zero implies the following equality 
\begin{equation}\label{NewNeweq4_9}
\mid f_{cd\bar{i}} \mid^2 = \mid f_{ad\bar{j}} \mid^2 .
\end{equation}
In fact, this relation follows from the properties already imposed in the cancellation of $2$ to $2$ off-diagonal processes, as follows. Looking at the three on-shell diagrams of
figure~\ref{fig5amplit_flipping} we note that the following relations hold,
\begin{equation}
\label{double_equality_following_the_flip_flow}
f_{a b \bar{c}} f_{c d \bar{i}} f_{\bar{c} \bar{d}i} \underset{\text{\tiny{type I flip}}}{=}
f_{a d \bar{j}} f_{jb\bar{i}} f_{\bar{c} \bar{d}i}\underset{\text{\tiny{type II flip}}}{=} f_{a d \bar{j}} f_{\bar{a} \bar{d} j} f_{a b \bar{c}},
\end{equation}
where we used the sign rules satisfied by the different types of flips summarized in figure~\ref{2_to_2_scattering_tree_level_simultaneous_poles_changing_sign}. Simplifying the common factor $f_{ab\bar{c}}$ in the first and in the third terms in~\eqref{double_equality_following_the_flip_flow} and using~\eqref{New_eq4_11} we obtain exactly the relation~\eqref{NewNeweq4_9} that is therefore simply a consequence of the flipping rule.
From now on we refer to these singularities as
`flipped singularities' since they cancel simply using the flipping rule.

The relation~\eqref{NewNeweq4_9} is consistent with the area rule~\eqref{Area_rule_abs_f_value}. This consistency is due to the fact that we are assuming, as part of the simply-laced scattering conditions, that the flipping rule happens between pairs of graphs. If we violated this condition, having an $s$-,
$t$- and $u$-channel particle
simultaneously on-shell in the quadrilateral
defined by $a(p_1)$, $b(p_2)$, $i(p_3+p_4)$, $d(p_3)$ in figure
\ref{fig5amplit_flipping}, not only $c$ and $j$ would contribute to
the pole but also another on-shell particle, $l$ say, a bound state of
$b(p_2)$ and $d(p_3)$. In this situation 
an extra contribution $\frac{\mid C^{(3)}_{bd\bar{l}} \mid^2}{\Delta^2_{b d l}}$ would be present in the square brackets of   \eqref{NewNeweq4_8}, providing another term in the relation \eqref{NewNeweq4_9} that would become
\begin{equation}\label{NewNeweq4_9bis}
\mid f_{cd\bar{i}} \mid^2 = \mid f_{ad\bar{j}} \mid^2+\mid f_{bd\bar{l}} \mid^2 .
\end{equation}
In this situation we see that for a given $d$ particle the absolute value of its corresponding $f$-function depends on the particles with which it couples according with a set of constraints of the form expressed in \eqref{NewNeweq4_9bis}. This is what generally happens in non simply-laced affine Toda theories. 

However, though the relation~\eqref{NewNeweq4_9} agrees with the area rule~\eqref{Area_rule_abs_f_value}, it is  not enough to prove it. At a first glance it may seem
from the expression in \eqref{NewNeweq4_9}
that given a generic particle $d$, the absolute value of its
$f$-function $f_{d i j}$ does not depend on the particles $i$ and
$j$ with which it couples. Such a conclusion is too hasty, since the indices $\{c, \bar{i}\}$ and $\{a, \bar{j}\}$ appearing on the LHS and on the RHS of~\eqref{NewNeweq4_9} are not arbitrary but correspond to particular on-shell channels inside the amplitudes $M_{cd}^{(4)}$ and $M_{ad}^{(4)}$. To show that the area rule is universally satisfied by all the models respecting property~\ref{Simply_laced_scattering_conditions} we need to study also the second kind of singularities, happening when the momenta
of two particles become collinear; we refer to these as
`collinear singularities'.

This second situation can arise only when
we have at least three equal particles in the scattering. Indeed suppose we
take the limit $a_3 \to a_4$; in this situation the term in front of
$M^{(4)}_{cd}(a_3,a_4)$ in equation 
\eqref{Equation_Added_eq4_7}
becomes infinity but at
the same time, if $c$ and $d$ are particles of different types, we
also have $M^{(4)}_{cd}(a_3,a_4) \to 0$ . This is indeed the limit in which
the transmitting and the reflecting processes involving the $2$ to $2$
scattering of the particles $c$ and $d$ become equal. Since the latter
event is forbidden in any integrable theory in this limit both the
processes need to be zero and collinear
singularities are not allowed. 

\begin{figure}
\begin{tikzpicture}
\draw[directed] (3.5,0) -- (4.5,2);
\draw[directed] (4.5,-1) -- (3.5,0);
\draw[directed] (4.5,-1) -- (4.5,2);
\draw[directed] (4.5,-4) -- (4.5,-1);
\draw[directed] (4.6,-4) -- (4.6,-1);
\draw[directed] (4.6,-1) -- (4.6,2);
\filldraw[color=black!60, fill=black!5, very thick](4.55 ,-1) circle (0.02);
\filldraw[color=black!60, fill=black!5, very thick](4.55 ,2) circle (0.02);

\filldraw[black] (3,1.2)  node[anchor=west] {$b(p_2)$};
\filldraw[black] (3,-.7)  node[anchor=west] {$a(p_1)$};
\filldraw[black] (3.5,0.2)  node[anchor=west] {$c(p_4)$};
\filldraw[black] (4.6,0.2)  node[anchor=west] {$c(p_5)$};
\filldraw[black] (3.5,-2.5)  node[anchor=west] {$c(p_3)$};
\filldraw[black] (4.6,-2.5)  node[anchor=west] {$c(p_4)$};

\draw[directed] (8.5,0) -- (9.5,2);
\draw[directed] (9.5,-1) -- (8.5,0);
\draw[directed] (9.5,-1) -- (9.5,2);
\draw[directed] (9.5,-4) -- (9.5,-1);
\draw[directed] (8.5,-3) -- (8.5,0);
\draw[directed] (9.5,-4) -- (8.5,-3);
\draw[directed][red] (8.5,-3) -- (9.5,-1);

\filldraw[black] (8,1.2)  node[anchor=west] {$b(p_2)$};
\filldraw[black] (8.5,0)  node[anchor=west] {$a(p_1)$};
\filldraw[black] (9.5,0.2)  node[anchor=west] {$c(p_4)$};
\filldraw[black] (9.6,-2.5)  node[anchor=west] {$c(p_5)$};
\filldraw[black] (7.5,-1.5)  node[anchor=west] {$c(p_3)$};
\filldraw[black] (8.2,-4)  node[anchor=west] {$a(p_1)$};
\filldraw[red] (8.5,-1.3)  node[anchor=west] {$b(p_2)$};

\draw[directed] (13.5,-3) -- (14.5,-1);
\draw[directed] (14.5,-4) -- (13.5,-3);
\draw[directed] (14.5,-4) -- (14.5,-1);
\draw[directed] (14.5,-1) -- (14.5,2);
\draw[directed] (13.5,-3) -- (13.5,0);
\draw[directed] (13.5,0) -- (14.5,2);
\draw[directed][red] (14.5,-1) -- (13.5,0);

\filldraw[black] (13,1.2)  node[anchor=west] {$b(p_2)$};
\filldraw[red] (13.5,0)  node[anchor=west] {$a(p_1)$};
\filldraw[black] (14.5,0.2)  node[anchor=west] {$c(p_5)$};
\filldraw[black] (14.6,-2.5)  node[anchor=west] {$c(p_4)$};
\filldraw[black] (12.5,-1.5)  node[anchor=west] {$c(p_3)$};
\filldraw[black] (13.2,-4)  node[anchor=west] {$a(p_1)$};
\filldraw[black] (13.5,-1.3)  node[anchor=west] {$b(p_2)$};

\end{tikzpicture}
\caption{Infinite contributions to the residue of the amplitude $a + b + c \to c + c$ when collinear singularities are present. The infinities sum to zero so that in the end the residue is finite. }
\label{fig5amplit_collinear}
\end{figure}
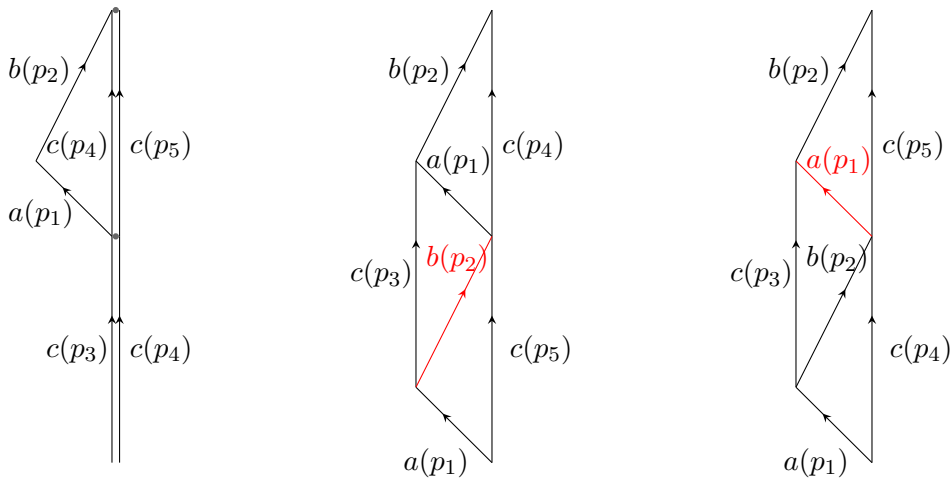

The story is different if the labels $c$ and $d$ are equal. In this
case  transmission cannot be distinguished from reflection and
$M^{(4)}_{cc}(a_3,a_4)$ (we give particles $c$ and $d$  the same
label since we are analysing the situation in which they are equal)
does not go to zero for $a_3 \to a_4$; so it is possible to have
singularities due to collinear momenta. In particular if $d=c$, in the
limit $a_3 \to a_4$ the amplitude $M_{cc}^{(4)}(a_4,a_3)$ is nonzero
while the term in front of it in equation 
\eqref{Equation_Added_eq4_7}
goes to
infinity. However in this case other poles appear in the amplitudes
$M_{bc}^{(4)}$ and $M_{ac}^{(4)}$ that cancel this singularity. A
picture of this situation is shown in figure
\ref{fig5amplit_collinear} where we highlight how the collinear
singularity in the first graph is accompanied by two poles due to two on-shell  
particles propagating in $M_{bc}^{(4)}$ and $M_{ac}^{(4)}$.
An explicit computation of the residue is reported in appendix \ref{App:0_2}. In order to cancel the residue for this kind of pole we need to require that in the collinear limit the scattering amplitude for a process of the form $c+c \to c+c$ is given by
\begin{equation}\label{eq4_8}
M^{(4)}_{cc}\Bigl|_{collinear}=-\frac{i}{4} m_c^2 \mid f_{\bar{a} \bar{b} c} \mid^2 .
\end{equation}
Such relations allow the $4$-point coupling
$C^{(4)}_{cc\bar{c}\bar{c}}$ to be fixed in terms of the $3$-point couplings also
in those cases in which the procedure described in
\ref{4pointsby3points} cannot be applied (i.e. when the $4$-point
amplitude is diagonal). 
We note that the absolute value of a function
$f_{\bar{a} \bar{b} c}$ containing a particle of type $c$ does not
depend on the particles $a$ and $b$ with which it couples, being the LHS  
of the equality~\eqref{eq4_8} independent on $a$ and $b$.
In contrast to~\eqref{NewNeweq4_9}, the particles $a$ and $b$ are completely arbitrary, being any pair of particles coupling with $c$. For this
reason we can state that any theory satisfying the simply-laced
scattering conditions reported in~\ref{Simply_laced_scattering_conditions} respects an area rule of the form \eqref{Area_rule_abs_f_value}. Indeed, given the arbitrariness of the interacting particles, from the relation in \eqref{eq4_8} we can have only two possible situations: either there exist two decoupled sectors of the theory that do not interact each other and can have two possible different absolute value of their $f$-functions, or, if all the particles are connected, there exists only one possible value of $|f_{ijk}|$ that needs to be common to all the $3$-point couplings. This implies that the expression in \eqref{eq4_8} can consistently be written as 
\begin{equation}\label{eq4_8bis0}
M^{(4)}_{cc}\Bigl|_{collinear}=-\frac{i}{4} m_c^2 \mid f \mid^2 
\end{equation}
where $|f|$ is the common value to all the absolute values of the $f$-functions.
The combination of \eqref{eq4_8bis0} with the area rule~\eqref{Area_rule_abs_f_value} ensures, in theories satisfying simply-laced scattering conditions, that the residue of the amplitude at the pole $a_5\to a_3$ is equal to zero, independently of the value of $a_3$.  Therefore the amplitude has no singularities in $a_5$ and it is a constant in this variable. The entire discussion can be repeated identically for the variable $a_3$ so that we have
$$
\frac{\partial M^{(5)}}{\partial a_3}=\frac{\partial M^{(5)}}{\partial a_5}=0,
$$
and $M^{(5)}$ is constant everywhere.

As a simple check that everything is working correctly,
we consider a theory of a single real scalar field of mass $m$. In this case the $3$-point coupling is given by the relation in \eqref{New_eq4_6} and can be written as
\begin{equation}\label{eq4_8_bis1}
C^{(3)}= \frac{\sqrt{3}}{4} m^2 \lambda
\end{equation}
where on the right hand side we have written the value of the area of an equilateral triangle of side $m$ and $\lambda$ corresponds to the value of the $f$-function in equation \eqref{New_eq4_6}.
By a direct calculation of the scattering amplitude for a $2$ to $2$ collinear process in this theory we obtain
\begin{equation}\label{eq4_8_bis2}
M^{(4)}_{cc}\Bigl|_{collinear}= (-i C^{(3)})^2 \frac{i}{4m^2-m^2}-(-i C^{(3)})^2 \frac{i}{m^2}-(-i C^{(3)})^2 \frac{i}{m^2}-iC^{(4)} .
\end{equation}
Comparing this expression with \eqref{eq4_8bis0}, where in this case $\mid f \mid^2=\lambda^2$, we find that the value of the $4$-point coupling cancelling poles in $5$-point events needs to be
\begin{equation}\label{eq4_8_bis3}
C^{(4)}= \frac{9}{16} m^2 \lambda^2 .
\end{equation}
If we write the first few terms of the Lagrangian that we are constructing 
\begin{equation}\label{eq4_8_bis4}
\mathcal{L}=\frac{1}{2}\partial_\mu \phi \partial^\mu \phi-\frac{m^2}{2}\phi^2 -\frac{1}{3!}\frac{\sqrt{3}}{4} m^2 \lambda \phi^3-\frac{1}{4!} \frac{9}{16}m^2 \lambda^2 \phi^4 + \ldots
\end{equation}
we note that they are the lower orders in the expansion of the following Bullough-Dodd theory
\begin{equation}\label{eq4_8_bis5}
\mathcal{L}=\frac{1}{2}\partial_\mu \phi \partial^\mu \phi-\frac{8}{9} \frac{m^2}{\lambda^2} \Bigl( e^{2 \frac{\sqrt{3}}{4} \lambda \phi} + 2 e^{-\frac{\sqrt{3}}{4} \lambda \phi} - 3 \Bigr) .
\end{equation}
All the other couplings can be obtained from \eqref{eq4_8_bis1} and \eqref{eq4_8_bis3} acting iteratively with \eqref{eq0_6} and it can be shown that they match with the expansion of \eqref{eq4_8_bis5}.

\subsection{No-particle production in 6-point processes and factorisation}\label{No particles production in 6-point processes}

\begin{figure}
\begin{tikzpicture}

%Image 1
\draw[directed] (6,0) -- (7,0);
\draw[directed] (6.5,-1.6) -- (7.2,-0.6);
\draw[directed] (7.9,0.5) -- (8.6,1.5);

\draw[directed] (8.3,0) -- (9.5,0);
\filldraw[black] (8.8,0.4)  node[anchor=west] {$c$};

\draw[directed] (10.2,0.7) -- (10.2,1.7);
\draw[directed] (10.9,0) -- (11.9,0);
\draw[directed] (10.2,-1.8) -- (10.2,-0.6);

\filldraw[color=red!60, fill=red!5, very thick](7.6,0) circle (0.7);
\filldraw[color=red!60, fill=red!5, very thick](10.2,0) circle (0.7);

\filldraw[black] (5,0)  node[anchor=west] {$c(p_3)$};
\filldraw[black] (6,-2)  node[anchor=west] {$a(p_1)$};
\filldraw[black] (7.8,1.8)  node[anchor=west] {$a(p_4)$};

\filldraw[black] (11.8,0.1)  node[anchor=west] {$c(p_6)$};
\filldraw[black] (10.2,1.4)  node[anchor=west] {$b(p_5)$};
\filldraw[black] (10.2,-1.7)  node[anchor=west] {$b(p_2)$};

%Image 2
\draw[directed] (6+9,0) -- (7+9,0);
\draw[directed] (9+9,-1.6) -- (9.7+9,-0.6);
\draw[directed] (10.4+9,0.5) -- (11.1+9,1.5);

\draw[directed] (8.3+9,0) -- (9.5+9,0);
\filldraw[black] (8.8+9,0.4)  node[anchor=west] {$c$};

\draw[directed] (7.6+9,0.7) -- (7.6+9,1.7);
\draw[directed] (10.9+9,0) -- (11.9+9,0);
\draw[directed] (7.6+9,-1.8) -- (7.6+9,-0.6);

\filldraw[color=red!60, fill=red!5, very thick](7.6+9,0) circle (0.7);
\filldraw[color=red!60, fill=red!5, very thick](10.2+9,0) circle (0.7);

\filldraw[black] (5+9,0)  node[anchor=west] {$c(p_3)$};
\filldraw[black] (8.8+9,-2)  node[anchor=west] {$a(p_1)$};
\filldraw[black] (10.2+9,1.7)  node[anchor=west] {$a(p_4)$};

\filldraw[black] (11.8+9,0.1)  node[anchor=west] {$c(p_6)$};
\filldraw[black] (6.6+9,1.4)  node[anchor=west] {$b(p_5)$};
\filldraw[black] (6.6+9,-1.7)  node[anchor=west] {$b(p_2)$};

\end{tikzpicture}
\caption{Allowed poles in the $6$-point amplitude $a+b+c \to a+b+c$. The sum of these two contributions is nonzero only in a subregion of the momentum space where the sets of incoming and outgoing momenta are equal.}
\label{fig6amplit_pairs_cancellation}
\end{figure}
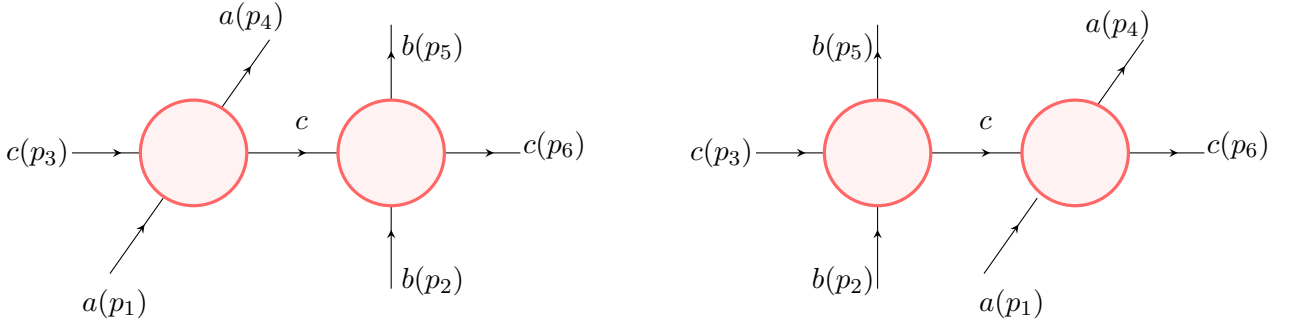

Events involving $6$ external particles are relatively simple to analyse
once we know that non-diagonal scattering is not allowed in $4$- and $5$-point processes. In this case the most general process not trivially equal to zero is given by
\begin{equation}\label{eq4_9}
a(p_1)+b(p_2)+c(p_3) \rightarrow a(p_4)+b(p_5)+c(p_6).
\end{equation}
Indeed, if there were more than three different particles then any time an internal propagator goes on-shell it would factorise the amplitude into two processes of which at least one is inelastic, generating a zero residue. For the event represented in equation \eqref{eq4_9}, all the poles also cancel. The reason is that such poles appear always in copies as shown in figure \ref{fig6amplit_pairs_cancellation}. The two diagrams in figure \ref{fig6amplit_pairs_cancellation} are equal except for the fact that in the limit $a_4 \to a_1$ (i.e. on the pole) they present two propagators with opposite sign, that are given respectively by
\begin{equation}\label{eq4_10}
G_1=\frac{i}{m_a m_c} \frac{-a_1^2 a_3}{a_1 - a_4} \frac{1}{a_3^2-a_1^2}
\end{equation}
and
\begin{equation}\label{eq4_11}
G_2=\frac{i}{m_a m_c} \frac{a_1^2 a_3}{a_1 - a_4} \frac{1}{a_3^2-a_1^2}
\end{equation}
This suffices to prove that the sum of the two singularities in the diagrams is equal to zero. 

The situation is different if we maintain the $i \epsilon$ prescription in the propagators. In this case each propagator can be written in terms of its principal value and its delta contribution. While the former cancels once it is summed to all the other diagrams (indeed the divergent parts of the principal values of $G_1$ and $G_2$ sum to zero while the remaining finite contribution is cancelled by all the other non-divergent diagrams) the latter gives a nonzero result. We can prove this using the distribution formula

\begin{equation}
\lim_{\epsilon \to 0^+} \Bigl( \frac{1}{x+i\epsilon} - \frac{1}{x-i\epsilon}\Bigr)= (-2 \pi i) \delta(x) .
\end{equation}
By a direct sum of the propagators in \eqref{eq4_10}, \eqref{eq4_11} and considering the extra $i \epsilon$ factor in the denominators we obtain
\begin{equation}
\label{additional_delta_function_coming_in_six_point_scattering}
\begin{split}
G_1+G_2&=\frac{i}{m_a m_c} a_1^2 a_3 \lim_{\epsilon \to 0^+} \Biggl( \frac{1}{(a_1-a_4)(a_1^2-a_3^2)+\frac{i a_1^2 a_3 \epsilon}{m_a m_c}} - \frac{1}{(a_1-a_4)(a_1^2-a_3^2)-\frac{i a_1^2 a_3 \epsilon}{m_a m_c}} \Biggr)\\
&=\frac{2 \pi}{m_a m_c} \frac{a_1^2 a_3}{|a_1^2-a_3^2|} \delta(a_1-a_4) = \frac{\pi}{m_a m_c} \frac{\delta(\theta_1 - \theta_4)}{|\sinh \theta_{13}|}
\end{split}\end{equation}
where we define $\theta_{13}$ the difference between the rapidities of the $a$- and $c$-particle, having respectively momenta $p_1$ and $p_3$.
Using such result by a direct sum of the two diagrams in figure \ref{fig6amplit_pairs_cancellation} and multiplying by the extra factor
\begin{equation*}
(2 \pi)^2 \delta^{(2)} \Bigl(\sum_{i=1,2,3} p_i-\sum_{k=4,5,6} p_k \Bigr)\equiv (2 \pi)^2 \delta \Bigl( \sum_{i=1,2,3} p^0_i-\sum_{k=4,5,6} p^0_k \Bigr) \ \delta \Bigl( \sum_{i=1,2,3} p^1_i-\sum_{k=4,5,6} p^1_k \Bigr)
\end{equation*}
coming from the conservation of the total momentum we obtain
\begin{equation}
\label{6_point_interaction_in_which_propagates_c_in_the_middle}
4\pi^3 \ \frac{1}{m_a m_c |\sinh \theta_{13}|} M^{(4)}_{ac}(a_1,a_3) \frac{1}{m_b m_c |\sinh \theta_{23} |} M^{(4)}_{bc}(a_2,a_3) \ \delta(\theta_1 - \theta_4) \delta(\theta_2 - \theta_5) \delta(\theta_3 - \theta_6).
\end{equation}
In \eqref{6_point_interaction_in_which_propagates_c_in_the_middle} we exploited the fact that the additional delta function arising from \eqref{additional_delta_function_coming_in_six_point_scattering} constraints the possible space of outgoing momenta to a smaller subregion. In particular we used the following equality
$$
\delta(\theta_1-\theta_4) \delta^{(2)} \Bigl(\sum_{i=1,2,3} p_i-\sum_{k=4,5,6} p_k \Bigr) =\delta(\theta_1-\theta_4) \delta^{(2)} \bigl( p_2 + p_3-p_5 - p_6 \bigr) = \delta (\theta_1 - \theta_4) \frac{\delta (\theta_2 - \theta_5) \delta (\theta_3 - \theta_6)}{m_b m_c |\sinh \theta_{23} |} 
$$
to factorise the $6$-point scattering into a product of two $4$-point amplitudes. 
Inserting then the normalisation factor (a multiplicative term $\frac{1}{\sqrt{4 \pi}}$ for each external particle) and adding the contribution \eqref{6_point_interaction_in_which_propagates_c_in_the_middle} to the other two pairs of diagrams similar to the one shown in figure~\ref{fig6amplit_pairs_cancellation}, but with particles $a$ and $b$ propagating in the middle, we find that the final $6$-point S-matrix is given by
\begin{equation}
\label{total_6_point_S_matrix_abc_to_abc}
\begin{split}
S^{tree}_{abc}(\theta_1, \theta_2, \theta_3)&= \bigl[S^{tree}_{ac}(\theta_{13}) S^{tree}_{bc}(\theta_{23}) + S^{tree}_{ab}(\theta_{12}) S^{tree}_{bc}(\theta_{23}) + S^{tree}_{ab}(\theta_{12}) S^{tree}_{ac}(\theta_{13}) \bigr] \\
& \times \delta(\theta_1 - \theta_4) \delta(\theta_2 - \theta_5) \delta(\theta_3 - \theta_6).
\end{split}
\end{equation}
In the equality above the tree-level part of the $2$ to $2$ S-matrix is bound to the $4$-point amplitude through the relation \eqref{Tree_level_bootstrap_conversion_between_S_matrix_in_theta_momentum_space} which is valid at all the orders in perturbation theory.

It is interesting to note how equation \eqref{total_6_point_S_matrix_abc_to_abc} exactly matches the factorisation requirement we expect to see at the tree level. This fact needs to be valid at any order in the coupling if the theory is integrable and therefore must hold order by order in perturbation theory
\begin{equation}
 \begin{split}
S_{abc}(\theta_1, \theta_2, \theta_3)&= S_{ab} (\theta_{12}) S_{bc} (\theta_{23}) S_{ac} (\theta_{13})\\
&=\bigl(1+ S^{tree}_{ab} (\theta_{12}) + \ldots \bigr) \bigl(1+S^{tree}_{bc} (\theta_{23}) + \ldots \bigr)  \bigl(1+S^{tree}_{ac} (\theta_{13}) + \ldots \bigr).
\end{split}
\end{equation}
Summarising, all the non-diagonal $6$-point processes are null if we fix the $6$-point vertices appropriately through \eqref{eq0_6}, while the diagonal processes are nonzero only in a small  region of the momentum space, exactly when the amplitude is factorised into the product of $2$ to $2$ interactions.

\section{Tree level integrability from root systems in affine Toda field theories} \label{sect1}

We now move on from general considerations to
show how the requirements found previously to forbid inelastic scattering at tree level
are universally satisfied in all of the affine Toda field theories. In these models the masses and nonzero couplings are related to the geometry of the underlying root systems and we are able to use this to provide a general tree level proof of their perturbative integrability.
We use the conventions and properties recorded in appendix \ref{App:1}.

These theories have been known from many years to be integrable, at least classically \cite{a1,a2}, since they possess an infinite number of higher-spin conserved charges in involution. Their S-matrix elements can be conjectured through the bootstrap principle \cite{a16,a18,a19,a20,aa20,a21,a22,a23,a25,a26,Corrigan:1993xh,Dorey:1993np,Oota:1997un,aa24,Fring:1991gh}, and have passed many consistency checks \cite{Braden:1990qa,Braden:1991vz,Braden:1990wx,Sasaki:1992sk,Braden:1992gh}. At loop level, the behaviour of the twisted and untwisted non simply-laced theories is complicated by the fact that the mass ratios become coupling-dependent, as discussed in  \cite{a25,a26,Corrigan:1993xh,Dorey:1993np,Oota:1997un}. But even at tree level, a complete proof of the tree-level perturbative integrability of general affine Toda field theories, in the sense of this paper, has so far been missing.

\subsection{Affine Toda Lagrangian and couplings}
Affine Toda field theories are a particular class of (1+1)-dimensional quantum field theories describing the interaction of $r$ bosonic scalar fields $\phi=(\phi_1,\ldots,\phi_r)\in\mathbb{R}^r$.
Each theory is defined in terms of a set of $r+1$ vectors $\{\alpha'_0\dots\alpha'_r\}$, all lying in $\mathbb{R}^r$, whose inner products are encoded in one of the twisted or untwisted affine Dynkin diagrams, together with a mass scale $m$ and a coupling\footnote{The choice of using $\g$ instead of the more standard notation $\beta$ to label the interaction scale of the model is because $\beta$ will be used below as a root label.}  $\g$. The Lagrangian is 
\begin{equation}\label{eq2_1}
\mathcal{L}=\frac{1}{2} \bigl(\partial_\mu \phi, \partial^\mu \phi \bigr)-V
\end{equation}
where the potential is given by
\begin{equation}
    \label{the_standard_affine_Toda_potential}
V=\frac{m^2}{\g^2}\sum_{i=0}^r n_i e^{\g \cdot (\phi, \alpha_i')}.
\end{equation}
The integers $n_i$ appearing in the Lagrangian are characteristic for each algebra. The set $\{\alpha'_i\}_{i=1}^r$ comprises the simple roots of the algebra and the additional root $\alpha'_0$, necessary to have a stable vacuum, is defined by
$$
\alpha'_0=-\sum_{i=0}^r n_i \alpha'_i,
$$
in such a way that, having imposed $n_0=1$, we have
$$
\sum_{i=0}^r n_i \alpha'_i=0.
$$
The last condition ensures that the Taylor expansion of the potential around $\phi=0$ does not have a linear term in $\phi$ and therefore the point $\phi=0$ is a minimum around which it is possible to perform standard perturbation theory.

An important observation of Freeman \cite{Freeman:1991xw} is that is possible to write the potential of a generic untwisted affine Toda field theory in a way that makes many perturbative properties of the model more explicit, as
\begin{equation}\label{eq2_2}
V=\frac{m^2}{\g^2} \Bigl(e^{\g \cdot ad_\phi} z_1, z_{h-1} \Bigr).
\end{equation}
The elements $z_1$, $z_{h-1}$ are defined in \eqref{eq1_22} in terms of the generators $\{E_{\pm \alpha'_i}\}_{i=0}^r$ associated to a particular Cartan subalgebra $\mathcal{H'}$; it is possible to check that the two expressions in \eqref{the_standard_affine_Toda_potential} and \eqref{eq2_2} are 
the same by substituting \eqref{eq1_22} into \eqref{eq2_2} and using
\begin{equation}
\label{An_adjoint_representation_relation}
e^{\g \cdot ad_\phi} E_{\alpha'}=e^{\g \cdot (\phi,\alpha')} E_{\alpha'}
\end{equation}
together with \eqref{eq1_20_1}. This is enough to prove that the potential in \eqref{eq2_2} is exactly the same as the potential appearing, with a more standard notation, in~\eqref{the_standard_affine_Toda_potential}.

On the other hand the elements $z_1$ and $z_{h-1}$ are the eigenvectors of the Coxeter element associated to the root system $\Phi$, composed by the roots $\{ \alpha \}$ living in a different Cartan subalgebra $\mathcal{H}$ spanned by the set of generators $\{ h_a \}_{a=1}^r$ in \eqref{eq1_0}.
By the fact that $\{ \frac{1}{\sqrt{h}}y_a \}_{a=1}^r$ is an orthonormal basis in $\mathcal{H}'$  (see \eqref{eq1_17})
the field can be expanded on this basis in the following way
\begin{equation}\label{eq2_0}
\phi=\Bigl( \phi , \frac{1}{\sqrt{h}} y^{\dagger}_a \Bigr)  \frac{1}{\sqrt{h}} y_a \equiv \phi_a \frac{1}{\sqrt{h}} y_a
\end{equation}
and the kinetic term of the Lagrangian can be written as
\begin{equation}\label{eq2_3}
\frac{1}{2} \bigl(\partial_\mu \phi, \partial^\mu \phi \bigr)=\frac{1}{2} \partial_\mu \phi_a \partial^\mu \phi_{\bar{a}} 
\end{equation}
 where, as we previously said, we are working with the convention $\phi_{\bar{a}}=\phi_{a}^*$.
 Taylor expanding the operator $e^{\g \cdot ad_{\phi}}$ and using the relation in \eqref{eq2_0} the potential can be written in the following form
\begin{equation}\label{eq2_5}
V= \sum_{n=0}^{\infty}\frac{1}{n!} C^{(n)}_{a_1 a_2 \ldots a_n} \phi_{a_1}\ldots \phi_{a_n}
\end{equation}
with the couplings given by
\begin{equation}\label{eq2_6}
C^{(n)}_{a_1 a_2 \ldots a_n}= \frac{m^2}{h^{\frac{n}{2}}} \g^{n-2}  \Bigl( [y_{a_1}, [y_{a_{2}} ,[\ldots [y_{a_n}, z_1] \ldots ]], z_{h-1} \Bigr) .
\end{equation}
In order to have a potential formulation making more explicit the properties of the theory, we substitute the explicit form of $\{ y_a \}_{a=1}^r$ \eqref{eq1_16} in \eqref{eq2_6}.  After having used the fact that $z_1$ and $z_{h-1}$ belong to the Cartan subalgebra $\mathcal{H}$ and the properties in \eqref{Coxeter_geometry_projections_of_phi_in_complex_notation} we obtain the following expression for the couplings 
\begin{equation}\label{eq2_11}
\begin{split}
&C^{(n)}_{a_1, a_2, \ldots , a_n}=\\
&-i^n\frac{ m^2}{h^{\frac{n}{2}}} \g^{n-2} 2 h \frac{\gamma_{a_1}^2 q_1^{a_1}\ \gamma_{a_n}^2 q_1^{a_n}}{|\textbf{Q}_1|^2} \sum_{\alpha_1 \in \Gamma_{a_1} \ldots \alpha_n \in \Gamma_{a_n}} \bigl( [e_{\alpha_{2}},[e_{\alpha_{3}},[\ldots [e_{\alpha_{n-1}}, e_{\alpha_{n}}]\ldots]]], e_{\alpha_1} \bigr)  e^{i(U_{\alpha_n}-U_{\alpha_1})} .
\end{split}
\end{equation}
The sum runs over the roots inside the different orbits $\Gamma_{a_1}, \ldots , \Gamma_{a_n}$ and $\vec{Q}_1$ is defined as in \eqref{definition_of_big_Q_in_terms_of_q_and_alphasquares}.  $\vec{q}_1$ is the eigenvector of the transpose of the Cartan matrix with smallest eigenvalue; it contains $r$ entries $q_1^a$ ($a=1,\dots, r$), one for each orbit of the root system. $U_{\alpha_1}$ and $U_{\alpha_n}$ are the angles that the different roots $\alpha_1 \in \Gamma_{a_1}$ and $\alpha_n \in \Gamma_{a_n}$ generate when are projected on the spin-1 eigenplane of the Coxeter element $w$.

We now proceed to investigate this relation for various values of $n$ so as to find useful expressions for the different couplings of the theory.   
 
\subsection{Masses}
Writing the general formula \eqref{eq2_11} in the case $n=2$ we have
\begin{equation}\label{eq2_12}
C^{(2)}_{a b}=2m^2 \gamma_a^2 \gamma_b^2 \frac{q_1^{a} \ q_1^{b}}{|\textbf{Q}_1|^2} \sum_{\alpha \in \Gamma_{a} \beta \in \Gamma_{b}} \bigl(e_{\beta}, e_{\alpha} \bigr)  \ e^{i(U_{\beta}-U_{\alpha})} .\end{equation}
 The only terms in the sum different from zero are those corresponding to $\beta=-\alpha$. This is possible only if the particle $b$ is the complex conjugate of particle $a$, for which we have $\Gamma_a=-\Gamma_b$. In that case for each root $\alpha$ in the orbit $\Gamma_a$ there exists exactly one root $\beta$ such that $U_\beta=U_{-\alpha}=U_\alpha + \pi$. Therefore the exponential in \eqref{eq2_12} is equal to $-1$ for any nonzero term in the sum and there are exactly $h$ terms contributing to the sum (one for each element of the Coxeter orbit). The final result is 
\begin{equation}\label{eq2_13}
C^{(2)}_{a b}= 2 h  m^2 (\gamma_a^2)^2 \delta_{\bar{a} b}  \frac{(q_1^{a})^2}{|\textbf{Q}_1|^2} .
\end{equation}
The second order expansion of the potential associated to such coupling is 
\begin{equation}\label{eq2_14}
\frac{1}{2}C^{(2)}_{a b} \phi_a \phi_b = \frac{1}{2} \phi_a \phi^*_a 2 h m^2 (\gamma_a^2)^2 \frac{(q_1^{a})^2}{|\textbf{Q}_1|^2} 
\end{equation}
from which we can read the values of the masses
\begin{equation}\label{eq2_15}
m_a=\sqrt{2 h} m \gamma_a^2 \frac{q_1^{a}}{|\textbf{Q}_1|} .
\end{equation}
We note that in the chosen basis the mass-matrix is diagonal, therefore the next orders in the potential expansion exactly correspond to the interaction-couplings of the theory. Comparing the result in \eqref{eq2_15} with the formulas in \eqref{Coxeter_geometry_projections_of_phi_in_complex_notation} we see that the masses of the theory are directly connected to the absolute values of root projections onto the $s=1$ eigenplane of the Coxeter element, a fact which played a role both in the formulation of the fusing rule in \cite{a24}, and in Freeman's mass formula proof in \cite{Freeman:1991xw}.

 \subsection{3-point couplings}\label{3-point couplings}
 To study other couplings we substitute the expression of the masses \eqref{eq2_15} in \eqref{eq2_11}.
 Defining 
 \begin{equation}\label{eq3_(-1)}
\tilde{C}^{(n)}_{\alpha_1, \alpha_2, \ldots , \alpha_n} \equiv e^{i(U_{\alpha_n}-U_{\alpha_1})} \bigl( [e_{\alpha_{2}},[e_{\alpha_{3}},[\ldots [e_{\alpha_{n-1}}, e_{\alpha_{n}}]\ldots]]], e_{\alpha_1} \bigr)
\end{equation}
we can write the $n$-point coupling as
\begin{equation}\label{eq3_0}
C^{(n)}_{a_1, a_2, \ldots , a_n}=- i^n \frac{ \g^{n-2}}{h^{\frac{n}{2}}} m_{a_1} m_{a_n} \sum_{\alpha_1 \in \Gamma_{a_1} \ldots \alpha_n \in \Gamma_{a_n}} \tilde{C}^{(n)}_{\alpha_1, \alpha_2, \ldots , \alpha_n} .
\end{equation}
Writing the couplings in this way we see that a given coupling $C^{(n)}_{a_1, a_2, \ldots , a_n}$ is nonzero only if there exist $n$ roots $\alpha_1 \in \Gamma_{a_1} , \ldots , \alpha_n \in \Gamma_{a_n}$ such that $\alpha_1+\alpha_2+\ldots+\alpha_n=0$. In other words there exists a (possibly non-planar) `polygon' 
with $n$ sides whose projection on $s=1$ plane are the masses of the particles $a_1,a_2,\dots,a_n$. Moreover the nonzero terms of the sum in \eqref{eq3_0} are those for which any partial sum of their roots
\begin{equation}\begin{split}
&\alpha_{n-1}+\alpha_n ,\\
&\alpha_{n-2}+\alpha_{n-1}+\alpha_n ,\\
&\vdots \\
&\alpha_2+\alpha_3+\ldots +\alpha_n
\end{split}\end{equation}
is either a root or zero.
This is a simple consequence of the second commutation relation in \eqref{eq1_6}.

Writing the equation \eqref{eq3_0} in the case $n=3$ we obtain
\begin{equation}\label{eq3_1}
C^{(3)}_{a b c}=-i^3\frac{ \g}{h^{\frac{3}{2}}} m_{a} m_{c} \sum_{\alpha \in \Gamma_{a}, \beta \in \Gamma_{b}, \gamma \in \Gamma_{c}} \tilde{C}_{\alpha \beta \gamma} .
\end{equation}
In this case the only nonzero terms in the sum are those in which $\alpha+\beta+\gamma=0$, which is the fusing rule of \cite{a24}. Suppose we fix the root $\alpha \in \Gamma_a$ and search for all the roots $\gamma \in \Gamma_c$ and $\beta \in \Gamma_b$ such that $\gamma+\beta=-\alpha$. There are exactly two roots triangles satisfying this relation composed by $\{ \alpha$, $\beta$ , $\gamma \}$ and $\{ \alpha$, $\beta'$ , $\gamma' \}$ (see figure \ref{fig:1_pr6_sym_roots} in appendix \ref{App:1}).
Referring to figure \ref{fig:1_pr6_sym_roots} we note that $U_{\gamma}-U_{\alpha}=-(U_{\gamma'}-U_{\alpha})$. Moreover as explained in \ref{last_subappendix} we have $N_{\beta, \gamma}=-N_{\beta', \gamma'}$ and we obtain
\begin{equation}\label{eq3_2}
\tilde{C}_{\alpha \beta \gamma} + \tilde{C}_{\alpha \beta' \gamma'}= e^{i(U_{\gamma}-U_{\alpha})} \bigl( [e_\beta, e_\gamma], e_{\alpha} \bigr)+e^{i(U_{\gamma'}-U_{\alpha})} \bigl( [e_{\beta'}, e_{\gamma'}], e_{\alpha} \bigr)= -2i N_{\beta, \gamma} \sin(U_{\gamma}-U_{\alpha}) .
\end{equation}
There are exactly $h$ copies of this term connected by the Coxeter element, so that the final result for the $3$-point coupling is 
\begin{equation}\label{eq3_3}
C^{(3)}_{a b c}=2 \frac{ \g}{\sqrt{h}} m_{a} m_{c} N_{\beta \gamma} \sin(U_{\gamma}-U_{\alpha}).
\end{equation}
Since $U_{\alpha}$ and $U_{\gamma}$ are the imaginary values of the rapidities of the fusing particles $a$ and $c$, this relation can also be written as
\begin{equation}\label{eq3_4}
C^{(3)}_{a b c}=4  \frac{ \g}{\sqrt{h}} \Delta_{abc} N_{\beta \gamma} sign \Bigl(\sin(U_{\gamma}-U_{\alpha}) \Bigr) .
\end{equation}
For simply-laced cases, for which all structure constants have the same absolute value,
this gives exactly the area relation for the $3$-point couplings. This result was found on a case-by-case basis in \cite{a23} and  proved in a universal way in \cite{Fring:1991me}. While the proof in the latter paper a little different from that given here, both rely fundamentally on Freeman's re-writing of the potential in the form \eqref{eq2_2}.

The relation \eqref{eq3_4}  is of fundamental importance in the pole cancellation of $2$ to $2$ scattering interactions. Indeed we can associate to the $f$-functions found in the previous section the following values
\begin{equation}
\label{f_function_values_in_affine_Toda_theories_from_root_system}
f_{a b c}=4  \frac{ \g}{\sqrt{h}} N_{\beta \gamma} sign \Bigl(\sin(U_{\gamma}-U_{\alpha}) \Bigr),
\end{equation}
where $\alpha$, $\beta$ and $\gamma$ are three arbitrary roots in the orbits $\Gamma_a$, $\Gamma_b$ and $\Gamma_c$ such that $\alpha+\beta+\gamma=0$. It is then possible  to check that the constraint~\eqref{New_eq4_12} directly follows from \eqref{structure_constants_relation_property}, implying the absence of singularities in $2$ to $2$ non-diagonal processes in all the affine Toda theories. 
Let verify this in more detail. 

The different momenta $p_j=m_j e^{i U_j}$ evaluated on the pole position in figure~\ref{figure_t/u_channel_poles}  correspond to the root projections onto the spin-$1$ eigenplane of the Coxeter element, making a correspondence between the imaginary rapidities of the momenta and the  arguments of the projected roots.  If $a$ is a  particle of the theory corresponding to the orbit $\Gamma_a$ then on the pole position its rapidity is a purely imaginary number $iU_\alpha$ corresponding to the argument of a projected root $\alpha \in \Gamma_a$ on the spin-$1$ eigenplane of $w$. 
In the computation of the couplings it is however important to distinguish between particles and antiparticles. In a physical process, by crossing symmetry, we can always convert incoming particles into outgoing antiparticle (and vice versa) leaving the amplitude invariant; the ones can be converted into the others by changing the sign of their momenta and all their quantum numbers in all the Feynman diagrams. This fact is taken into account by the root system where, for a given particle $a$ associated to the orbit $\Gamma_a$ containing a set of roots $\{ \alpha, w \alpha, \dots, w^{h-1} \alpha \}$, we have an antiparticle $\bar{a}$ associated to $-\Gamma_a$ containing the roots $\{ -\alpha, -w \alpha, \dots, -w^{h-1} \alpha \}$. If the particle is real then the two orbits $\Gamma_a$ and $-\Gamma_a$ coincide and both contain the same set of roots.
In the process~\eqref{eq4_0} two particles, $a$ and $b$, are annihilated and another two, $c$ and $d$, are created; this means that the Feynman diagrams contributing to the scattering have to contain the pair of indices $(a,b)$ , to annihilate the incoming particles and $(\bar{c},\bar{d})$ to create the outgoing ones. The associated angles of the projected roots, that enter in the couplings, are $U_{\alpha}$ and $U_{\beta}$ for the incoming states and $U_{-\gamma}=U_{\gamma}+\pi$, $U_{-\delta}=U_{\delta}+\pi$ for the two outgoing states.

Suppose that the momenta $p_1$, $p_2$, $p_3$ and $p_4$, presenting the geometry at the pole shown in figure~\ref{figure_t/u_channel_poles}, are the projections of some roots $\alpha \in \Gamma_a$, $\beta \in \Gamma_b$, $\gamma \in \Gamma_c$ and $\delta \in \Gamma_d$. Then from figure~\ref{figure_t/u_channel_poles}, measuring the angles following the counterclockwise convention, the values of signum-functions corresponding to the different channels assume the values  
\begin{subequations}
\label{New_eq4_7}
\begin{align}
\label{Neweq4_7_1}
&sign \bigl(\sin(U_{\beta}-U_{\alpha}) \bigr) \ sign \bigl(\sin(U_{-\delta}-U_{-\gamma}) \bigr)=(-1) \times (+1)=-1\\
\label{Neweq4_7_2}
&sign \bigl(\sin(U_{-\gamma} -U_{\alpha}) \bigr) \ sign \bigl(\sin(U_{-\delta} - U_\beta) \bigr) = (+1) \times (-1)=-1 \\
\label{Neweq4_7_3}
&sign \bigl(\sin(U_{-\delta} -U_\alpha) \bigr) \ sign \bigl(\sin(U_{-\gamma} - U_\beta) \bigr) = (+1) \times (-1)=-1.
\end{align}
\end{subequations}
If now we substitute the values of the $f$-functions~\eqref{f_function_values_in_affine_Toda_theories_from_root_system} into the constraint~\eqref{New_eq4_12} and use the relations \eqref{New_eq4_7} we end up with the identity~\eqref{structure_constants_relation_property}. This relation is universally satisfied by the structure constants of any semi-simple Lie algebra and it is the reason for the absence of inelastic scattering in $2$ to $2$ processes.

Let us also check that the remaining properties of the $3$-point couplings are satisfied. 
First we note that if three roots $\alpha \in \Gamma_a$, $\beta \in \Gamma_b$ and $\gamma \in \Gamma_c$ satisfy $\alpha+\beta+\gamma=0$ then the following equality holds
\begin{equation}\label{New_eq4_9bis}
sign \Bigl(\sin(U_\alpha-U_\beta) \Bigr)=sign \Bigl(\sin(U_\beta-U_\gamma) \Bigr)=sign \Bigl(\sin(U_\gamma-U_\alpha) \Bigr).
\end{equation}
This cyclic relation together with property~\ref{App:1_pr2} implies
\begin{equation}\label{New_eq4_10}
f_{ijk}=f_{jki}=f_{kij}.
\end{equation}
If we add to this the fact that both the structure constant $N_{\alpha \beta}$ and $sign \bigl(\sin(U_\alpha  - U_\beta) \bigr)$ are antisymmetric under the exchange $\alpha \to \beta$ we obtain that the $3$-point couplings are symmetric under exchange of any pair of indices.
The condition~\eqref{New_eq4_11}, required to have a real Lagrangian (and therefore a unitary theory) follows from   property~\ref{App:1_pr3}.

Note that in simply-laced Toda models the nonzero
structure constants of the Lie algebra share 
the same absolute value 
\begin{equation}
\label{connection_between_root_normalisation_and_structure_constant_normalisation}
|N_{\alpha \beta}|^2=\frac{\Lambda^2}{2}
\end{equation}
where $\Lambda$ is the normalisation chosen for the root lengths. The relation \eqref{connection_between_root_normalisation_and_structure_constant_normalisation} follows from \eqref{norm_coupling_property4} noting that in simply-laced models $\alpha+\beta$ and $\alpha-\beta$ cannot both be roots at the same time, since otherwise we would have $2\Lambda^2=(\alpha+\beta)^2+(\alpha-\beta)^2=2\alpha^2+2\beta^2=4\Lambda^2$.
This implies that the absolute value of $f_{abc}$, if not zero, does not depend on the choice of the coupled particles and we can write
\begin{equation}\label{mod_f_simply_laced}
|f_{a b c}|=|f|=4  \frac{ \g}{\sqrt{h}} \sqrt{\frac{\Lambda^2}{2}} .
\end{equation}
Thus all the $f$-functions in simply-laced theories have
the same absolute value, as expected given the general
considerations of the last section. Indeed in the simply-laced theories, as explained in appendix
\ref{structure_constants_properties_appendix}, if we have four
different roots $\alpha$, $\beta$, $\gamma$ and $\delta$ such that
$\alpha+\beta=\gamma+\delta=\epsilon$, where $\epsilon$ is another
allowed root, then we have that or $\alpha-\gamma$ or $\alpha-\delta$
is a root, but not both. Since the space of momenta producing poles is
the projection of the root space, in such theories the cancellation of
singularities in $4$-point non-diagonal scattering always happens between pairs of diagrams and we never have $3$ propagators on-shell simultaneously. On the other hand we also know, as just remarked, that if $\alpha+\beta$ is a root, $\alpha-\beta$ cannot be a root too. This means that we cannot have more than one on-shell propagating particle at a time in $2$ to $2$ diagonal scattering. These two conditions satisfied by the simply-laced Toda theories are exactly the simply-laced scattering conditions studied in the previous section, through which, by imposing the cancellation of poles in inelastic processes, we came to the conclusion that $|f_{a b c}|$ cannot depend on the choice of indices $a$, $b$ and $c$. 
Thus imposing the absence of particle production in models satisfying property~\ref{Simply_laced_scattering_conditions}, and extracting the $3$-point couplings of simply-laced affine Toda theories starting from their Lie algebra properties, we find the same area rule. 

However simply-laced affine Toda theories are not the only integrable models respecting the simply-laced scattering conditions highlighted in the previous section. To such theories we should add the twisted theories, obtained by folding the affine extension of certain simply-laced Dynkin diagrams. These models live in a subalgebra of their simply-laced parents and therefore inherit all the properties of the ADE series. We will discuss these cases separately in section~\ref{Generalization_to_twisted_Lie_algebras}.

We proceed now to the study of higher-point couplings in affine Toda theories.

\subsection{4-point couplings and generalisation}

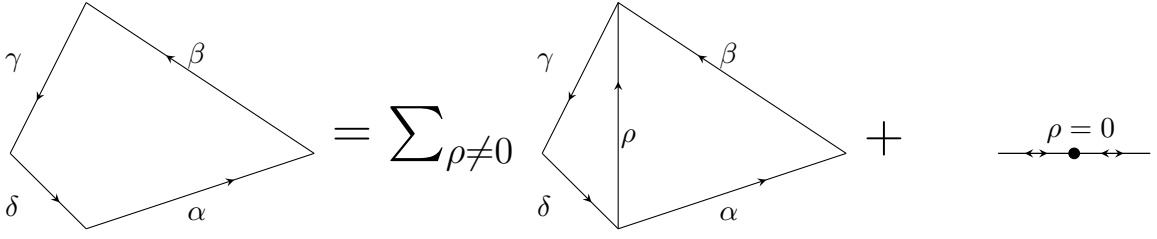
\begin{figure}
\begin{tikzpicture}
 \filldraw[black] (6.6,0.2)  node[anchor=west]{\huge{$=\sum_{\rho\ne0}$}};
\draw[directed] (3.5,2) -- (2.5,0);
\draw[directed] (6.5,0) -- (3.5,2);
\draw[directed] (2.5,0) -- (3.5,-1);
\draw[directed] (3.5,-1) -- (6.5,0);
\filldraw[black] (2.3,1.2)  node[anchor=west] {$\gamma$};
\filldraw[black] (2.3,-.7)  node[anchor=west] {$\delta$};
\filldraw[black] (4.7,1.3)  node[anchor=west] {$\beta$};
\filldraw[black] (4.7,-0.8)  node[anchor=west] {$\alpha$};

\draw[directed] (10.5,2) -- (9.5,0);
\draw[directed] (13.5,0) -- (10.5,2);
\draw[directed] (9.5,0) -- (10.5,-1);
\draw[directed] (10.5,-1) -- (10.5,2);
\draw[directed] (10.5,-1) -- (13.5,0);
\filldraw[black] (9.3,1.2)  node[anchor=west] {$\gamma$};
\filldraw[black] (9.3,-.7)  node[anchor=west] {$\delta$};
\filldraw[black] (10.4,0.2)  node[anchor=west] {$\rho$};
\filldraw[black] (11.7,1.3)  node[anchor=west] {$\beta$};
\filldraw[black] (11.7,-0.8)  node[anchor=west] {$\alpha$};

 \filldraw[black] (13.6,0.2)  node[anchor=west]{\huge{$+$}};

\draw[directed] (16.5,0) -- (15.5,0);
\draw[directed] (15.5,0) -- (16.5,0);
\draw[directed] (16.5,0) -- (17.5,0);
\draw[directed] (17.5,0) -- (16.5,0);
\filldraw[black] (16.5,0) circle (2pt) node[anchor=west] {};
\filldraw[black] (16,0.3)  node[anchor=west] {$\rho=0$};
\end{tikzpicture}
\caption{Pictorial representation of the $4$-point coupling as a sum over roots $\rho$.}
\label{ZANTE}
\end{figure}

The next step is to compute the $4$-point couplings. From \eqref{eq3_0} we can write them as
 \begin{equation}\label{eq3_12}
 C^{(4)}_{a b c d}=-\frac{ \g^{2}}{h^2} m_a m_d \sum_{\alpha \in \Gamma_{a} \ldots \delta \in \Gamma_{d}} e^{i(U_{\delta}-U_{\alpha})} \bigl([e_\gamma , e_\delta], [e_\alpha, e_\beta] \bigr).
 %=C^{(4,1)}_{a b c d}+C^{(4,0)}_{a b c d}
\end{equation} 
By inserting the identity between $[e_\gamma , e_\delta]$ and $[e_\alpha, e_\beta]$, written as
\begin{equation}\label{eq3_13}
I=\sum_{a=1}^r h_a \otimes h_a -\sum_{l=1}^r \sum_{\rho \in \Gamma_l}  e_\rho \otimes e_{-\rho}
\end{equation}
we can split~\eqref{eq3_12} in two terms
\begin{equation}
\label{split_of_4_point_coupling_in_two_terms}
C^{(4)}_{a b c d}=C^{(4,0)}_{a b c d}+C^{(4,1)}_{a b c d}.
\end{equation}
$C^{(4,0)}_{a b c d}$ corresponds to inserting the piece of the identity given by the basis of the Cartan subalgebra.
It is reproduced by summing over the configurations $\beta=-\alpha$ and $\gamma=-\delta$ in~\eqref{eq3_12}.
$C^{(4,1)}_{a b c d}$ is instead obtained by summing over the roots $\alpha+\beta=\rho=-\gamma-\delta$, where $\rho$ is still a root of the system. A geometrical picture is shown in figure~\ref{ZANTE}.

We start from the computation of $C^{(4,0)}_{a b c d}$ and then we move to $C^{(4,1)}_{a b c d}$.
Summing over all the roots in the orbits for which we have $\gamma=-\delta$ and $\beta=-\alpha$ we obtain
\begin{equation}\label{eq3_14}
C^{(4,0)}_{a b c d}=\frac{ \g^{2}}{h^2} m_a m_d \delta_{a \bar{b}} \delta_{c \bar{d}} \sum_{\alpha \in \Gamma_{a}, \delta \in \Gamma_{d}} e^{i(U_{\delta}-U_{\alpha})} \bigl( \alpha, \delta \bigr) .
\end{equation}
We can expand the scalar product $\bigl( \alpha, \delta \bigr)$ on the eigenvectors $z_s$ of the Coxeter element and express the projections in terms of the masses using \eqref{eq2_15}, \eqref{Coxeter_geometry_projections_of_phi_in_complex_notation}. We then find the following result
\begin{equation}\label{eq3_15}
C^{(4,0)}_{a b c d}=\frac{ \g^{2}}{h^2} \frac{m_a^2 m_d^2}{h m^2} \delta_{a \bar{b}} \delta_{c \bar{d}} \sum_s \sum_{\alpha \in \Gamma_{a}} e^{iU_{\alpha}(s-1)} \sum_{\delta \in \Gamma_{d}}  e^{iU_{\delta}(-s+1)}.
\end{equation}
Both the sums over $\alpha \in \Gamma_a$ and $\delta \in \Gamma_d$ constitute closed paths in the complex plane for any $s$ different from one, meaning that $s=1$ is the only value that returns nonzero results. Developing for example the second sum we obtain  
\begin{equation}\label{eq3_16}
\sum_{\delta \in \Gamma_{d}}  e^{iU_{\delta}(-s+1)}=h \ \delta_{s1}.
\end{equation}
Using this fact the final expression for \eqref{eq3_15} is
 \begin{equation}\label{eq3_17}
C^{(4,0)}_{a b c d}=\frac{ \g^{2}}{h} \frac{m_a^2 m_d^2}{m^2} \delta_{a \bar{b}} \delta_{c \bar{d}} .
\end{equation}
We compute now $C_{abcd}^{(4,1)}$. To avoid to many indices we omit to indicate the orbits in the sum; it is clear we intend the sum performed over $\alpha \in \Gamma_a$, $\beta \in \Gamma_b$, $\gamma \in \Gamma_c$ and $\delta \in \Gamma_d$, such that $\alpha+\beta=\rho=-\gamma-\delta$. We can write this term in the following form
\begin{equation}\label{eq3_18}
C^{(4,1)}_{a b c d}=-\frac{ \g^{2}}{h^2} m_a m_d \sum_l \sum_{\rho \in \Gamma_l} \biggl( \sum_{\substack{\alpha, \ \beta \\ \alpha+\beta=\rho}}  e^{i(U_{-\rho}-U_{\alpha})} \bigl([e_\beta, e_{-\rho}],e_\alpha \bigr) \sum_{\substack{\gamma, \ \delta \\ \gamma+\delta=-\rho}} e^{i(U_{\delta}-U_{\rho})} \bigl( [e_\gamma , e_\delta], e_\rho \bigr)\biggr) .
\end{equation}
The two sums on the right hand side of \eqref{eq3_18} are evaluated at the same value of $\rho$ and run over all the $\{ \alpha,\beta \}$ and $\{ \gamma,\delta \}$ in their respective orbits such that $\alpha+\beta=\rho=-\gamma-\delta$, where $\rho$ are roots running in different orbits $\Gamma_l$ ; we have separated the sum over the roots $\rho$ into a sum over orbits $l$ and a sum over the roots inside these orbits.
However it is possible to check that the two sums 
\begin{equation}\label{eq3_19}
\sum_{{\substack{\gamma, \delta \\ \gamma+\delta=-\rho}}} e^{i(U_{\delta}-U_{\rho})} \bigl( [e_\gamma , e_\delta], e_\rho \bigr)
\end{equation}
and
\begin{equation}\label{eq3_20}
\sum_{{\substack{\alpha, \beta \\ \alpha+\beta=\rho}}} e^{i(U_{-\rho}-U_{\alpha})} \bigl([e_\beta, e_{-\rho}],e_\alpha \bigr)
\end{equation}
are separately invariant moving $\rho$ inside the orbit $\Gamma_l$, so we can choose one of them and substitute it with the average over all the roots $\rho$ in the orbit. The expression in \eqref{eq3_18} can therefore be written as
 \begin{equation}\label{eq3_21}
 \begin{split}
&C^{(4,1)}_{a b c d}=\\
&-\frac{ \g^{2}}{h^2} m_a m_d \sum_l \biggl(\frac{1}{h} \sum_{\rho' \in \Gamma_l} \sum_{\alpha+\beta=\rho'} e^{i(U_{-\rho'}-U_{\alpha})} \bigl([e_\beta, e_{-\rho'}],e_\alpha \bigr) \sum_{\rho \in \Gamma_l} \sum_{\gamma+\delta=-\rho} e^{i(U_{\delta}-U_{\rho})} \bigl( [e_\gamma , e_\delta], e_\rho \bigr)  \biggr).
\end{split}
\end{equation}
 We recognise in this last relation the $3$-point couplings that we found in \eqref{eq3_1}, giving us the result
 \begin{equation}\label{eq3_22}
C^{(4,1)}_{a b c d}=\sum_l C^{(3)}_{ab\bar{l}} \ \frac{1}{m_l^2} \ C^{(3)}_{lcd}
\end{equation}
Combining now \eqref{eq3_22} and \eqref{eq3_17} we obtain
 \begin{equation}\label{eq3_23}
C^{(4)}_{a b c d}= \frac{ \g^{2}}{h} \frac{m_a^2 m_d^2}{m^2} \delta_{a \bar{b}} \delta_{c \bar{d}}+\sum_l C^{(3)}_{ab\bar{l}} \ \frac{1}{m_l^2} \ C^{(3)}_{lcd} .
\end{equation}
This result has been obtained assuming a particular order for the indices $a$, $b$, $c$ and $d$. However, by the symmetry of the coupling, it needs to be valid for any order and we have
\begin{equation}\begin{split}
\sum_l C^{(3)}_{ab\bar{l}} \ \frac{1}{m_l^2} \ C^{(3)}_{lcd} + \frac{ \g^{2}}{h} \frac{m_a^2 m_d^2}{m^2} \delta_{a \bar{b}} \delta_{c \bar{d}}&=\sum_r C^{(3)}_{ac\bar{r}} \ \frac{1}{m_r^2} \ C^{(3)}_{rbd} + \frac{ \g^{2}}{h} \frac{m_a^2 m_d^2}{m^2} \delta_{a \bar{c}} \delta_{b \bar{d}}\\
&=\sum_j C^{(3)}_{ad\bar{j}} \ \frac{1}{m_j^2} \ C^{(3)}_{jbc} + \frac{ \g^{2}}{h} \frac{m_a^2 m_b^2}{m^2} \delta_{a \bar{d}} \delta_{c \bar{b}}.
\end{split}\end{equation}
This relation exactly corresponds to the value that the $4$-point coupling must assume to set all the non-diagonal $2$ to $2$ scattering processes to zero, as shown in \eqref{New_eq4_13_fixed}. The missing term containing the Kronecker delta functions in~\eqref{New_eq4_13_fixed} is not in contradiction with~\eqref{eq3_23} since in the scattering event we assumed the initial and the final particles to be different.

For a generic $n$-point process we can make a similar decomposition of the coupling into subcouplings. In particular the number of decompositions depends in a certain sense on the number of possible partitions of the $n$-sided polygon formed by the roots associated to the $n$-point coupling considered. For each permutation in which we write the indices of the coupling we obtain a polygon composed of a particular set of roots. We consider different partitions of such
a polygon as the different ways in which we split it into subpolygons, drawing diagonals emerging always from the same vertex.
Any time we split the $n$-gon in two subpolygons drawing a diagonal
we obtain one term given by propagators connecting two lower-order
couplings and one term corresponding to the null value of the diagonal
(in expression \eqref{eq3_23} it is given by the Kronecker delta
function). The recursion relation obtained in \eqref{eq0_6}
corresponds to summing three suitable partitions of the coupling so to
cancel the values given by the zero-diagonals and leave only the
`propagator-terms'. What we do is to write 
\begin{equation}\label{eq3_24}
C_{a_1 \ldots a_n}^{(n)}=C_{a_1 \ldots a_n}^{(n)}+C_{a_1 \ldots a_n}^{(n)}-C_{a_1 \ldots a_n}^{(n)}
\end{equation}
where on the right hand side the three terms are equal but written in different ways accordingly with their partitions. 
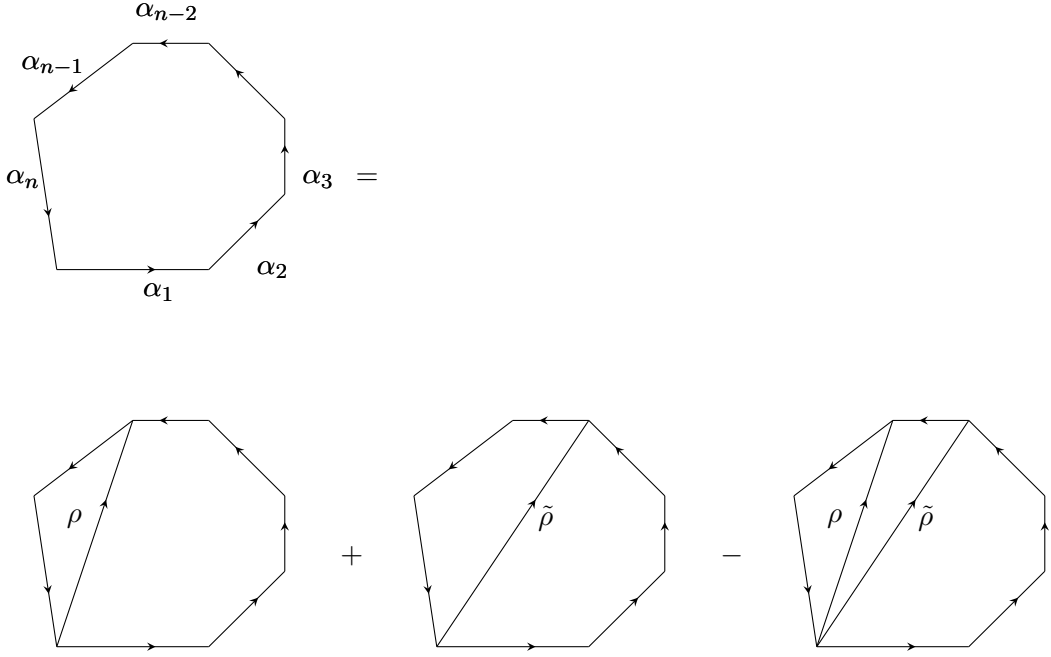
\begin{figure} 
  
\begin{tikzpicture}
\draw[directed] (0,0) -- (2,0);
\draw[directed] (2,0) -- (3,1);
\draw[directed] (3,1) -- (3,2);
\draw[directed] (3,2) -- (2,3);
\draw[directed] (2,3) -- (1,3);
\draw[directed] (1,3) -- (-0.3,2);
\draw[directed] (-0.3 , 2) -- (0,0);

\filldraw[black] (1, -0.3)  node[anchor=west] {$\alpha_{1}$};
\filldraw[black] (2.5, 0.)  node[anchor=west] {$\alpha_{2}$};
\filldraw[black] (3.1 , 1.2)  node[anchor=west] {$\alpha_{3}$};
\filldraw[black] (0.9 , 3.4)  node[anchor=west] {$\alpha_{n-2}$};
\filldraw[black] (-0.6 , 2.7)  node[anchor=west] {$\alpha_{n-1}$};
\filldraw[black] (-0.8 , 1.2)  node[anchor=west] {$\alpha_{n}$};

\filldraw[black] (3.8,1.2)  node[anchor=west] {$=$};

\draw[directed] (0,-5) -- (2,-5);
\draw[directed] (2,-5) -- (3,1-5);
\draw[directed] (3,1-5) -- (3,2-5);
\draw[directed] (3,2-5) -- (2,3-5);
\draw[directed] (2,3-5) -- (1,3-5);
\draw[directed] (1,3-5) -- (-0.3,2-5);
\draw[directed] (-0.3 , 2-5) -- (0,0-5);
\draw[directed] (0,0-5) -- (1,3-5);
\filldraw[black] (0, -3.3)  node[anchor=west] {$\rho$};

\filldraw[black] (3.6,1.2-5)  node[anchor=west] {$+$};

\draw[directed] (0+5,-5) -- (2+5,-5);
\draw[directed] (2+5,-5) -- (3+5,1-5);
\draw[directed] (3+5,1-5) -- (3+5,2-5);
\draw[directed] (3+5,2-5) -- (2+5,3-5);
\draw[directed] (2+5,3-5) -- (1+5,3-5);
\draw[directed] (1+5,3-5) -- (-0.3+5,2-5);
\draw[directed] (-0.3+5 , 2-5) -- (0+5,0-5);
\draw[directed] (0+5,0-5) -- (2+5,3-5);
\filldraw[black] (6.2, -3.3)  node[anchor=west] {$\tilde{\rho}$};

\filldraw[black] (3.6+5,1.2-5)  node[anchor=west] {$-$};

\draw[directed] (0+10,-5) -- (2+10,-5);
\draw[directed] (2+10,-5) -- (3+10,1-5);
\draw[directed] (3+10,1-5) -- (3+10,2-5);
\draw[directed] (3+10,2-5) -- (2+10,3-5);
\draw[directed] (2+10,3-5) -- (1+10,3-5);
\draw[directed] (1+10,3-5) -- (-0.3+10,2-5);
\draw[directed] (-0.3+10 , 2-5) -- (0+10,0-5);
\draw[directed] (0+10,0-5) -- (1+10,3-5);
\draw[directed] (0+10,0-5) -- (2+10,3-5);

\filldraw[black] (0+10, -3.3)  node[anchor=west] {$\rho$};
\filldraw[black] (6.2+5, -3.3)  node[anchor=west] {$\tilde{\rho}$};

\filldraw[black] (1, -0.3)  node[anchor=west] {$\alpha_{1}$};
\filldraw[black] (2.5, 0.)  node[anchor=west] {$\alpha_{2}$};
\filldraw[black] (3.1 , 1.2)  node[anchor=west] {$\alpha_{3}$};
\filldraw[black] (0.9 , 3.4)  node[anchor=west] {$\alpha_{n-2}$};
\filldraw[black] (-0.6 , 2.7)  node[anchor=west] {$\alpha_{n-1}$};
\filldraw[black] (-0.8 , 1.2)  node[anchor=west] {$\alpha_{n}$};

\end{tikzpicture}
\caption{Decomposition of a $n$-point coupling as a sum of different partitions in such a way to obtain the recursion relation \eqref{eq0_6}. }
\label{fig:partitionMultiregge}
\end{figure}
Referring to figure \ref{fig:partitionMultiregge} we write all the contributions for the different partitions. They are given by
\begin{equation}\label{eq3_25}
C^{(n)}_{a_1, a_2, \ldots , a_n}=\frac{\g^2}{h} \frac{m^2_{a_n}}{m^2} \delta_{\bar{a}_n, a_{n-1}} C^{(n-2)}_{a_1 , \ldots, a_{n-2}} +\sum_l C^{(n-1)}_{a_1 , \ldots, a_{n-2},l} \frac{1}{m_l^2} C^{(3)}_{\bar{l},a_{n-1},a_n} ,
\end{equation}
\begin{equation}\label{eq3_26}
C^{(n)}_{a_1, a_2, \ldots , a_n}=\frac{\g^2}{h} \frac{1}{m^2} C^{(n-3)}_{a_1 , \ldots, a_{n-3}} C^{(3)}_{a_{n-2}, a_{n-1}, a_n} +\sum_s C^{(n-2)}_{a_1 , \ldots, a_{n-3},s} \frac{1}{m_s^2} C^{(4)}_{\bar{s},a_{n-2}, a_{n-1},a_n} ,
\end{equation}
\begin{equation}\begin{split}\label{eq3_27}
C^{(n)}_{a_1, a_2, \ldots , a_n}&=\frac{\g^2}{h} \frac{m^2_{a_n}}{m^2} \delta_{\bar{a}_n, a_{n-1}} C^{(n-2)}_{a_1 , \ldots, a_{n-2}} + \frac{\g^2}{h} \frac{1}{m^2} C^{(n-3)}_{a_1 , \ldots, a_{n-3}} C^{(3)}_{a_{n-2}, a_{n-1}, a_n}\\
&+\sum_{l,s} C^{(n-2)}_{a_1 , \ldots, a_{n-3},s} \frac{1}{m_s^2} C^{(3)}_{\bar{s},a_{n-2},l}\frac{1}{m_l^2}C^{(3)}_{\bar{l},a_{n-1},a_n} .
\end{split}\end{equation}
Summing the two expressions in \eqref{eq3_25}, \eqref{eq3_26} and then taking the difference with \eqref{eq3_27} we obtain the following value for the $n$-point coupling
\begin{equation}\begin{split}\label{eq3_28}
C^{(n)}_{a_1, a_2, \ldots , a_n}&=\sum_l  C^{(n-1)}_{a_1, \ldots , a_{n-2}, l} \frac{1}{m_l^2} C^{(3)}_{\bar{l}, a_{n-1}, a_{n}} + \sum_s  C^{(n-2)}_{a_1, \ldots , a_{n-3}, s} \frac{1}{m_s^2} C^{(4)}_{\bar{s}, a_{n-2}, a_{n-1},a_{n}}\\
&- \sum_{l,s} C^{(n-2)}_{a_1, \ldots , a_{n-3}, s} \frac{1}{m_s^2} C^{(3)}_{\bar{s}, a_{n-2}, l} \frac{1}{m_l^2}C^{(3)}_{\bar{l}, a_{n-1},a_{n}} .
\end{split}\end{equation}
This expression is exactly the same result as
obtained in \eqref{eq0_6} by imposing the absence of particle production in the multi-Regge limit in all the untwisted affine Toda models. 

\subsection{From root systems to elastic S-matrices in affine Toda theories} \label{From_root_system_to_elastic_S_matrices_in_affine_Toda_theories}

To conclude a general proof of tree-level perturbative integrability in affine Toda theories, it remains to check that the bootstrap requirements \eqref{tree_level_bootstrap_constraint_on_the_S_matrix_coming_from_5_point_pole_cancellation_equation} are universally satisfied. As mentioned in the previous section, this fact, together with the impossibility of $2$ to $2$ off-diagonal scattering, imposes the absence of poles in $5$-point processes. 
We will see soon that the constraint \eqref{tree_level_bootstrap_constraint_on_the_S_matrix_coming_from_5_point_pole_cancellation_equation} follows directly from root system properties in a  geometrical way. 

Let focus on the allowed process
$$
a(p_1)+b(p_2) \to a(p_1) + b(p_2).
$$
In terms of the rapidity difference $\theta=\theta_1 - \theta_2$ the Mandelstam variables take the values
\begin{equation}
\label{the_Mandelstam_variables_for_a_2_to_2_allowed_process}
\begin{split}
s&=(p_1+p_2)^2 = m_a^2+m_b^2 + 2 m_a m_b \cosh \theta\\
t&=(p_1-p_2)^2 = m_a^2+m_b^2 - 2 m_a m_b \cosh \theta\\
u&=(p_1-p_1)^2=0
\end{split}
\end{equation}
Since $u=0$, the sum of the Feynman diagrams coming from the $u$-channel is cancelled by the piece of the four point coupling \eqref{eq3_23} not containing the Kronecker delta functions. 
The amplitude is then given by
\begin{equation}
\label{first_expression_for_the_allowed_amplitude_ab_to_ab_still_couplings_not_substituted}
M_{ab}=-i \sum_{i} \frac{C^{(3)}_{ab \bar{i}} C^{(3)}_{i \bar{a} \bar{b}}}{s-m^2_i} -i \sum_{j} \frac{C^{(3)}_{a \bar{b} \bar{j}} C^{(3)}_{j \bar{a} b}}{t-m^2_j} - i \frac{\g^2}{h} \frac{m_a^2 m_b^2}{m^2} .
\end{equation}
We can write the masses of the propagating particles in terms of the angles formed by the momenta $p_1$, $p_2$ on the pole positions 
$m_i^2=m_a^2+m_b^2 + 2 m_a m_b \cos U_{ab}^i$ , $m_j^2=m_a^2+m_b^2 - 2 m_a m_b \cos U_{ab}^j$. 
To find these angles we run over the inequivalent mass triangles in the orbit. We keep the root $\gamma_a \in \Gamma_a$, corresponding to the $a$-particle, fixed and we move over the roots $\beta=w^{-p} \gamma_b \in \Gamma_b$. Since to each propagating particle there correspond two mass triangles, one the reflection of the other, to find all the inequivalent scattering channels we need to move $w^{-p} \gamma_b$ over half of the orbit. 
Defining $\tilde{\Gamma}_b$ to be
half of the orbit corresponding to the $b$ particle and plugging (\eqref{eq3_3}, \eqref{the_Mandelstam_variables_for_a_2_to_2_allowed_process}) into \eqref{first_expression_for_the_allowed_amplitude_ab_to_ab_still_couplings_not_substituted} we can write the $s$ and $t$ channels in the more compact form
$$
M_{ab}=\frac{2 i \g^2}{h} m_a m_b \sum_{\beta \in \tilde{\Gamma}_b} \bigl( |N_{\gamma_a , \beta}|^2 - |N_{\gamma_a , -\beta}|^2 \bigr) \frac{\sinh^2 i U_{\gamma_a \beta}}{\cosh \theta - \cos U_{\gamma_a \beta}} - i \frac{\g^2}{h} \frac{m_a^2 m_b^2}{m^2}.
$$
We use the convention of writing the difference between the
angles of the root projections on the spin-$1$ eigenplane of $w$
as $U_{\alpha \beta}=U_\alpha - U_\beta$.
Surprisingly looking at the relation in \eqref{norm_coupling_property4} we note that the structure constants disappear completely from the formula above, leaving the place to a scalar product between the roots associated to the interacting particles
\begin{equation}
\label{second_expression_for_the_allowed_amplitude_ab_to_ab_couplings_substituted_and_N_substituted_with_scalar_product}
M_{ab}=- \frac{2 i \g^2}{h} m_a m_b \sum_{\beta \in \tilde{\Gamma}_b} (\gamma_a, \beta) \frac{\sinh^2 i U_{\gamma_a \beta}}{\cosh \theta - \cos U_{\gamma_a \beta}} - i \frac{\g^2}{h} \frac{m_a^2 m_b^2}{m^2} .
\end{equation}
For the ADE series, only one of $ |N_{\gamma_a , \beta}|^2$ and $ |N_{\gamma_a , -\beta}|^2$ can be different from zero at a time and indeed the scalar product $(\gamma_a, \beta)$ along $\tilde{\Gamma}_b$ can only assume the values $\mp \frac{\Lambda^2}{2}$ corresponding to the propagation of particles in the $s/t$ channels (we remember that $\Lambda$ indicates the normalization chosen for the roots; it is convenient to assume $\Lambda=\sqrt{2}$ in simply-laced theories in such a way that the structure constants are phases and the residues for the $s/t$ channels, up to a prefactor, are $\pm1$). 
In the case of non simply-laced theories it is possible that for a given choice of $\beta$ in $\tilde{\Gamma}_b$ there is a propagating particle both in the $s$- and in the $t$-channel at the same time leaving open the possibility for other 
values of the scalar product $(\gamma_a, \beta)$.
It is temping to promote the relation in \eqref{second_expression_for_the_allowed_amplitude_ab_to_ab_couplings_substituted_and_N_substituted_with_scalar_product} to the more compact result
\begin{equation}
\label{third_expression_for_the_allowed_amplitude_ab_to_ab_couplings_substituted_and_N_substituted_with_scalar_product_sum_over_the_full_orbit}
M_{ab}=-\frac{i \g^2}{h} m_a m_b \sinh^2 \theta \sum_{\beta \in \Gamma_b} \frac{ (\gamma_a, \beta)}{\cosh \theta - \cos U_{\gamma_a \beta}} .
\end{equation}
In \eqref{third_expression_for_the_allowed_amplitude_ab_to_ab_couplings_substituted_and_N_substituted_with_scalar_product_sum_over_the_full_orbit}
we are summing over the roots $\beta$ in the full orbit $\Gamma_b$, and
for this reason we have removed the prefactor $2$ compared to the expression in \eqref{second_expression_for_the_allowed_amplitude_ab_to_ab_couplings_substituted_and_N_substituted_with_scalar_product} since now we are double counting all the fusing triangles. The other modification has been to remove the surviving piece of the $4$-point coupling and promoting the angles at the numerator, corresponding to the poles positions, to be $\theta$ dependent , $\sinh i U_{\gamma_a \beta} \to \sinh \theta$.

In order to check that the two expressions in \eqref{second_expression_for_the_allowed_amplitude_ab_to_ab_couplings_substituted_and_N_substituted_with_scalar_product} and \eqref{third_expression_for_the_allowed_amplitude_ab_to_ab_couplings_substituted_and_N_substituted_with_scalar_product_sum_over_the_full_orbit} are actually the same we need to study their behaviour in the neighbourhood of all the poles and at $\theta \to \infty$. 
The verification that the residues of \eqref{second_expression_for_the_allowed_amplitude_ab_to_ab_couplings_substituted_and_N_substituted_with_scalar_product} and \eqref{third_expression_for_the_allowed_amplitude_ab_to_ab_couplings_substituted_and_N_substituted_with_scalar_product_sum_over_the_full_orbit} are the same at any singularity $\theta=i U_{\gamma_a \beta}$ is a simple check that we leave to the reader. The only fact that deserves attention is that for any pole, there are two terms in the sum~\eqref{third_expression_for_the_allowed_amplitude_ab_to_ab_couplings_substituted_and_N_substituted_with_scalar_product_sum_over_the_full_orbit} contributing to it; such terms have to be summed to  reproduce the residues of \eqref{second_expression_for_the_allowed_amplitude_ab_to_ab_couplings_substituted_and_N_substituted_with_scalar_product}.

Let focus on what happens when $\theta \gg1$.
If we write the roots $\beta \in \Gamma_b$ as $\beta=w^{-p} \gamma_b$ with $p=0, \ldots h-1$
then in the limit $\theta \to \infty$ the expression in \eqref{third_expression_for_the_allowed_amplitude_ab_to_ab_couplings_substituted_and_N_substituted_with_scalar_product_sum_over_the_full_orbit} becomes
\begin{equation}
\label{proof_starting_that_for_big_theta_the_two_different_amplitudes_reproduce_both_the_four_point_coupling}
\begin{split}
M_{ab}\Bigl|_{\theta \gg 1}&=-\frac{i \g^2}{h} m_a m_b \tanh^2 \theta \Bigl[ \cosh \theta \sum_{p=0}^{h-1}  (\gamma_a, w^{-p} \gamma_b)\\
& -   \sum_{p=0}^{h-1} (\gamma_a, w^{-p} \gamma_b) \sum_{\substack{q=0 \\ q \neq p}}^{h-1} \cos U_{\gamma_a , w^{-q} \gamma_b} + o((\cosh \theta)^{-1} ) \Bigr] .
\end{split}
\end{equation}
Since the sum over all the roots in an orbit is equal to zero, the term proportional to $\cosh \theta$ in the square brackets is zero, avoiding a divergence as $\theta \to \infty$. 
The coefficient in front to the term of order $(\cosh \theta)^{0}$ is instead given by
\begin{equation}
\label{Temporary_equation_to_verify_the_high_theta_limit_of_M_summed_over_the_entire_orbitt}
\begin{split}
&\sum_{p=0}^{h-1} (\gamma_a, w^{-p} \gamma_b) \sum_{\substack{q=0 \\ q \neq p}}^{h-1} \cos U_{\gamma_a , w^{-q} \gamma_b} = - \sum_{p=0}^{h-1} (\gamma_a, w^{-p} \gamma_b) \cos U_{\gamma_a , w^{-p} \gamma_b}\\
&= - \gamma_a^2 \ \gamma_b^2 \sum_s \frac{q_s^a \ q_s^b}{|{\bf Q_s}|^2} \sum_{p=0}^{h-1} \Bigl( e^{i (s+1) (U_{\gamma_a} - U_{w^{-p} \gamma_b})} + e^{i (s-1) (U_{\gamma_a} - U_{w^{-p} \gamma_b})} \Bigr) .
\end{split}
\end{equation}
The first equality above has been obtained considering that the sum of $\cos U_{\gamma_a , w^{-q} \gamma_b}$ performed over the entire orbit is equal to zero. The second equality is derived by decomposing the root $w^{-p} \gamma_b$ over the basis of Coxeter eigenvectors and writing the projections using the relation \eqref{Coxeter_geometry_projections_of_phi_in_complex_notation}.
Finally the remaining sum in the second line of \eqref{Temporary_equation_to_verify_the_high_theta_limit_of_M_summed_over_the_entire_orbitt} generates two Kronecker deltas $h \delta_{s ,-1}$ and $h \delta_{s ,1}$ leading to
\begin{equation}
\begin{split}
&\sum_{p=0}^{h-1} (\gamma_a, w^{-p} \gamma_b) \sum_{\substack{q=0 \\ q \neq p}}^{h-1} \cos U_{\gamma_a , w^{-q} \gamma_b} = - 2 h \gamma_a^2 \ \gamma_b^2  \frac{q_1^a \ q_1^b}{|{\bf Q}_1|^2} .
\end{split}
\end{equation}
If we substitute the mass expression \eqref{eq2_15} into this last relation and plug the result into \eqref{proof_starting_that_for_big_theta_the_two_different_amplitudes_reproduce_both_the_four_point_coupling} we discover that in the large rapidity limit the amplitude reduces to
\begin{equation}
 M_{ab} (\theta) \Bigl|_{\theta \gg 1}=-\frac{i \g^2}{h} \frac{m^2_a m^2_b}{m^2} \tanh^2 \theta.
\end{equation}
Interestingly the amplitude written in \eqref{third_expression_for_the_allowed_amplitude_ab_to_ab_couplings_substituted_and_N_substituted_with_scalar_product_sum_over_the_full_orbit} for $\theta \gg 1$ reproduces exactly the delta term coming from the $4$-point coupling and explicitly expressed in \eqref{second_expression_for_the_allowed_amplitude_ab_to_ab_couplings_substituted_and_N_substituted_with_scalar_product}, implying the equality between the two expressions in \eqref{second_expression_for_the_allowed_amplitude_ab_to_ab_couplings_substituted_and_N_substituted_with_scalar_product} and \eqref{third_expression_for_the_allowed_amplitude_ab_to_ab_couplings_substituted_and_N_substituted_with_scalar_product_sum_over_the_full_orbit}. 

We may also ask about the behaviour of \eqref{third_expression_for_the_allowed_amplitude_ab_to_ab_couplings_substituted_and_N_substituted_with_scalar_product_sum_over_the_full_orbit} in the collinear limit $\theta \ll 1$. 
In this situation the amplitude is only nonzero  if along the orbit $\Gamma_b$ there is one root $\beta$ forming a projected angle $U_{\gamma_a , \beta}=0$ on the spin-$1$ plane for which at the denominator of \eqref{third_expression_for_the_allowed_amplitude_ab_to_ab_couplings_substituted_and_N_substituted_with_scalar_product_sum_over_the_full_orbit} we have a term $\cosh \theta -1$ going to zero as fast as the $\sinh^2 \theta$ term in the numerator.
This situation  is realised if the roots $\gamma_a$ and $\gamma_b$ identifying the orbits have the same colour, $\circ$ or $\bullet$, according with the definition given in appendix \ref{App:1}, i.e. they belong to the same orthogonal root set in the Dynkin diagram. If this holds we obtain
\begin{equation}
\lim_{\theta \to 0} M_{ab} (\theta) = - (\gamma_a , \gamma_b)  \frac{i \g^2}{h} m_a m_b\lim_{\theta \to 0} \frac{\sinh^2 \theta}{\cosh \theta-1} =  -(\gamma_a , \gamma_b)  \frac{2i \g^2}{h} m_a m_b .
\end{equation}
Since $\gamma_a$ and $\gamma_b$ have the same colour, the only scalar product $(\gamma_a , \gamma_b)$ that is different from zero is when the two roots are exactly the same. This implies that in the collinear limit the amplitude is given by
\begin{equation}
\label{collinear_limit_obtained_sending_theta_to_zero_in_the_amplitude_formula_in_terms_of_root_orbits}
 M_{ab} (\theta) \Bigl|_{\theta \ll 1} = -\frac{2 i \g^2}{h}  \gamma_a^2 \ m_a^2 \delta_{ab}.
\end{equation}
In simply-laced models all the roots have the same length $\gamma_a^2=\Lambda^2$. In this case, taking into account \eqref{mod_f_simply_laced}, the expression in \eqref{collinear_limit_obtained_sending_theta_to_zero_in_the_amplitude_formula_in_terms_of_root_orbits} coincides exactly with the requirement \eqref{eq4_8bis0} for the cancellation of $5$-point amplitudes in theories satisfying simply-laced scattering conditions. Since simply-laced affine Toda theories belong to this class of models, this concludes the proof of their tree-level integrability.

\subsection{Tree-level bootstrap from Coxeter geometry}

The constraint~\eqref{eq4_8bis0} is not valid in non simply-laced theories; indeed it is a requirement for the cancellation of $5$-point processes only for theories satisfying property~\ref{Simply_laced_scattering_conditions}.
Despite that, we can still prove the cancellation of $5$-point processes in non simply-laced models by checking that the tree-level bootstrap relations~\eqref{tree_level_bootstrap_constraint_on_the_S_matrix_coming_from_5_point_pole_cancellation_equation} are satisfied. This is indeed the requirement for having zero residues in $5$-point amplitudes, as discussed
in section~\ref{No particles production in 5-point processes}. 

We can write the tree level S-matrix elements dividing the amplitude \eqref{third_expression_for_the_allowed_amplitude_ab_to_ab_couplings_substituted_and_N_substituted_with_scalar_product_sum_over_the_full_orbit} by the flux and normalisation factor $4m_am_b \sinh \theta$. We obtain
\begin{equation}
\label{S_matrix_obtained_summing_over_the_entire_orbit_for_a_process_ab_to_ab}
S^{tree}_{ab} (\theta)=-\frac{i \g^2}{4 h} \sum_{\beta \in \Gamma_b}(\gamma_a, \beta) \frac{ \sinh \theta}{\cosh \theta - \cos U_{\gamma_a \beta}}.
\end{equation}
In \eqref{S_matrix_obtained_summing_over_the_entire_orbit_for_a_process_ab_to_ab} all the mass dependence has disappeared and the scattering properties are all encoded in the scalar products between $\gamma_a$ and the different roots $\beta$ in the orbit $\Gamma_b$. From \eqref{S_matrix_obtained_summing_over_the_entire_orbit_for_a_process_ab_to_ab} we see that the building blocks necessary to construct the different S-matrix elements of the theory are given by $ \frac{\sinh \theta }{\cosh \theta - \cos U_{\gamma_a,\beta}}$, where $U_{\gamma_a,\beta}$ are the different fusing angles formed by the projections of $\gamma_a$ and $\beta$ on the spin-$1$ eigenplane of the Coxeter element. To prove the bootstrap relations in \eqref{tree_level_bootstrap_constraint_on_the_S_matrix_coming_from_5_point_pole_cancellation_equation}  such building blocks need to be rewritten in a clever way, since at the moment translations of $\theta$ and $U_{\gamma_a,\beta}$ do not correspond inside a single building block. To that end we note that the following identity holds
$$
\frac{\sinh \theta}{\cosh \theta -\cos U}=\frac{1}{2} \biggl[ \coth \Bigl( \frac{\theta}{2}-i\frac{U}{2} \Bigr) + \coth \Bigl( \frac{\theta}{2}+i\frac{U}{2} \Bigr)\biggr].
$$
Moreover given a generic root $\beta$ inside the orbit $\Gamma_b$ for which $(\gamma_a,\beta)\ne 0$, then there exists another root $\beta'$, whose projection on the spin-$1$ plane is obtained by reflecting $P_1(\beta)$ with
respect to $P_1(\gamma_a)$ and satisfying $(\gamma_a,\beta')=(\gamma_a,\beta)$ (we label with $P_1(x)$ the projection of a generic vector $x$ living in the root space on the spin-$1$ eigenplane of $w$ as defined in \eqref{new_notation_5}).
If we define $U_{\gamma_a \beta}=U$ then  $U_{\gamma_a \beta'}=-U$ and  the sum of the associated building blocks can be written as
$$
\frac{\sinh \theta}{\cosh \theta -\cos U}+\frac{\sinh \theta}{\cosh \theta -\cos (-U)}= \coth \Bigl( \frac{\theta}{2}-i\frac{U}{2} \Bigr) + \coth \Bigl( \frac{\theta}{2}+i\frac{U}{2} \Bigr)
$$
From this fact we see that we can substitute the functions 
$\frac{\sinh \theta}{\cosh \theta -\cos U}$ in \eqref{S_matrix_obtained_summing_over_the_entire_orbit_for_a_process_ab_to_ab} with either $\coth \Bigl( \frac{\theta}{2}-i\frac{U}{2} \Bigr)$ or $\coth \Bigl( \frac{\theta}{2}+i\frac{U}{2} \Bigr)$. As long as we sum over the entire orbit the choice of sign in front of the fusing angles $U$ it is not important since any time there is a pair of roots whose projections form an angle $U$ then there exists another pair of roots presenting the opposite angle and having the same scalar product.
Therefore the $S$-matrix can be written as
\begin{equation}
\label{S_matrix_obtained_summing_over_the_entire_orbit_for_a_process_ab_to_ab_coth_formulation_with_plusminus}
S^{tree}_{ab} (\theta)=-\frac{i \g^2}{4 h} \sum_{\beta \in \Gamma_b} (\gamma_a, \beta) \coth \Bigl( \frac{\theta}{2}\pm i\frac{U_{\gamma_a \beta}}{2} \Bigr),
\end{equation}
where the the sign in front of the angles $U_{\gamma_a,\beta}$ can be freely chosen. 
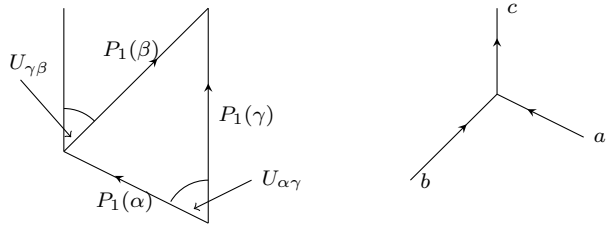
\begin{figure}
\begin{center}
\begin{tikzpicture}
\tikzmath{\y=1.9;}

\draw[][] (-1*\y,0.8*\y) arc(90:45:0.3*\y);
\draw[][] (0*\y,0.3*\y) arc(90:150:0.3*\y);

\draw[directed] (0*\y,0*\y) -- (0*\y,1.5*\y);
\draw[directed] (0*\y,0*\y) -- (-1*\y,0.5*\y);
\draw[directed] (-1*\y,0.5*\y) -- (0*\y,1.5*\y);
\draw[] (-1*\y,0.5*\y) -- (-1*\y,1.5*\y);
\draw[->] (-1.3*\y,1*\y) -- (-0.95*\y,0.6*\y);
\draw[->] (0.3*\y,0.3*\y) -- (-0.1*\y,0.1*\y);

\filldraw[black] (0.3*\y,0.3*\y)  node[anchor=west] {\scriptsize{$U_{\alpha \gamma}$}};
\filldraw[black] (-1.45*\y,1.1*\y)  node[anchor=west] {\scriptsize{$U_{\gamma \beta}$}};
\filldraw[black] (0.0*\y,0.75*\y)  node[anchor=west] {\scriptsize{$P_1(\gamma)$}};
\filldraw[black] (-0.85*\y,0.15*\y)  node[anchor=west] {\scriptsize{$P_1(\alpha)$}};
\filldraw[black] (-0.8*\y,1.2*\y)  node[anchor=west] {\scriptsize{$P_1(\beta)$}};

\draw[directed] (2*\y,0.9*\y) -- (2*\y,1.5*\y);
\draw[directed] (2.6*\y,0.6*\y) -- (2*\y,0.9*\y);
\draw[directed] (1.4*\y,0.3*\y) -- (2*\y,0.9*\y);

\filldraw[black] (1.4*\y,0.3*\y)  node[anchor=west] {\scriptsize{$b$}};
\filldraw[black] (2.6*\y,0.6*\y)  node[anchor=west] {\scriptsize{$a$}};
\filldraw[black] (2*\y,1.5*\y)  node[anchor=west] {\scriptsize{$c$}};

\end{tikzpicture}
\caption{The image on the left reports the projections of the roots $\alpha \in \Gamma_a$, $\beta \in \Gamma_b$, $\gamma \in \Gamma_c$ on the momentum plane (spin-$1$ Coxeter eigenplane) forming the fusing triangle corresponding to the coupling $C^{(3)}_{ab\bar{c}}$. On the RHS it is shown the corresponding fusing process $a+b \to c$. }
\label{Triangular_relation_between_the_roots_alpha_beta_and_gamma_projected_to_prove_tree_level_bootstrap}
\end{center}
\end{figure}
Using this new building block notation the bootstrap relations follow directly by the linearity of the scalar product, since now we can match
translations of $\theta$ with translations of the fusing angles.

Suppose there exists a $3$-point coupling $C^{(3)}_{ab \bar{c}} \ne 0$. Then there must be three roots $\alpha \in \Gamma_a$, $\beta \in \Gamma_b$ and $\gamma \in \Gamma_c$ satisfying $\alpha+\beta=\gamma$. Projecting these vectors on the momentum eigenplane we obtain the mass fusing triangle $\Delta_{abc}$. We refer to figure \ref{Triangular_relation_between_the_roots_alpha_beta_and_gamma_projected_to_prove_tree_level_bootstrap} to label the angles of the projected triangle $\bar{U}_{ac}^b=U_{\alpha, \gamma}$ and $\bar{U}_{bc}^a=U_{\gamma, \beta}$. 
In this manner, using the building block convention \eqref{S_matrix_obtained_summing_over_the_entire_orbit_for_a_process_ab_to_ab_coth_formulation_with_plusminus} with the choice of plus sign in front of the fusing angles $U$ the bootstrap equality \eqref{tree_level_bootstrap_constraint_on_the_S_matrix_coming_from_5_point_pole_cancellation_equation} is verified as follows
\begin{equation}
\label{Checking_Tree_Level_Bootstrap_with_Plus_sign}
\begin{split}
&S^{tree}_{da} (\theta-i\bar{U}_{ac}^b) + S^{tree}_{db} (\theta+i\bar{U}_{bc}^a) =S^{tree}_{da} (\theta-iU_{\alpha, \gamma}) + S^{tree}_{db} (\theta+iU_{\gamma, \beta})=\\
& -\frac{i \g^2}{4 h}  \sum_{p=0}^{h-1} \biggl[ (\alpha, w^{-p} \gamma_d) \coth \Bigl( \frac{\theta}{2}+\frac{i}{2} U_{\alpha  w^{-p}\gamma_{d}}  - \frac{i}{2} U_{\alpha \gamma} \Bigr)  + (\beta, w^{-p} \gamma_d) \coth \Bigl( \frac{\theta}{2}+\frac{i}{2} U_{\beta  w^{-p}\gamma_{d}}  + \frac{i}{2} U_{\gamma \beta} \Bigr) \biggr]=\\
&-\frac{i \g^2}{4 h} \sum_{p=0}^{h-1}  \biggl[ (\alpha, w^{-p} \gamma_d) \coth \Bigl( \frac{\theta}{2}+\frac{i}{2} U_{\gamma  w^{-p}\gamma_{d}} \Bigr) + (\beta, w^{-p} \gamma_d)\coth \Bigl( \frac{\theta}{2}+\frac{i}{2} U_{\gamma  w^{-p}\gamma_{d}} \Bigr) \biggr]=\\
&-\frac{i \g^2}{4 h} \sum_{p=0}^{h-1}  (\gamma, w^{-p} \gamma_d) \coth \Bigl( \frac{\theta}{2}+\frac{i}{2} U_{\gamma  w^{-p}\gamma_{d}} \Bigr)=S^{tree}_{dc} (\theta) .
\end{split}
\end{equation}
A general tree-level proof of the absence of inelastic scattering in all the untwisted affine Toda theories has therefore been completed.
It is worth noting that the fusing triangle in figure \ref{Triangular_relation_between_the_roots_alpha_beta_and_gamma_projected_to_prove_tree_level_bootstrap} can be reflected with
respect to the side corresponding to $P_1(\gamma)$,
generating an equivalent mass triangle. The bootstrap relation associated to this new triangle can be proved analogously to what we did in \eqref{Checking_Tree_Level_Bootstrap_with_Plus_sign} but using the formula \eqref{S_matrix_obtained_summing_over_the_entire_orbit_for_a_process_ab_to_ab_coth_formulation_with_plusminus} with the minus sign convention in front of the fusing angles.

To recover the bootstrapped S-matrix proposed in \cite{a24,aa24,a23} for the ADE series of affine Toda theories it is useful to write the root scalar product using the fundamental weights as shown in \eqref{wirting_roots_labeling_the_orbits_in_terms_of_fundamental_weights}
$$
(\gamma_a, w^{-p} \gamma_b)= ((1-w^{-1}) \lambda_a , w^{-p} \gamma_b)=( \lambda_a , w^{-p} \gamma_b) - ( \lambda_a , w^{-p+1} \gamma_b)
$$
The two equivalent expressions for the tree-level S-matrix \eqref{S_matrix_obtained_summing_over_the_entire_orbit_for_a_process_ab_to_ab} and \eqref{S_matrix_obtained_summing_over_the_entire_orbit_for_a_process_ab_to_ab_coth_formulation_with_plusminus} can then be written as follows
\begin{subequations}
\begin{equation}
\label{S_matrix_obtained_summing_over_the_entire_orbit_for_a_process_ab_to_ab_expressed_in_terms_of_weights}
S^{tree}_{ab} (\theta)=\frac{i \g^2}{4 h} \sum_{p=0}^{h-1}   ( \lambda_a , w^{-p} \gamma_b) \biggl(- \frac{\sinh \theta}{\cosh \theta - \cos \bigl( \frac{\pi}{h} u(\gamma_a ,w^{-p} \gamma_b) \bigr)} +\frac{\sinh \theta}{\cosh \theta - \cos \bigl( \frac{\pi}{h} u(\gamma_a, w^{-(p+1)} \gamma_b) \bigr)}   \biggr),
\end{equation}
\begin{equation}
\label{S_matrix_obtained_summing_over_the_entire_orbit_for_a_process_ab_to_ab_expressed_in_terms_of_weights_Patrick_tree_level_expansion}
S^{tree}_{ab} (\theta)=\frac{i \g^2}{4 h} \sum_{p=0}^{h-1} (\lambda_a, w^{-p} \gamma_b) \biggl(- \coth \Bigl( \frac{\theta}{2}\pm \frac{i\pi}{2h}u(\gamma_a, w^{-p}\gamma_b) \Bigr)+\coth \Bigl( \frac{\theta}{2}\pm \frac{i\pi}{2h}u(\gamma_a, w^{-(p+1)}\gamma_b) \Bigr) \biggr),
\end{equation}
\end{subequations}
where we have defined the angles in units of $\frac{\pi}{h}$, $U_{\alpha, \beta}=\frac{\pi}{h} u(\alpha, \beta)$.
Formula \eqref{S_matrix_obtained_summing_over_the_entire_orbit_for_a_process_ab_to_ab_expressed_in_terms_of_weights}, that has been completely determined starting from perturbation theory, represents the order $\g^2$ expansion of the quantum S-matrix 
\begin{equation}
\label{Fring_Olive_notation_for_the_bootstrapped_S_matrix}
S_{ab}(\theta)=1+S_{ab}^{tree}(\theta)+\ldots=\prod_{p=0}^{h-1} \{ 1+u(\gamma_a,w^{-p}\gamma_b)\}^{\frac{1}{2}(\lambda_a,w^{-p}\gamma_b)}
\end{equation}
in terms of the building blocks \cite{a23}
$$
\{ x\} = 1+\frac{i \g^2}{2 h}  \biggl(- \frac{\sinh \theta}{\cosh \theta - \cos \bigl( \frac{\pi}{h} (x-1) \bigr)} +\frac{\sinh \theta}{\cosh \theta - \cos \bigl( \frac{\pi}{h} (x+1) \bigr)}   \biggr)+o(\g^4).
$$
The factor $\frac{1}{2}$ in the exponents of \eqref{Fring_Olive_notation_for_the_bootstrapped_S_matrix} arises because we are writing the S-matrix as a  product of the building blocks over the full orbit, a convention adopted in \cite{Fring:1991gh}. For each brick $\{ x \}$ in the physical strip having a positive exponent $\frac{1}{2}( \lambda_a , w^{-p} \gamma_b)$ there exists a companion $\{ x' \}=\{ 2h- x \}$ with exponent $\frac{1}{2}( \lambda_a , w^{-p'} \gamma_b)=-\frac{1}{2}( \lambda_a , w^{-p} \gamma_b)$. The product of the two partners is obtained using general properties of the building blocks and is given by
$$
\{ x \}^{\frac{1}{2}( \lambda_a , w^{-p} \gamma_b)} \{ 2h - x \}^{-\frac{1}{2}( \lambda_a , w^{-p} \gamma_b)}= \{ x \}^{\frac{1}{2}( \lambda_a , w^{-p} \gamma_b)} \{ x \}^{\frac{1}{2}( \lambda_a , w^{-p} \gamma_b)} = \{ x \}^{( \lambda_a , w^{-p} \gamma_b)}.
$$
On the other hand \eqref{S_matrix_obtained_summing_over_the_entire_orbit_for_a_process_ab_to_ab_expressed_in_terms_of_weights_Patrick_tree_level_expansion} is the direct tree-level expansion of the equivalent formula
\begin{equation}
\label{Patrick_Dorey_notation_for_the_bootstrapped_S_matrix}
S_{ab}(\theta)=1+S_{ab}^{tree}(\theta)+\ldots=\prod_{p=0}^{h-1} \{ 1+u(\gamma_a,w^{-p}\gamma_b)\}_{\pm}^{(\lambda_a,w^{-p}\gamma_b)}.
\end{equation}
as first given in \cite{a24,aa24}, with the  plus or minus signs in front of the different scattering angles $u(\gamma_a,\beta)$ corresponding to the two possible choices of building blocks $\{1+u(\gamma_a,\beta)\}_+$ and $\{1+u(\gamma_a,\beta)\}_{-}$.

\subsection{Generalisation to twisted affine Toda theories}
\label{Generalization_to_twisted_Lie_algebras}
In this short subsection we show how to extend the analysis carried out so far to a generic twisted Toda theory (which is a model whose Dynkin diagram is obtained by folding the affine extension of a particular simply-laced diagram). A deeper discussion of twisted Lie algebras and their properties is reported in appendix~\ref{sub_appendix_on_twisted_Cox}; here we limit ourselves to explaining why the scattering constraints and the bootstrap relations previously obtained are universally satisfied also by these models, implying their tree level perturbative integrability.

We start by considering a simply-laced theory with a potential of the form in \eqref{eq2_2} and a copy of Cartan subalgebras $\mathcal{H}$ and $\mathcal{H'}$ associated to a pair of root systems $\Phi$ (containing a set of roots $\{ \alpha \}$) and $\Phi'$ (containing the roots $\{ \alpha' \}$). We can write the Lagrangian of the model both in terms of the generators associated to the root system $\Phi$, or in terms of the roots in $\Phi'$. If we follow the latter approach, by substituting \eqref{eq1_22} in \eqref{eq2_2} and using the relation~\eqref{An_adjoint_representation_relation}
we find the standard formulation for the Toda potential~\eqref{the_standard_affine_Toda_potential}.

Now we can apply the folding procedure on the extended set of simple roots, as explained in appendix \ref{sub_appendix_on_twisted_Cox}. In this way we reduce the set to $\{\alpha'^\parallel_i\}_{i=0}^r$, composed by the projections of the roots on the eigenspace invariant under the automorphism $\sigma$ of the Dynkin diagram. The potential, after the reduction, becomes
\begin{equation}
\label{reduced_potential_after_the_folding}
\frac{m^2}{\g^2}\sum_{i=0}^r n_i e^{\g \cdot (\phi, \alpha'^\parallel_i)}.
\end{equation}
Since many of the $\alpha'^\parallel_i$ are equal each other (as it happens in the example reported in \ref{sub_appendix_on_twisted_Cox_example} where $\alpha'_0, \alpha'_1, \alpha'_2$ and $\alpha'_3$ have the same projection on the $\sigma$-invariant space) many terms in \eqref{reduced_potential_after_the_folding} can be summed and we obtain the potential associated to the twisted model.
On the other hand, as discussed in \ref{sub_appendix_on_twisted_Cox}, the effect of the reduction is the same as setting to zero all the vectors of the basis \eqref{eq1_16} that are not invariant under the action of $\sigma$. This corresponds to suppressing all the couplings in \eqref{eq2_6} that contain one or more of the vectors $y_a$ that we set equal to zero.
The set of masses and couplings defining the twisted theory is therefore the subset of the masses and couplings of the simply-laced theory that survives the folding. 
Moreover the root orbits in $\Phi$ defining the set of vectors \eqref{eq1_16} invariant under the action of $\sigma$ form a subalgebra of the initial simply-laced root system. All the scattering properties are then satisfied within this subalgebra. If we consider for example figure \ref{fig:partitionMultiregge}, that corresponds to a pictorial representation of the relation \eqref{eq3_28}, we note that all the roots $\rho$ and $\tilde{\rho}$ that are expressed as sums of different roots in the $\sigma$-invariant subalgebra need to belong themselves to this subalgebra. This implies that for a given set of $\sigma$-invariant indices $a_1, \ldots , a_n$ (that means $y_{a_1},\dots,y_{a_n}$ are invariant under the action of $\sigma$) the associated coupling $C^{(n)}_{a_1 \dots a_n}$ is defined by summing on the RHS of \eqref{eq3_28} only over the  intermediate indices inside the subalgebra. The relation \eqref{eq0_6}, together with the other scattering constraints, is therefore completely satisfied internally to such a subalgebra ensuring the perturbative integrability of twisted Toda theories.

\section{Conclusions}\label{conclusions}

In this paper we have given necessary and sufficient conditions for a theory with a Lagrangian of the form~\eqref{eq0_1} to be perturbatively integrable, with a purely elastic S-matrix, at tree level. We proved that the theory is completely defined once the mass ratios and the $3$-point couplings are known. Once this fundamental data is given, the $4$- and the higher-point couplings can be uniquely determined using the equalities in \eqref{New_eq4_13_fixed}, \eqref{eq0_6} and by requiring that a set of tree-level bootstrap relations  \eqref{tree_level_bootstrap_constraint_on_the_S_matrix_coming_from_5_point_pole_cancellation_equation} is satisfied. The tree-level bootstrap approach is particularly useful to find the $4$-point couplings $C^{(4)}_{cccc}$, in which four real equal particles 
fuse together; indeed such couplings contribute to $4$-point elastic processes and the  relation~\eqref{New_eq4_13_fixed} cannot be applied.

Probing the space of tree level perturbatively integrable theories corresponds of searching for the mass ratios and $3$-point couplings from which the entire Lagrangian can be constructed iteratively. This problem can be addressed by two different routes. On the one hand it is possible to select the possible candidates by searching for what masses and $3$-point couplings allow for the cancellation of poles in all 2 to 2 inelastic processes, according to what has been discussed in section~\ref{sect2_2to2notallowed}. Subsequently only the models having S-matrices satisfying the relations~\eqref{tree_level_bootstrap_constraint_on_the_S_matrix_coming_from_5_point_pole_cancellation_equation} are actually integrable models at tree level. On the other hand it is possible to address the problem in the opposite direction, by first imposing the tree level bootstrap relations~\eqref{tree_level_bootstrap_constraint_on_the_S_matrix_coming_from_5_point_pole_cancellation_equation}. This second approach is in principle simpler: matching the pole structure of the different S-matrix elements entering the bootstrap corresponds to setting the values of the masses, while matching the residues at the poles corresponds to defining the absolute values of the $3$-point couplings. To determine the relative signs among the different $3$-point couplings we need then to require the cancellation of 4-point inelastic processes from which we obtain a set of relations of the form in~\eqref{New_eq4_12}.

Although we did not address the problem of classifying all perturbatively integrable theories at the tree level, we have provided tools to pursue that goal.
Hints for such a classification can be found in \cite{aa20,Koubek:1991jj} where the first steps were performed for non-perturbative S-matrices of minimal models using a bootstrap approach. Reproducing the classification at tree level should be simpler than studying the quantum exact S-matrices, since the fact that the RHS of~\eqref{tree_level_bootstrap_constraint_on_the_S_matrix_coming_from_5_point_pole_cancellation_equation} involves a sum instead of a product does not introduce higher order singularities that often generate inconsistencies. 
Restricting  attention to theories satisfying the simply-laced scattering conditions
% ~\ref{Simply_laced_scattering_conditions} 
we point out that the number of degrees of freedom is further restricted. Once we fix the masses, the $3$-point couplings (if they are not null) must obey the area rule
\eqref{Area_rule_abs_f_value}. The only relevant information we need
to know is therefore what masses we can have and what $3$-point
couplings are not zero. If we want to classify all the possible
integrable theories with $N$ different particle types and satisfying
simply-laced scattering conditions the set of such systems lives in 
\begin{equation}
\mathbb{R}^{N-1} \ \times \ (\mathbb{Z}_2)^{\frac{N(N+1)(N+2)}{6}}
\end{equation}
the product between the space of mass ratios and the $3$-point couplings. 
We have shown how simply-laced and twisted affine Toda field theories belong to this class of models, but it remains unclear if they are the only integrable models in two dimensions respecting property~\ref{Simply_laced_scattering_conditions}, or it is possible to find some integrable theory verifying such conditions that is not an affine Toda theory.

In the second part of the paper we verified that the conditions imposed by the absence of non-diagonal processes are satisfied for all the affine Toda field theories, exploiting universal properties of their underlying Lie algebras and root systems. 
The results we found show a beautiful connection between the absence of production
in affine Toda theories with the properties of
their associated Lie algebras. The on-shell momenta contributing to
singularities are the projections of the higher dimensional polytopes
composed by the roots and the entire no particle production mechanism
comes from the hidden geometry. These results provide the first universal proof of tree-level perturbative integrability valid for any affine Toda theory. 

For a complete
proof of perturbative integrability in affine Toda models, a loop investigation of the same
phenomenon is needed. In affine Toda
theories based on simply-laced algebras, one-loop corrections do not
modify the mass ratios \cite{aa20,a23} and the
nature of the theory emerging from tree level diagrams is
preserved. In \cite{Braden:1990wx} second and third
order anomalous threshold singularities were found perturbatively, displaying perfect agreement with  bootstrap
results. These results revealed a number of universal features and it would be interesting to relate them more closely to the properties of root systems, as done in this paper at tree level.
A similarly simple behaviour has also been seen in some
supersymmetric extensions of non simply-laced theories
\cite{Delius:1990ij,Delius:1991sv}. Also in these cases, though in a
more complicated manner, we expect to find a cancellation mechanism
coming from a connection between the amplitude singularities and tiled polytopes living in a higher-dimensional space. However when we turn to the non simply-laced and twisted affine Toda theories, the inclusion of loops complicates the picture considerably.
In particular, the mass ratios in general shift in a coupling-dependent way, and some three-point couplings vanish. Full quantum S-matrices for these  
theories have been proposed \cite{a25,a26,Corrigan:1993xh,Dorey:1993np}, and a construction of these S-matrices based on a deformed Coxeter element was found in
\cite{Oota:1997un}, but a full geometrical understanding of their properties remains an open problem.

Finally the no particle production approach may be used to study integrability preserving boundaries and defects. In this respect boundary \cite{b3,b6} and defect \cite{d1,d3,d4,d7} potentials have been studied at classical and quantum level by imposing the conservation of the higher-spin charges. However it would also be interesting to construct such interactions starting from the no particle production requirement. It is possible that also in these situations the proof could emerge from some root system geometry contained in the boundary and defect potentials.

\vskip 20pt
\noindent
{\bf Acknowledgments}

\noindent
This work has received funding from the  European Union's  Horizon  2020  research  and  innovation programme under the Marie  Sk\l odowska-Curie  grant  agreement  No.  764850 \textit{``SAGEX''}, and from the STFC under consolidated grant ST/T000708/1 “Particles, Fields and Spacetime”.

\vskip 30pt

\appendix

\section{The Cayley-Menger determinant}  \label{App:2}
This appendix collects some elementary
geometrical properties of simplices used in the paper to prove the absence of singularities in inelastic processes. 
\subsection{The basic formula}\label{BasicFormula}
Let $\vec{x}^0$,
$\vec{x}^1$, \dots,
$\vec{x}^n$ be $n+1$ points in $\mathbb{R}^n$, and set
$X_{ij}=\norm{\vec{x}^i-\vec{x}^j}^2$ for $0\le i,j\le n$. 
Let $v_n$ be the $n$-dimensional simplex with vertices 
$\vec{x}^0$, $\vec{x}^1$, \dots, $\vec{x}^n$. Then a classic result from distance geometry states that $\mbox{vol}(v_n)^2$,
the square of the volume of $v_n$, 
is given by the Cayley-Menger determinant:
\begin{equation}\label{Cayley_Menger_determinant}
\mbox{vol}(v_n)^2=
\frac{(-1)^{n+1}}{(n!)^22^n}\,
\begin{vmatrix}
~0 & X_{01} & X_{02} & \cdots & X_{0n} & 1~     \\[3pt]
~X_{10} & 0 & X_{12} & \cdots & X_{1n} & 1~     \\[3pt]
~X_{20} & X_{21} & 0 & \cdots & X_{2n} & 1~     \\[1pt]
~\vdots & \vdots & \vdots & \ddots & \vdots & \vdots~ \\[3pt]
~X_{n0} & X_{n1} & X_{n2} & \cdots & 0 & 1~     \\[2.5pt]
~1 & 1 & 1 & \cdots & 1 & 0~
\end{vmatrix}~.
\end{equation}

\subsection{A generalisation}\label{A generalisation}

Now let $\vec{x}^0$,
$\vec{x}^1$, \dots,
$\vec{x}^{n+1}$ be $n+2$ points in $\mathbb{R}^n$, and set
$X_{ij}=\norm{\vec{x}^i-\vec{x}^j}^2$ for $0\le i,j\le n+1$. 
Let $v_n$ be the simplex with vertices
$\vec{x}^0$, $\vec{x}^1$, \dots, %$\vec{x}^{n-1}$,
$\vec{x}^n$, and $v_{n+1}$ the 
simplex with vertices
$\vec{x}^0$, $\vec{x}^1$, \dots, $\vec{x}^{n-1}$, $\vec{x}^{n+1}$.
Note that $v_n$ and $v_{n+1}$ have as a common face the $n-1$
dimensional simplex with vertices $\vec{x}^0$, $\vec{x}^1$, \dots,
$\vec{x}^{n-1}$. Then
\begin{equation}\label{Generalised_Cayley_Menger_determinant}
\mbox{vol}(v_n)
\mbox{vol}(v_{n+1})=
\frac{\varepsilon\,(-1)^{n+1}}{(n!)^22^n}\,
\begin{vmatrix}
~0 & X_{01} & \!\cdots & X_{0,n-1} & X_{0n} & 1~   \\[3pt]
~X_{10} & 0 & \!\cdots & X_{1,n-1} & X_{1n} & 1~    \\[3pt]
~\vdots & \vdots & \!\ddots & \vdots & \vdots & \vdots~ \\[3pt]
~X_{n-1,0} & X_{n-1,1} & \!\cdots & 0 & X_{n-1,n} & 1~     \\[2.5pt]
~X_{n+1,0} & X_{n+1,1} & \!\cdots & 
X_{n+1, n-1} & X_{n+1,n} & 1~   \\[2.5pt]
~1 & 1 & \cdots & 1 & 1 & 0~
\end{vmatrix}
\end{equation}
where $\varepsilon = {+}1 \slash {-}1$ according to whether
 $\vec{x}^n$ and $\vec{x}^{n+1}$ lie on the same / opposite 
sides of the codimension one hyperplane occupied by the common face.
In the special case where $\vec{x}^{n+1}=\vec{x}^n$, we have $\varepsilon=+1$, $v_{n+1}=v_{n}$,
$X_{n+1,i}=X_{n,i}$ for $i=1,
\dots, n$ and $X_{n+1,n}=0$, and this reduces to the Cayley-Menger
determinant.

\bigskip\noindent
To prove this result, we first set $\vec{x}^0=\vec{0}$. Then $\mathit{RHS}$, the
right-hand side of the
above formula, is equal to $\frac{\varepsilon(-1)^{n+1}}{(n!)^22^n}\,D$, where $D=$
\[
\!\!\!\!\!\!\!\!\!\!
\begin{vmatrix}
~0 & \norm{\vec{x}^1}^2 & \norm{\vec{x}^2}^2 & \!\cdots & \norm{\vec{x}^n}^2 & 1~     \\[3pt]
~\norm{\vec{x}^1}^2 & 0 & \norm{\vec{x}^1}^2{-}2\vec{x}^1{\cdot}\vec{x}^2{+}\norm{\vec{x}^2}^2 & \!\cdots &
\norm{\vec{x}^1}^2{-}2\vec{x}^1{\cdot}\vec{x}^n{+}\norm{\vec{x}^n}^2 & 1~     \\[3pt]
~\norm{\vec{x}^2}^2 & \norm{\vec{x}^2}^2{-}2\vec{x}^2{\cdot}\vec{x}^1{+}\norm{\vec{x}^1}^2 & 0 & \!\cdots &
\norm{\vec{x}^2}^2{-}2\vec{x}^2{\cdot}\vec{x}^n{+}\norm{\vec{x}^n}^2 & 1~     \\[1pt]
~\vdots & \vdots & \vdots & \!\ddots & \vdots & \vdots~ \\[3pt]
~\norm{\vec{x}^{n+1}}^2 & \norm{\vec{x}^{n+1}}^2{-}2\vec{x}^{n+1}{\cdot}\vec{x}^1{+}\norm{\vec{x}^1}^2 &
\norm{\vec{x}^{n+1}}^2{-}2\vec{x}^{n+1}{\cdot}\vec{x}^2{+}\norm{\vec{x}^2}^2 & \!\cdots &
\norm{\vec{x}^{n+1}}^2{-}2\vec{x}^{n+1}{\cdot}\vec{x}^n{+}\norm{\vec{x}^n}^2 & 1~     \\[2.5pt]
~1 & 1 & 1 & \cdots & 1 & 0~
\end{vmatrix}~.
\]
Subtracting the first row from every other row and the first column from every other column except for the last ones,
\[
D=
\begin{vmatrix}
~0 & \norm{\vec{x}^1}^2 & \norm{\vec{x}^2}^2 & \!\cdots & \norm{\vec{x}^n}^2 & 1~     \\[3pt]
~\norm{\vec{x}^1}^2 & -2\norm{\vec{x}^1}^2 & -2\vec{x}^1{\cdot}\vec{x}^2 & \!\cdots & 
-2\vec{x}^1{\cdot}\vec{x}^n & 0~     \\[3pt]
~\norm{\vec{x}^2}^2 & -2\vec{x}^2{\cdot}\vec{x}^1 & -2\norm{\vec{x}^2}^2 & \!\cdots & 
-2\vec{x}^2{\cdot}\vec{x}^n & 0~     \\[1pt]
~\vdots & \vdots & \vdots & \!\ddots & \vdots & \vdots~ \\[3pt]
~\norm{\vec{x}^{n+1}}^2 & -2\vec{x}^{n+1}{\cdot}\vec{x}^1 &
-2\vec{x}^{n+1}{\cdot}\vec{x}^2 & \!\cdots &
-2\vec{x}^{n+1}{\cdot}\vec{x}^n & 0~     \\[2.5pt]
~1 & 0 & 0 & \cdots & 0 & 0~
\end{vmatrix}~.
\]
Expanding by the last row and column,
\[
D
=
-\begin{vmatrix}
-2\vec{x}^1{\cdot}\vec{x}^1 & -2\vec{x}^1{\cdot}\vec{x}^2 & \!\cdots & 
-2\vec{x}^1{\cdot}\vec{x}^n      \\[3pt]
 -2\vec{x}^2{\cdot}\vec{x}^1 &-2\vec{x}^2{\cdot}\vec{x}^2 & \!\cdots & 
-2\vec{x}^2{\cdot}\vec{x}^n      \\[1pt]
~\vdots & \vdots & \!\ddots & \vdots  \\[3pt]
-2\vec{x}^{n+1}{\cdot}\vec{x}^1 &
-2\vec{x}^{n+1}{\cdot}\vec{x}^2 & \!\cdots &
-2\vec{x}^{n+1}{\cdot}\vec{x}^n      
\end{vmatrix} =
-(-2)^n\begin{vmatrix}
\vec{x}^1{\cdot}\vec{x}^1 & \vec{x}^1{\cdot}\vec{x}^2 & \!\cdots &
\vec{x}^1{\cdot}\vec{x}^n      \\[3pt]
\vec{x}^2{\cdot}\vec{x}^1 & \vec{x}^2{\cdot}\vec{x}^2& \!\cdots &
\vec{x}^2{\cdot}\vec{x}^n      \\[1pt]
~\vdots & \vdots & \!\ddots & \vdots  \\[3pt]
\vec{x}^{n+1}{\cdot}\vec{x}^1 &
\vec{x}^{n+1}{\cdot}\vec{x}^2 & \!\cdots &
\vec{x}^{n+1}{\cdot}\vec{x}^n
\end{vmatrix}~.
\]
Now without loss of generality we choose coordinates so that
$\vec{x}^1=(x^1_1,0,0,\dots)$,
$\vec{x}^2=(x^2_1,x^2_2,0,\dots)$, \dots ,
$\vec{x}^n=(x^n_1,x^n_2,\dots x^n_n)$,
$\vec{x}^{n+1}=(x^{n+1}_1,x^{n+1}_2,\dots x^{n+1}_n)$.
Notice that in these coordinates the hyperplane inhabited by the common face is 
$\mbox{span}(\vec{e}_1,\dots ,\vec{e}_{n-1})$, and so $\varepsilon=\mbox{sign}(x^n_nx^{n+1}_n)$.
Then 
\begin{align*}
\mathit{RHS}=\frac{\varepsilon(-1)^{n+1}}{(n!)^22^n}\,D 
& = \frac{\varepsilon}{(n!)^2}\,
\begin{vmatrix}
~x^1_1x^1_1 & x^1_1x^2_1 & \!\cdots & x^1_1x^n_1      \\[4pt]
~x^2_1x^1_1 & x^2_1x^2_1 + x^2_2x^2_2 & \!\cdots & x^2_1x^n_1 + x^2_2x^n_2~  \\[1pt]
~\vdots & \vdots & \!\ddots & \vdots  \\[3pt]
~x^{n+1}_1x^1_1 & x^{n+1}_1x^2_1 + x^{n+1}_2x^2_2 & \!\cdots & x^{n+1}_1x^n_1+x^{n+1}_2x^n_2+\dots~
\end{vmatrix}\\[6pt]
& = \frac{\varepsilon\,x_1^1}{(n!)^2}\,\begin{vmatrix}
~x^1_1 & x^1_1x^2_1 & \!\cdots & x^1_1x^n_1      \\[3pt]
~x^2_1 & x^2_1x^2_1 + x^2_2x^2_2 & \!\cdots & x^2_1x^n_1 + x^2_2x^n_2~  \\[1pt]
~~\vdots & \vdots & \!\ddots & \vdots  \\[3pt]
~x^{n+1}_1 & x^{n+1}_1x^2_1 + x^{n+1}_2x^2_2 & \!\cdots & x^{n+1}_1x^n_1+x^{n+1}_2x^n_2+\dots~
\end{vmatrix}~.
\end{align*}
Subtracting $x^2_1$ times the first column from the second, 
$x^3_1$ times the first column
from the third, and so on, and then expanding by the first row,
\[
\mathit{RHS} \,=\, 
\frac{\varepsilon\,x_1^1}{(n!)^2}\,
\begin{vmatrix}
~x^1_1 & 0 & \!\cdots & 0      \\[3pt]
~x^2_1 & x^2_2x^2_2 & \!\cdots & x^2_2x^n_2~  \\[1pt]
~~\vdots & \vdots & \!\ddots & \vdots  \\[3pt]
~x^{n+1}_1 & x^{n+1}_2x^2_2 & \!\cdots & x^{n+1}_2x^n_2+\dots~
\end{vmatrix}
\,=\,
\frac{\varepsilon\,(x_1^1)^2}{(n!)^2}\,
\begin{vmatrix}
~x^2_2x^2_2 & \!\cdots & x^2_2x^n_2~  \\[1pt]
~~\vdots & \!\ddots & \vdots  \\[3pt]
~x^{n+1}_2x^2_2 & \!\cdots & x^{n+1}_2x^n_2+\dots~
\end{vmatrix}~.
\]
Now we repeat the procedure until, as final step, we obtain
\[
\mathit{RHS}\,=\,\frac{\varepsilon\,(x^1_1)^2(x^2_2)^2\dots (x^{n-2}_{n-2})^2}{(n!)^2}\,
\begin{vmatrix}
\,x^{n-1}_{n-1}x^{n-1}_{n-1} & x^{n-1}_{n-1}x^n_{n-1}\,  \\[3pt]
\,x^{n+1}_{n-1}x^{n-1}_{n-1} & x^{n+1}_{n-1}x^n_{n-1} + x^{n+1}_nx^n_n\,
\end{vmatrix}
\,=\,\frac{\varepsilon\,(x^1_1)^2(x^2_2)^2\dots
(x^{n-1}_{n-1})^2x^n_nx^{n+1}_n}{(n!)^2}\,.
\]
Since $x^n_nx^{n+1}_n = \varepsilon |x^n_nx^{n+1}_n|$, this
is equal to $\mbox{vol}(v_n)\mbox{vol}(v_{n+1})$, as required.

\subsection{An application}

Let $\vec{x}^{i}$, $i=0,1,2,3$, be vectors identifying four points on a plane. Embedding the plane in $\mathbb{R}^3$, the Cayley-Menger determinant for the volume of the simplex with these four vertices must vanish:
\[
0=
\begin{vmatrix}
~0 & X_{01} & X_{02} & X_{03} & 1~ \\
~X_{10} & 0 & X_{12} & X_{13} & 1~ \\
~X_{20} & X_{21} & 0 & X_{23} & 1~ \\
~X_{30} & X_{31} & X_{32} & 0 & 1~ \\
~1 & 1 & 1 & 1 & 0~
\end{vmatrix}
\]
Differentiating the above equation,
\begin{align*}
0=
d
\begin{vmatrix}
~0 & X_{01} & X_{02} & X_{03} & 1~ \\
~X_{10} & 0 & X_{12} & X_{13} & 1~ \\
~X_{20} & X_{21} & 0 & X_{23} & 1~ \\
~X_{30} & X_{31} & X_{32} & 0 & 1~ \\
~1 & 1 & 1 & 1 & 0~
\end{vmatrix}
&=
\begin{vmatrix}
~0 & dX_{01} & dX_{02} & dX_{03} & 0~ \\
~X_{10} & 0 & X_{12} & X_{13} & 1~ \\
~X_{20} & X_{21} & 0 & X_{23} & 1~ \\
~X_{30} & X_{31} & X_{32} & 0 & 1~ \\
~1 & 1 & 1 & 1 & 0~
\end{vmatrix} 
+
\begin{vmatrix}
~0 & X_{01} & X_{02} & X_{03} & 1~ \\
~dX_{10} & 0 & dX_{12} & dX_{13} & 0~ \\
~X_{20} & X_{21} & 0 & X_{23} & 1~ \\
~X_{30} & X_{31} & X_{32} & 0 & 1~ \\
~1 & 1 & 1 & 1 & 0~
\end{vmatrix}
\\[4pt]
&+
\begin{vmatrix}
~0 & X_{01} & X_{02} & X_{03} & 1~ \\
~X_{10} & 0 & X_{12} & X_{13} & 1~ \\
~dX_{20} & dX_{21} & 0 & dX_{23} & 0~ \\
~X_{30} & X_{31} & X_{32} & 0 & 1~ \\
~1 & 1 & 1 & 1 & 0~
\end{vmatrix}
+
\begin{vmatrix}
~0 & X_{01} & X_{02} & X_{03} & 1~ \\
~X_{10} & 0 & X_{12} & X_{13} & 1~ \\
~X_{20} & X_{21} & 0 & X_{23} & 1~ \\
~dX_{30} & dX_{31} & dX_{32} & 0 & 0~ \\
~1 & 1 & 1 & 1 & 0~
\end{vmatrix}  .
\end{align*}
Expanding this expression row by row and using the fact that the determinant of a matrix is equal to that of its transpose we obtain
\begin{align*}
0=& -2 \ dX_{01}
\begin{vmatrix}
~X_{10}  & X_{12} & X_{13} & 1~ \\
~X_{20} & 0 &  X_{23} & 1~ \\
~X_{30} & X_{32} & 0 & 1~ \\
~1 & 1 & 1  & 0~
\end{vmatrix} 
+ 2 \ dX_{02}
\begin{vmatrix}
~X_{10} & 0 & X_{13} & 1~ \\
~X_{20} & X_{21} & X_{23} & 1~ \\
~X_{30} & X_{31} & 0 & 1~ \\
~1 & 1 & 1  & 0~
\end{vmatrix}
- 2 \ dX_{03}
\begin{vmatrix}
~X_{10} & 0 & X_{12} & 1~ \\
~X_{20} & X_{21} & 0 & 1~ \\
~X_{30} & X_{31} & X_{32} & 1~ \\
~1 & 1 & 1 & 0~
\end{vmatrix}
\\[4pt]
&- 2 \ dX_{12}
\begin{vmatrix}
~0 & X_{01} & X_{03} & 1~ \\
~X_{20} & X_{21} & X_{23} & 1~ \\
~X_{30} & X_{31} & 0 & 1~ \\
~1 & 1 & 1 & 0~
\end{vmatrix}
+ 2 \ dX_{13}
\begin{vmatrix}
~0 & X_{01} & X_{02} & 1~ \\
~X_{20} & X_{21} & 0 & 1~ \\
~X_{30} & X_{31} & X_{32} & 1~ \\
~1 & 1 & 1 & 0~
\end{vmatrix} 
-2 \ dX_{23}
\begin{vmatrix}
~0 & X_{01} & X_{02} & 1~ \\
~X_{10} & 0 & X_{12} & 1~ \\
~X_{30} & X_{31} & X_{32} & 1~ \\
~1 & 1 & 1 & 0~
\end{vmatrix} 
\end{align*}
Swapping some judiciously-chosen rows and columns within the determinants and using the generalised Cayley-Menger determinant \eqref{Generalised_Cayley_Menger_determinant}  we obtain
\begin{equation}\label{relation_among_distances_of_quadrilaterals}
\begin{split}
&\varepsilon_{01}(2,3) \Delta_{230} \Delta_{231} \ dX_{01} \ + \ \varepsilon_{02}(1,3) \Delta_{130} \Delta_{132} \ dX_{02} \ + \\
&\varepsilon_{03}(1,2) \Delta_{120} \Delta_{123} \ dX_{03} \ +\ \varepsilon_{12}(0,3) \Delta_{031} \Delta_{032} \ dX_{12} \ + \\
&\varepsilon_{13}(0,2) \Delta_{021} \Delta_{023} \ dX_{13} \ +\ \varepsilon_{23}(0,1) \Delta_{012} \Delta_{013} \ dX_{23}=0 .
\end{split}
\end{equation}
where $\varepsilon_{ij}(m,n)=+1/{-}1$ according to whether $\vec{x}^{i}$ and $\vec{x}^{j}$ lie on the same / opposite sides of the line connecting $\vec{x}^{m}$ and $\vec{x}^{n}$.

Let us assume the lengths of the sides of the quadrilateral defined by the ordered vertices $\vec{x}^{i}$ ($i=0,1,2,3$) are kept fixed and move only its diagonals. This is equivalent to requiring  $dX_{01}=dX_{12}=dX_{23}=dX_{30}=0$. In this case the relation \eqref{relation_among_distances_of_quadrilaterals} becomes
\begin{equation}\label{relation_among_diagonal_of_quadrilaterals}
 \frac{dX_{13}}{dX_{02}} = - \frac{\varepsilon_{02}(1,3)}{\varepsilon_{13}(0,2)} \frac{\Delta_{130} \Delta_{132}}{ \Delta_{021} \Delta_{023}}
\end{equation}
If the quadrilateral is convex then $\varepsilon_{02}(1,3)=\varepsilon_{13}(0,2)<0$ giving the relation \eqref{New_eq4_3convex}. If the quadrilateral is concave then either $\varepsilon_{02}(1,3)=-\varepsilon_{13}(0,2)=1$ or $\varepsilon_{02}(1,3)=-\varepsilon_{13}(0,2)=-1$. In both cases the RHS of \eqref{relation_among_diagonal_of_quadrilaterals} is positive and we obtain \eqref{New_eq4_3concave}. 
Note that in the relation \eqref{relation_among_distances_of_quadrilaterals} we use the name of the vertices to define the triangles while in the rest of the paper we define triangles using the labels of their sides.

\section{Residues in 5-point amplitudes} \label{App:0_2}
In this appendix we show explicit calculations of 
$\underset{a_5=a_3}{\mathrm{Res}}M^{(5)}(a_3,a_5)$, defined in~\eqref{Equation_Added_eq4_7}, at its pole positions as a function of $a_3$.
We split the singularities of the residue in two kinds that we call respectively `flipped
singularities' and `collinear singularities'. We discuss the two different singularities  separately. 

\subsection{Flipped singularities}

Referring to the case studied in figure \ref{fig5amplit_flipping} we write only the divergent part in square brackets of \eqref{Equation_Added_eq4_7}. We keep the two momenta of the $d$-particles parallel (that correspond to the limit $a_5 = a_3$) and we move them simultaneously in such a way to be close to have the $i$ particle on-shell ($p_i^2=m_i^2$). In this way we are close to the pole of the residue.  As convention we choose the direction of the momentum $p_4$ to be the axis respect to which we measure the angles and we choose the counterclockwise as positive direction of angles. In this way looking at the first picture in figure \ref{fig5amplit_flipping} we have $a_4=1$, $a_1=e^{i\bar{U}^b_{ac}}$, $a_2=e^{-i\bar{U}^a_{bc}}$ and $a_3=a_5=e^{i(\pi-\bar{U}^i_{cd})}$. We define $\bar{U}^{A}_{BC}$ the angle in the triangle $\Delta_{ABC}$ between the sides $B$ and $C$. Substituting such quantities in the divergent part of the residue \eqref{Equation_Added_eq4_7} we obtain 
\begin{equation}\label{App:2_1}
\underset{a_5=a_3}{\mathrm{Res}}M^{(5)}(a_3,a_5) \sim   \frac{1}{m_c} \frac{1}{\sin(\bar{U}_{cd}^i)} \frac{\mid C^{(3)}_{cd\bar{i}} \mid^2}{p_i^2-m_i^2} - \frac{1}{m_a} \frac{1}{\sin(\bar{U}_{ad}^j)}\frac{\mid C^{(3)}_{ad\bar{j}} \mid^2}{p_j^2-m_j^2}
\end{equation}
Always referring to the first drawing in figure \ref{fig5amplit_flipping} now we move the inclination of $d(p_3)$ and $d(p_5)$ keeping them parallel. 
Expanding $p_j^2$ in terms of $p_i^2$ around the pole $p_i^2=m_i^2$ we obtain
\begin{equation} \label{App:2_3}
p_j^2= m_j^2+\frac{d p_j^2}{d p_i^2}\biggr|_{p_i^2=m_i^2} (p_i^2-m_i^2) + \ldots=m_j^2+\frac{\Delta_{adj}}{\Delta_{cdi}} (p_i^2-m_i^2) + \ldots
\end{equation}
The last equality has been obtained using the formula \eqref{relation_among_distances_of_quadrilaterals} on the quadrilateral defined by the sides $a(p_1)$, $b(p_2)$, $i(p_3+p_4)$, $d(p_3)$
and having as diagonal $j(p_1+p_3)$ in the diagram on the top right of figure \ref{fig5amplit_flipping}.
Expressing at this point the pole in terms of $p_i^2$ we see that~\eqref{App:2_1} is proportional to~\eqref{NewNeweq4_8}.

\subsection{Collinear singularities}

As before we keep $a_4=1$ the direction respect to which we measure
the angles following the counterclockwise convention. At this point we
choose to move $p_3$ and $p_5$ keeping them parallel and with fixed
length. So we choose $a_3=a_5=e^{i\theta}$.  The collinear singularity happens in the limit $\theta \to 0$. As before we have $a_1=e^{i\bar{U}^b_{ac}}$, $a_2=e^{-i\bar{U}^a_{bc}}$.
Expanding the red propagators in figure \ref{fig5amplit_collinear} respect to the angle $\theta$ around the value $\theta = 0$ we obtain 
\begin{equation}\label{App:2_5}
p_b^2(\theta)=m_b^2-2 m_a m_c \sin(\bar{U}_{ac}^b) \theta \hspace{4mm}, \hspace{4mm} p_a^2(\theta)=m_a^2+2 m_b m_c \sin(\bar{U}_{bc}^a) \theta
\end{equation}
Now we expand the expression between square brackets in equation \eqref{Equation_Added_eq4_7} isolating the pole $\frac{1}{\theta}$. After a straightforward calculation we obtain
\begin{equation}\label{App:2_6}
%Res(M_5)=
\underset{a_5=a_3}{\mathrm{Res}}M^{(5)}(a_3,a_5) \sim \frac{i}{2 m_c \theta} \Bigl[ M^{(4)}_{cc} (1,1) + \frac{i}{2m_a^2 \sin^2(\bar{U}_{ac}^b)} \mid C^{(3)}_{ab\bar{c}} \mid^2+\frac{i}{2m_b^2 \sin^2(\bar{U}_{bc}^a)} \mid C^{(3)}_{ab\bar{c}} \mid^2 \Bigr]
\end{equation}
The term $M^{(4)}_{cc} (1,1)$ represents the amplitude for a process $c+c \to c+c$ in which the momenta of the incoming and outgoing particles are collinear. 
Then we substitute in this formula the expression for the $3$-point coupling in \eqref{New_eq4_6} and we obtain that the residue is equal to zero if the equality in \eqref{eq4_8} is verified.

\section{Semisimple Lie algebras and Coxeter geometry}  \label{App:1}
The purpose of this appendix is to give the reader the tools
needed to construct the geometry of Toda theories associated to affine Dynkin-Kac diagrams. With this in mind we start with a reminder of some properties of Lie algebras and root systems. 

\subsection{The Weyl group and the Coxeter element} \label{Weyl group and Coxeter element}
Given a semisimple Lie algebra $\mathcal{G}$ it is always possible to find a Cartan subalgebra $\mathcal{H} < \mathcal{G}$ comprising a maximal set of commuting generators $\{h_a\}_{a=1}^r$, the dimension $r$ of $\mathcal{H}$  being equal to the rank of the algebra. Diagonalising the adjoint action of this subalgebra on the remaining generators of $\mathcal{G}$ we obtain the following basis in $\mathcal{G}$ :
\begin{equation}\label{eq1_0}
\{ e_\alpha \} \hspace{3mm},\hspace{3mm} \{h_a\}_{a=1}^r
\end{equation}
where
\begin{equation}\label{eq1_1}
[h_a,h_b]=0 \hspace{3mm}, \hspace{3mm} [h_a,e_\alpha]=(\alpha)_a e_\alpha .
\end{equation}
The index $a$ runs from $1$ to $r$ and labels the basis of the Cartan subalgebra while the set of vectors $\{ \alpha \}$ in $\mathbb{R}^r$ make up the root system of $\mathcal{G}$. 
We can always choose a basis such that the roots are real and
\begin{equation}\label{eq1_2}
e_\alpha^\dagger = -e_{-\alpha} \hspace{3mm} , \hspace{3mm} h_a^\dagger=h_a .
\end{equation}
Up to a multiplicative factor we can define a scalar product
\begin{equation}\label{eq1_3}
(X,Y)=\mathrm{Tr} (\text{ad}_X \  \text{ad}_Y)
\end{equation}
in such a way that $\{ e_\alpha, h_a\}$ is a basis in $\mathcal{H}$ satisfying
\begin{subequations}
\label{eq1_4}
\begin{align}
\label{eq:1_4a}
(h_a,h_b)&=\delta_{a b} , \\
\label{eq:1_4b}
(e_\alpha, e_{-\beta})&=-\delta_{\alpha \beta} .
 \end{align}
\end{subequations}
Working with these assumptions we have 
\begin{equation}\label{eq1_6}
[e_{\alpha},e_{-\alpha}]= - \alpha \hspace{3mm},\hspace{3mm} [e_{\alpha},e_{\beta}]=N_{\alpha \beta} \ e_{\alpha+\beta}
\end{equation}
where the last equality is different from zero only if $\alpha+\beta$ is a root; $N_{\alpha \beta}$ are the structure constants of the algebra.

For any root $\alpha$ we define the corresponding Weyl reflection to be the linear transformation acting as a reflection respect to the hyperplane orthogonal to  $\alpha$:
\begin{equation} \label{Weyl_reflection}
w_\alpha(x)=x-2 \frac{(x,\alpha)}{\alpha^2} \alpha .
\end{equation}
Such reflections map the root system to itself and the closed group $W$ that they generate is called the Weyl group. For a given choice of simple roots a Coxeter element of the Weyl group is the product of the Weyl reflections over that set of simple roots. Of course there are many possible choices depending on the choice of simple roots, but the Coxeter elements that can be constructed starting from these different choices, and from the orderings of the reflections in any choice, belong to the same conjugacy class of $W$. A convenient choice is the so-called Steinberg ordering~\cite{Steinberg_Paper}: split the simple roots in two sets, `black' and `white', indicated with the labels $\bullet$ and $\circ$. All the roots inside the same set are mutually orthogonal, and so are not  connected on the Dynkin diagram. We then write the Coxeter element $w$ as
\begin{equation} \label{Coxeter_element}
w=w_{\bullet} w_{\circ}=\prod_{\alpha \in \bullet} w_{\alpha} \prod_{\beta \in \circ} w_{\beta} .
\end{equation}
Kostant \cite{B. Kostant1} showed how to define
a particular set of $r$ roots $\{ \gamma_a \}_{a=1}^r$  from which the entire root system can be generated by the action of a given Coxeter element $w$. This action produces $r$ closed orbits each with $h$ elements, where $h$ is the Coxeter number of the algebra.
A convenient way of writing these roots uses the fundamental weights $\{ \lambda_a \}_{a=1}^r$ of the Lie algebra, defined by
\begin{equation}
\frac{2}{\alpha_a^2}(\lambda_a , \alpha_b)=\delta_{ab}.
\end{equation}
The root system can then be generated acting with $w$ on the roots
\begin{equation}
\label{wirting_roots_labeling_the_orbits_in_terms_of_fundamental_weights}
\gamma_a= (1-w^{-1}) \lambda_a=\begin{cases}
&\alpha_a \hspace{20mm} \text{if} \ a \in \circ \\
& - w^{-1} \alpha_a \hspace{10mm} \text{if} \ a \in \bullet 
\end{cases}
\end{equation}
(The first equality holds for any Coxeter element, while the second assumes the Steinberg ordering.)
We define the orbit $\Gamma_a$ to be the set of roots obtained by acting with the Coxeter element on the root $\gamma_a$ (after $h$ steps we come back to the starting root). In the context of Toda theories there is a one to one correspondence between such orbits and particle types. 

The relations in \eqref{Weyl_reflection}, \eqref{Coxeter_element} define the geometrical action of $w$ on the Cartan subalgebra, as a combination of  rotations of different subspaces of  $\mathcal{H}$, and it can be diagonalised respect to an orthonormal basis $\{ z_s \}$ in $\mathcal{H}$
\begin{equation} \label{eq1_9}
w  z_{s} = \omega^s z_s \hspace{3mm}, \hspace{3mm} \omega=e^{\frac{2 \pi i }{h}} .
\end{equation}
The structure of the eigenvalues is constrained by the fact that $w$ has periodicity $h$ while the numbers $s$ are the exponents of $\mathcal{G}$ and depend on the particular Lie algebra we are considering. They assume $r$ integer values in $\{1,\ldots, h-1\}$ where $s=1$ and $s=h-1$ are always exponents of the algebra. By the reality and periodicity of the Coxeter element it follows that we can define the eigenvectors so as to satisfy
\begin{equation} \label{eq1_9bis}
z_s^\dagger=z_{h-s} .
\end{equation}
It is evident from this that the two real combinations of $z_s$ and $z_{h-s}$ span a plane in the Cartan subalgebra on which $w$ acts as a rotation by $2\theta_s=2\pi s/h$. Note that if $h/2$ is a (possibly-repeated) exponent,  $z_{h/2}$ is real.

A natural way to find an orthonormal basis of eigenvectors of $w$ is to follow the argument presented at pp. 158-161 of~\cite{Carter_book} and discussed
for example in~\cite{a24,Corrigan:1994nd}. We define two elements
\begin{equation}
\label{Coxeter_geometry_first_definition_of_abullet_and_acirc_as_functions_of_the_roots}
a_s^\bullet= \sum_{b \in \bullet} q_s^b \alpha_b \hspace{10mm} a_s^\circ= \sum_{b \in \circ} q_s^b \alpha_b
\end{equation}
whose coefficients are the eigenvectors of the transpose of the Cartan matrix . Then using simple properties of Weyl reflections it can be proved, in all the cases in which $s \ne \frac{h}{2}$, that
\begin{equation}
\label{Coxeter_geometry_relations_between_acirc_and_abullet_as_presented_in_Corrigan_notes}
|a_{s}^\bullet|^2=|a_s^\circ|^2     \hspace{5mm} \text{and} \hspace{5mm}  (a_s^\bullet, a_s^\circ)= - |a_s^\bullet| |a_s^\circ| \cos \theta_s.
\end{equation}
We refer to \cite{Corrigan:1994nd} for a detailed derivation. 
For each exponent $s$ we can define  basis elements
\begin{equation}
\label{A_possible_basis_for_zs_and_zhminuss}
\begin{split}
&z_s= \frac{\sqrt{h}}{|a^\circ_s| \sqrt{2 \sin^2 \theta_s} } (e^{i \theta_s} a_s^\bullet +   a_s^\circ)\\
&z_{h-s}= \frac{\sqrt{h}}{|a^\circ_s| \sqrt{2 \sin^2 \theta_s} } (e^{-i \theta_s} a_s^\bullet +   a_s^\circ)
\end{split}
\end{equation}
on the plane spanned by $a_s^\bullet$, $a_s^\circ$; it can be shown that $w$ acts as a rotation of $\frac{2 \pi s}{h}$ on this plane, providing a set of eigenplanes of the Coxeter element.
By using~\eqref{Coxeter_geometry_relations_between_acirc_and_abullet_as_presented_in_Corrigan_notes} it can be checked that 
\begin{equation}\label{eq1_9bisbis}
\bigl(z^\dagger_{s_1}, z_{s_2}\bigr)= h \delta_{{s_1}{s_2}}.
\end{equation}
The only exception is $s=\frac{h}{2}$, in which case $a_{\frac{h}{2}}^\bullet=0$ and from~\eqref{A_possible_basis_for_zs_and_zhminuss},
$\bigl(z_{h/2}, z_{h/2}\bigr)= h/2$.
Using the expressions in~\eqref{A_possible_basis_for_zs_and_zhminuss} we can define the following projectors onto the different eigenspaces of $w$
\begin{equation}
\label{Definition_of_projection_valid_only_inside_the_Cartan}
P_s = \frac{1}{h}  z_s \otimes  z_{h-s}.
\end{equation}
The projections of the roots $\{ \gamma_a \}_{a=1}^r$ on such planes can be written using complex-number notation as
\begin{equation}
\label{Coxeter_geometry_projections_of_phi_in_complex_notation}
\begin{split}
&P_s(\gamma_a)= \sqrt{2}  \frac{\gamma_a^2 \ q_s^a}{|{\bf{Q}}_s|}  \hspace{19mm} \text{if $a \in \circ$}\\
&P_s(\gamma_a)= \sqrt{2}   \frac{\gamma_a^2 \ q^a_s}{|{\bf{Q}}_s|} \ e^{-i \theta_s} \hspace{10mm} \text{if $a \in \bullet$}
\end{split}\end{equation}
where ${\bf{Q}}_s$ is the vector defined by
\begin{equation}
\label{definition_of_big_Q_in_terms_of_q_and_alphasquares}
\textbf{Q}_s=(|\gamma_1| \ q_s^1, |\gamma_2| \ q_s^2,\ldots, |\gamma_r| \ q_s^r) .
\end{equation}
The phase in \eqref{Coxeter_geometry_projections_of_phi_in_complex_notation} is due to the fact that there is a $\theta_s$ angle between the two sets of black and white elements. From the equalities in \eqref{Coxeter_geometry_projections_of_phi_in_complex_notation} we can obtain the projections of all the roots on the different $w$-eigenplanes by acting with $w$
$$
P_s (w^{-p}\gamma_a)=e^{-2 i p \theta_s}P_s(\gamma_a) .
$$
Three roots that sum to zero in the root space $\mathbb{R}^r$ can therefore be projected on different two dimensional eigenplanes of the Coxeter element closing a set of $r$ triangular relations, one for each projector; such triangular relations represent conservation laws of charges with different spin, and are responsible for the simplicity of the S-matrix coming out from integrable theories. 

The action of the Coxeter element can be extended to the whole algebra $\mathcal{G}$  \cite{B. Kostant1} by requiring 
\begin{equation}\label{eq1_12}
w e_{\alpha}=A e_{\alpha} A^{-1}=e_{w \alpha}.
\end{equation}
This ensures that the structure constants of the algebra are connected by
\begin{equation}\label{eq1_12bis}
N_{\alpha \beta}=N_{w\alpha \ w\beta} .
\end{equation}
Using~\eqref{eq1_12} the projectors $P_s$ can also be extended to the whole algebra as~\cite{aa24}
\begin{equation}\label{new_notation_5}
P_s = \frac{1}{h} \sum_{p=0}^{h-1}  \omega^{-s p } w^{p}.
\end{equation}

\subsection{Some properties of structure constants} \label{structure_constants_properties_appendix}

Before proceeding to define the basis used to expand the field in section \ref{sect1} we show some simple properties of the structure constants that have interesting consequences on the couplings of the corresponding Toda theory. For a more detailed discussion we invite the interested reader to look at the chapter $4$ of \cite{Carter_book}.
\begin{mytheorem}\label{App:1_pr2}
If $\alpha$, $\beta$ and $\gamma$ are three roots satisfying $\alpha+\beta+\gamma=0$ then $N_{\alpha \beta}=N_{\beta \gamma}=N_{\gamma \alpha}$
\end{mytheorem}
\begin{proof}
Using the relations in \eqref{eq:1_4b} we obtain
\begin{equation*}
N_{\alpha \beta}= -\bigl([e_\alpha, e_\beta ],e_{\gamma}\bigr)= -\bigl(e_\alpha,[ e_\beta ,e_{\gamma}]\bigr)= N_{\beta \gamma}
\end{equation*}
The rest of the equalities are proved by cyclicity. 
\end{proof}

\begin{mytheorem}\label{App:1_pr3}
$N_{\alpha \beta}^*= N_{-\alpha , -\beta}$ 
\end{mytheorem}
\begin{proof}
From \eqref{eq1_2} we obtain 
\begin{equation*}
N_{\alpha \beta}^* e_{-\alpha-\beta}=-(N_{\alpha, \beta} \ e_{\alpha+\beta})^\dagger=-[e_\alpha, e_\beta ]^\dagger=-[e_{-\beta},e_{-\alpha}]=N_{-\alpha , -\beta} \ e_{-\alpha-\beta}
\end{equation*}
\end{proof}

\begin{mytheorem}\label{App:1_pr5}
Given four roots $\alpha$, $\beta$, $\gamma$, $\delta$ (with $\alpha \ne \gamma$, $\alpha \ne \delta$) such that $\alpha+\beta=\gamma+\delta=\epsilon$ where $\epsilon$ is another root, then
\begin{equation}\label{structure_constants_relation_property}
N_{\alpha , \beta} N_{-\gamma , -\delta} - N_{\alpha, -\gamma} N_{\beta,-\delta} + N_{\alpha, -\delta} N_{\beta, -\gamma}=0
\end{equation}
and at least one of the following conditions (or both) are satisfied
\begin{itemize}
\item $\gamma-\alpha=\beta-\delta=\rho$, with  $\rho$ root
\item $\delta-\alpha=\beta-\gamma=\tilde{\rho}$, with $\tilde{\rho}$ root
\end{itemize}
\end{mytheorem}

\begin{proof}
Since $\alpha+\beta$ and $\gamma+\delta$ correspond to the same root $\epsilon$ the two structure constants corresponding to $ \{ \alpha, \beta \}$ and $\{ \gamma,\delta \}$ are not zero and we can write their product in the following way
\begin{equation}
N_{\alpha , \beta} N_{-\gamma , -\delta} = -\bigl([e_\alpha , e_\beta] \ , [e_{-\gamma} , e_{-\delta}]\bigr)=-\bigl(e_\alpha \ , [e_\beta, [e_{-\gamma} , e_{-\delta}]\ ]\bigr)
\end{equation}
At this point we use the Jacobi identity and obtain
\begin{equation*}\begin{split}
N_{\alpha , \beta} N_{-\gamma , -\delta} &= \bigl(e_\alpha \ , [e_{-\gamma}, [e_{-\delta} , e_{\beta}]\ ]\bigr)+\bigl(e_\alpha \ , [e_{-\delta}, [e_{\beta} , e_{-\gamma}]\ ]\bigr)\\
&= \bigl([e_\alpha , e_{-\gamma}],  [e_{-\delta} , e_{\beta}]\ ]\bigr)+\bigl([e_\alpha , e_{-\delta}],  [e_{\beta} , e_{-\gamma}]\ ]\bigr)\\
&=N_{\alpha, -\gamma} N_{-\delta, \beta} (e_{-\rho}, e_\rho)+N_{\alpha, -\delta} N_{\beta, -\gamma} (e_{-\tilde{\rho}}, e_{\tilde{\rho}})
\end{split}\end{equation*}
From this fact using the convention in \eqref{eq:1_4b} and the antisymmetry of the structure constants we obtain~\eqref{structure_constants_relation_property}.

It is clear that if $\gamma-\alpha$ ($=\beta-\delta$) or $\delta-\alpha$ ($=\beta-\gamma$) is not a root we have that $N_{\alpha, -\gamma}$ or $N_{\alpha, -\delta}$ respectively is equal to zero. However since the equality in \eqref{structure_constants_relation_property} always holds and we know that the first term in the sum is nonzero then also the second or the third term in the sum (or both) need to be different from zero, that means that at least one of $\gamma-\alpha$ and $\delta-\alpha$ is a root.
\end{proof}
This fact is of fundamental importance in the cancellation of $2$ to $2$ non-diagonal processes in Toda models. 
We highlight that in simply-laced theories $\alpha-\gamma$ and $\alpha-\delta$ cannot both be roots at the same time. 
Indeed in this case, setting the root length equal to $\sqrt{2}$, and assuming that $\alpha-\gamma$ is a root, the equalities $(\alpha+\beta)^2=2 $ and $(\alpha-\gamma)^2=2$ imply
 $(\alpha,\beta)=-1$ and  
 $(\alpha,\gamma)=1$.
Hence
% \begin{equation}\label{simply_laced_contr2}
$
(\alpha, \delta)=(\alpha,\alpha+\beta-\gamma)=2-1-1=0$
% \end{equation}
and so $(\alpha-\delta)^2=4$, and $\alpha-\delta$ is not a root.

\begin{mytheorem}\label{App:1_pr4}
The ratios of the structure constants of any semisimple Lie algebra take specific values. In particular if the long roots of $\mathcal{G}$ are normalised to length $\sqrt{2}$ we have the following possibilities: if there are three roots $\alpha$, $\beta$ and $\gamma$ satisfying $\alpha+\beta+\gamma=0$ we can have
\begin{itemize}
\item $|N_{\alpha, \beta}|=1$ if $|\alpha|=|\beta|=|\gamma|=\sqrt{2}$ or there is one root of length $\sqrt{2}$ and the other two are shorter and with the same length.
\item $|N_{\alpha, \beta}|=\frac{1}{\sqrt{2}}$ if $|\alpha|=|\beta|=|\gamma|=1$.
\item $|N_{\alpha, \beta}|=\frac{2}{\sqrt{3}}$ if $|\alpha|=|\beta|=|\gamma|=\sqrt{\frac{2}{3}}$.
\end{itemize}
while in all the other cases $|N_{\alpha, \beta}|=0$.
\end{mytheorem}

\begin{proof}
We can write the absolute value of the structure constant corresponding to two roots $\alpha$, $\beta$ in the following form
\begin{equation*}\begin{split}
|N_{\alpha \beta}|^2 &= \bigl(N_{\alpha \beta} \ e_{\alpha+\beta} \ , \ N_{\alpha \beta}^* \  e_{\alpha+\beta}^\dagger \bigr) = \bigl([e_\alpha, e_\beta ] \ , \ [e_{-\beta}, e_{-\alpha} ] \bigr) = \bigl([[e_\alpha, e_\beta ] \ , e_{-\beta}] \ ,\ e_{-\alpha} \bigr)\\
 &= -([[ e_{-\beta}, e_\alpha ] \ , e_{\beta}] \ ,\ e_{-\alpha} )-([[e_\beta,e_{-\beta} ] \ , e_\alpha] \ ,\ e_{-\alpha} )=N_{-\beta,\alpha} N_{\alpha-\beta, \beta} - (\beta \ , \ \alpha)
\end{split}\end{equation*}
Combining properties \ref{App:1_pr2} and  \ref{App:1_pr3} we have
\begin{equation*}
N_{\alpha-\beta,\beta}=N_{\beta,-\alpha}= N^*_{-\beta,\alpha}
\end{equation*}
and therefore we obtain
\begin{equation}\label{norm_coupling_property4}
|N_{\alpha, \beta}|^2=|N_{-\beta,\alpha}|^2  - (\beta \ , \ \alpha) .
\end{equation}
All the absolute values of the structure constants can be obtained from this equality. We focus on the case in which $|\alpha|=|\beta|<|\gamma|=\sqrt{2}$; the other situations can be studied similarly. By the fact that $\alpha$, $\beta$ and $\gamma$ close a triangle and $\alpha^2=\beta^2$ the following relation holds
\begin{equation}
\alpha^2=\gamma^2+\beta^2+2 (\gamma, \beta) \rightarrow 2 (\gamma, \beta)=-\gamma^2 .
\end{equation}
We observe that the vector $\gamma-\beta$ cannot be a root since
\begin{equation}
(\gamma-\beta)^2=2\gamma^2+\beta^2 > 4
\end{equation}
which is bigger of the maximal allowed length. Therefore from the expression in \eqref{norm_coupling_property4} we obtain
\begin{equation}
|N_{\gamma,\beta}|=-(\gamma,\beta)=\frac{\gamma^2}{2}=1
\end{equation}
All the other situations can be analogously studied. 
\end{proof}

\subsection{A new Cartan subalgebra} \label{last_subappendix}

\begin{figure}
\begin{center}
\begin{tikzpicture}
\draw[directed] (3.5,-1) -- (4.5,1);
\draw[directed] (4.5,1) -- (7.5,-1);
\draw[directed] (3.5,-1) -- (4.5,-3);
\draw[directed] (4.5,-3) -- (7.5,-1);
\draw[directed] (7.5,-1) -- (3.5,-1);
\filldraw[black] (3.3,0.2)  node[anchor=west] {$\gamma$};
\filldraw[black] (3.3,-1.7)  node[anchor=west] {$\gamma'$};
\filldraw[black] (4.4,-0.8)  node[anchor=west] {$\alpha$};
\filldraw[black] (5.7,0.3)  node[anchor=west] {$\beta$};
\filldraw[black] (5.7,-2.3)  node[anchor=west] {$\beta'$};
\filldraw[black] (3.6,-0.8)  node[anchor=west] {$n$};
\filldraw[black] (3.6,-1.2)  node[anchor=west] {$n$};
\filldraw[black] (6.6,-0.8)  node[anchor=west] {$m$};
\filldraw[black] (6.6,-1.2)  node[anchor=west] {$m$};
\end{tikzpicture}
\caption{Two reflected triangles of roots $\alpha+\gamma+\beta=\alpha+w^{-n} \gamma+w^{m}\beta=0$.}
\label{fig:1_pr6_sym_roots}
\end{center}
\end{figure}
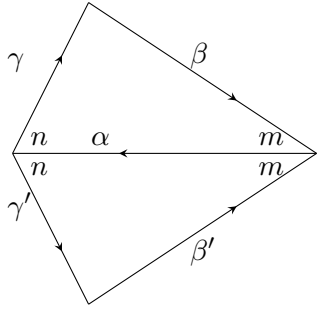

At this point we can continue with our discussion by defining a basis to expand the field appearing in the Lagrangian of an affine Toda model. Recalling a result from~\cite{B. Kostant1},
there is way to define a new Cartan subalgebra $\mathcal{H'}$, that is composed by a set of fixed elements under the action of $w$. Then $\mathcal{H}$ is said to be in apposition to $\mathcal{H'}$ respect to the Coxeter element $w$.
We define a basis for $\mathcal{H'}$ starting from our original set of roots as \cite{B. Kostant1} 
\begin{equation}\label{eq1_16}
y_a= i \sum_{p=0}^{h-1} e_{w^p \gamma_a}
\end{equation}
where $a=1,\dots,r$ and $\gamma_a$ are the roots labelling the orbits. From properties in \eqref{eq1_2} and \eqref{eq:1_4b} it is not difficult to prove that
\begin{equation}\label{eq1_17}
(y_a,y^{\dagger}_b)=h \delta_{ab}
\end{equation}
On the other hand it is also possible to prove that this set of elements commute; this is equivalent to asserting that any time there are two roots  $\gamma$ and $\beta$ belonging  to the orbits $\Gamma_c$, $\Gamma_b$ respectively such that their sum is a root $-\alpha$, then there are exactly two other  roots $\gamma'$ and $\beta'$ in the same orbits given by reflecting $\gamma$ and $\beta$ respect to $\alpha$ verifying $N_{\gamma', \beta'}=-N_{\gamma,\beta}$. A picture of the two reflected triangles whose angles are written in units of $\frac{\pi}{h}$ is shown in figure \ref{fig:1_pr6_sym_roots}. 

Diagonalising this new Cartan subalgebra $\mathcal{H'}$ we obtain a new set of generators $E_{\alpha'}$ that with an appropriate normalizations satisfy
\begin{subequations}
\label{eq1_20}
\begin{align}
\label{eq1_20_1}
(E_{\alpha'},E_{-\beta'})&=\delta_{\alpha',\beta'} ,\\
\label{eq1_20_1-2}
[E_{\alpha'},E_{-\alpha'}]&=\alpha' ,\\
\label{eq1_20_3}
[\frac{1}{\sqrt{h}}y_a,E_{\alpha'}]&=(\alpha',\frac{1}{\sqrt{h}} y_a)E_{\alpha'}.
\end{align}
\end{subequations}
Now we can choose a particular set of complex numbers $\{c_i\}_{i=0}^r$ such that the elements defined by
\begin{equation}\label{eq1_22}
z_1=\sum_{i=0}^{r} c_i E_{\alpha'_i} \hspace{3mm},\hspace{3mm} z_{h-1}=\sum_{i=0}^{r} c^*_i E_{-\alpha'_i}
\end{equation}
are exactly the Coxeter element eigenvectors with eigenvalues $\omega$ and $\omega^{h-1}$ of the old Cartan subalgebra $\mathcal{H}$. This is verified as long as we choose the constants  $\{c_i\}_{i=0}^r$ such to satisfy $|c_i|^2=n_i$, with $\sum_{i=0}^r n_i \alpha_i'=0$; here $\{ {\alpha_i}' \}_{i=0}^r$ are the new affine simple roots of the algebra \cite{Freeman:1991xw,B. Kostant1}.

\subsection{Folding and the twisted Coxeter element} \label{sub_appendix_on_twisted_Cox}

Until this point we have been working on a semisimple Lie algebra  constructed over an untwisted Dynkin diagram. In that case we have a copy of equivalent sets of simple roots $\{ \alpha_i \}_{i=1}^r$ and $\{ \alpha_i' \}_{i=1}^r$ associated to two different choices of the Cartan subalgebra, that we have called respectively $\mathcal{H}$ and $\mathcal{H}'$. 
Root systems associated to twisted Dynkin diagrams can be included in the previous analysis with little more effort as explained in \cite{Dorey:1992gr}. The Dynkin diagrams of these models can be obtained by `folding' one and only one of the extended simply-laced diagrams \cite{Olive:1982ye}. 

The idea is the following. Suppose there exists an automorphism $\sigma$ of a certain extended Dynkin diagram composed by a set of roots $\{ \alpha_i' \}_{i=0}^r$ acting as a permutation  of the points $\alpha' \to \sigma(\alpha')$. Here $\sigma$ is a symmetry of the diagram behaving as a linear map over the vector space $\mathcal{H}'$.
Therefore we can decompose each vector $a$ in  $\mathcal{H}'$ into a component living in the invariant space under the action of $\sigma$ ($a^{\parallel}$) and into a component perpendicular to such space ($a^{\perp}$)
$$
a=a^{\parallel}+a^{\perp}.
$$
To obtain the root system associated to the folded diagram starting from the simply-laced one we set all the components $\alpha'^{\perp}_i$ perpendicular to the subspace invariant under the action of $\sigma$ equal to zero. The new set of roots that is generated in this way lies entirely in the $\sigma$-invariant subspace, in other words all the roots in $\{ \alpha'^{\parallel}_i \}_{i=0}^r$ are eigenvectors of $\sigma$ with eigenvalue equal to one. Among them there are many roots that are equal each other after the reduction. In particular all the roots that are connected along a $\sigma$-orbit have the same projection on the $\sigma$-invariant eigenspace. This implies that the new root system has a smaller number of elements than one we started with, and now the roots can have in general different lengths since we have removed  their projection on the space not invariant under the action of $\sigma$. 

An equivalent way of seeing this is to look at the vectors $\{ \frac{1}{\sqrt{h}} y_a \}_{a=1}^r$ defined in \eqref{eq1_16} that form an orthonormal basis of $\mathcal{H}'$. Each root in $\mathcal{H}'$ can be written as a linear combination in terms of this basis
$$
\alpha'=\sum_{a=1}^r \alpha'_a  \frac{1}{\sqrt{h}} y_a.
$$
A case by case check \cite{a23} shows that for twisted Dynkin diagrams the space invariant under the action of $\sigma$ is generated by a subset of the vectors in \eqref{eq1_16}. From the point of view of affine Toda field theories, in which $\phi$ is an element of $\mathcal{H}'$, this corresponds to setting to zero all the field components along the directions $y_a$ not invariant under the automorphism $\sigma$. The root orbits $\tilde{\Gamma}_a \subset \mathcal{H}$ necessary to define the elements $y_a$ which survive the automorphism define a new root system $\tilde{\Phi}$ which is contained inside the root system $\Phi$ associated to the Cartan subalgebra $\mathcal{H}$. The generators associated to the roots in $\tilde{\Phi}$ define a subalgebra in the Lie algebra $\mathcal{G}$ and the action of $w$ in this embedded $\tilde{\Phi}$ is exactly that of a twisted Coxeter element of $\tilde{\Phi}$ \cite{Dorey:1992gr}. We now illustrate these considerations with a simple example.

\subsubsection{An example: the $A_2^{(2)}$ model from $D_4^{(1)}$ }
\label{sub_appendix_on_twisted_Cox_example}

\begin{figure}
\begin{center}
\begin{tikzpicture}

%D_4^{(1)} diagram
\filldraw[black] (-0.1,2.2)  node[anchor=west] {\bf{$D_4^{(1)}$}};

\draw (0,0) circle (0.2);
\filldraw[black] (0.1,-0.15)  node[anchor=west] {\tiny{$\alpha_4'$}};

\draw[] (0.2,0) -- (1.3,0);
\draw (1.5,0) circle (0.2);
\filldraw[black] (1.6,0)  node[anchor=west] {\tiny{$\alpha_3'$}};

\draw[] (-0.2,0) -- (-1.3,0);
\draw (-1.5,0) circle (0.2);
\filldraw[black] (-2.2,0)  node[anchor=west] {\tiny{$\alpha_1'$}};

\draw[] (0,0.2) -- (0,1.3);
\draw (0,1.5) circle (0.2);
\filldraw[black] (0.05,1.6)  node[anchor=west] {\tiny{$\alpha_0'$}};

\draw[] (0,-0.2) -- (0,-1.3);
\draw (0,-1.5) circle (0.2);
\filldraw[black] (0.05,-1.6)  node[anchor=west] {\tiny{$\alpha_2'$}};

\draw[->] (1.2,0.5)--(0.4,1.3) ;
\draw[->] (-1.2,-0.5)--(-0.4,-1.3) ;
\draw[->] (0.4,-1.3)--(1.2,-0.5) ;
\draw[->] (-0.4,1.3)--(-1.2,0.5) ;

\draw[->] (3.3,0)--(4.5,0) ;

%A_2^{(2)} diagram

\filldraw[black] (6.1,1.2)  node[anchor=west] {\bf{$A_2^{(2)}$}};
\draw (6,0) circle (0.2);
\draw[] (6.2,0.15) -- (7.3,0.15);
\draw[] (6.2,0.05) -- (7.3,0.05);
\filldraw[black] (6.5,0)  node[anchor=west] {\bf{$\Bigl<$}};
\draw[] (6.2,-0.05) -- (7.3,-0.05);
\draw[] (6.2,-0.15) -- (7.3,-0.15);
\draw[] (7.5,0) circle (0.2);

\filldraw[black] (5.2,0)  node[anchor=west] {\tiny{$\alpha_1'^\parallel$}};
\filldraw[black] (7.6,0)  node[anchor=west] {\tiny{$\alpha_4'^\parallel$}};

\end{tikzpicture}
\caption{Example of $A_2^{(2)}$ Dynkin diagram obtained by folding $D_4^{(1)}$.}
\label{Example_of_folding_in_Twisted_theories}
\end{center}
\end{figure}
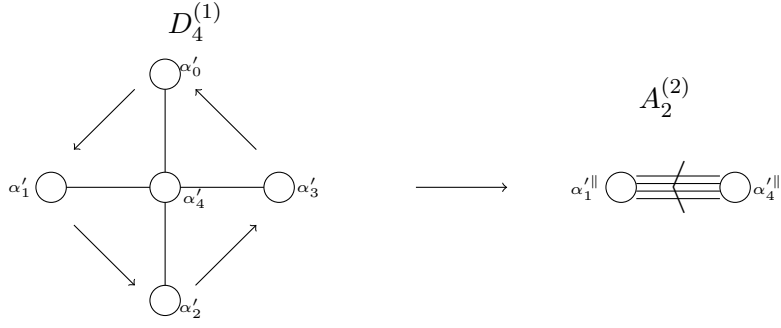

Let $\alpha'_1, \alpha'_2 , \alpha'_3, \alpha'_4$ be the simple roots making up the $D_4^{(1)}$ Dynkin diagram shown in figure~\ref{Example_of_folding_in_Twisted_theories}:
\begin{equation*}
\begin{split}
&\alpha'_1=(\frac{1}{\sqrt{3}}, \frac{1}{\sqrt{6}} , -1 , -\frac{1}{\sqrt{2}}) \hspace{3mm},\hspace{3mm} \alpha'_2=(\frac{1}{\sqrt{3}}, \frac{1}{\sqrt{6}} , 1 , -\frac{1}{\sqrt{2}}),\\
&\alpha'_3=(-\frac{2}{\sqrt{3}}, \frac{1}{\sqrt{6}} , 0 , -\frac{1}{\sqrt{2}}) \hspace{3mm},\hspace{3mm} \alpha'_4=(0, 0, 0 , \sqrt{2}),
\end{split}
\end{equation*}
and let $\alpha'_0$ be the lowest root defined by imposing
$\sum_{i=0}^4 n_i \alpha'_i=0$ with $(n_0, \ldots , n_4)= (1,1,1,1,2)$. 
The affine Toda theory constructed starting from this Dynkin diagram, up to an overall multiplicative factor, has a squared mass matrix of the form
$$
M^2=\sum_{i=0}^4 n_i \alpha'_i \otimes \alpha'_i=diag(2, 2, 2, 6),
$$
once the simple roots have been written in a  basis for which $M^2$ is diagonal. 

The diagram automorphism $\sigma$ defines a linear function by $\sigma \alpha'_4=\alpha'_4$, $\sigma \alpha'_1=\alpha'_2$ and so on for the other roots, as shown by the arrows connecting the different spots on the left hand side of figure \ref{Example_of_folding_in_Twisted_theories}.
The invariant space under the action of $\sigma$ is spanned by the component $(0,0,0,1)$ along which the root $\alpha'_4$ lies. The reduction procedure necessary to define the new root system associated to $A_2^{(2)}$ starting from the root system of $D_4^{(1)}$ corresponds to projecting the roots onto the direction invariant under sigma, which is $(0,0,0,1)$.
If we do this we obtain a new set of roots $\{ \alpha'^\parallel_i \}_{i=0}^4$, where $\alpha'^\parallel_0$, $\alpha'^\parallel_1$, $\alpha'^\parallel_2$ and $\alpha'^\parallel_3$ are all equal to $(0,0,0,-\frac{1}{\sqrt{2}})$ and $\alpha'^\parallel_4=\alpha_4$. The new Dynkin diagram obtained after the folding is therefore composed by just two roots, that we simply call $\alpha'^\parallel_1$ and $\alpha'^\parallel_4$ on the right hand side of figure \ref{Example_of_folding_in_Twisted_theories}, having lengths connected by $(\alpha'^\parallel_4)^2=4(\alpha'^\parallel_1)^2$.

On the other hand the invariant eigenvector $(0,0,0,1)$ corresponds to 
the element $\frac{1}{\sqrt{h}}y_4$ in \eqref{eq1_16} defined by summing all the generators with roots living in the orbit $\Gamma_4$ (the one associated to the central spot of the $D_4^{(1)}$ Dynkin diagram) of the second Cartan subalgebra $\mathcal{H}$. This orbit contains six elements, which form a set of roots for $A_2^{(1)}$. This implies that $A_2^{(1)}$ is a closed subspace of $D_4^{(1)}$ living entirely on the orbit $\Gamma_4$. The Coxeter element $w$ of $D_4^{(1)}$ corresponds then to a twisted Coxeter element in this  subspace $A^{(1)}_2$ \cite{Dorey:1992gr}. 

The same analysis holds for any twisted affine Dynkin diagram. In each case we start from an extended simply-laced diagram and apply the folding procedure. The space invariant under $\sigma$ is spanned by a subset of the vectors in \eqref{eq1_16} and such vectors contain in the sum the generators of a subalgebra of the original simply-laced root system. The Coxeter element $w$ of the initial simply-laced diagram behaves as a twisted Coxeter element defined on this subspace. In the example just presented we started from two equivalent root systems $\Phi$ and $\Phi'$ both associated to the simply-laced diagram $D_4^{(1)}$; after the folding $\Phi'$ reduced to the root systems of $A^{(2)}_2$ and $\Phi$ reduced to the root system of $A_2^{(1)}$.


\vskip 40pt



\begin{thebibliography}{99}

%Factorized S-matrix b2

%\cite{Zamolodchikov:1978xm}
\bibitem{a3}
  A.~B.~Zamolodchikov and A.~B.~Zamolodchikov,
  ``Factorized s Matrices in Two-Dimensions as the Exact Solutions of Certain Relativistic Quantum Field Models,''
  Annals Phys.\  {\bf 120} (1979) 253.
  \href{https://www.sciencedirect.com/science/article/pii/0003491679903919}{doi:10.1016/0003-4916(79)90391-9}.
  %%CITATION = doi:10.1016/0003-4916(79)90391-9;%%
  %1351 citations counted in INSPIRE as of 21 Feb 2019

%\cite{Parke:1980ki}
\bibitem{a4}
  S.~J.~Parke,
  ``Absence of Particle Production and Factorization of the $S$ Matrix in (1+1)-dimensional Models,''
  Nucl.\ Phys.\ B {\bf 174} (1980) 166.
  \href{https://www.sciencedirect.com/science/article/pii/0550321380901960}{doi:10.1016/0550-3213(80)90196-0}.
  %%CITATION = doi:10.1016/0550-3213(80)90196-0;%%
  %49 citations counted in INSPIRE as of 21 Feb 2019
  
  %\cite{Dorey:1996gd}
\bibitem{Dorey:1996gd}
  P.~Dorey,
  ``Exact S matrices,''
  \href{https://arxiv.org/abs/hep-th/9810026}{hep-th/9810026}.
  %%CITATION = HEP-TH/9810026;%%
  %114 citations counted in INSPIRE as of 01 May 2019


%\cite{Hoare:2018jim}
\bibitem{Hoare:2018jim}
B.~Hoare, N.~Levine and A.~A.~Tseytlin,
``On the massless tree-level S-matrix in 2d sigma models,''
J. Phys. A \textbf{52} (2019) no.14, 144005
\href{https://iopscience.iop.org/article/10.1088/1751-8121/ab0b79}{doi:10.1088/1751-8121/ab0b79}
\href{https://arxiv.org/pdf/1812.02549.pdf}{[arXiv:1812.02549 [hep-th]]}.
%15 citations counted in INSPIRE as of 18 Apr 2021

%\cite{Nappi:1979ig}
\bibitem{Nappi:1979ig}
C.~R.~Nappi,
``Some Properties of an Analog of the Nonlinear $\sigma$ Model,''
Phys. Rev. D \textbf{21} (1980), 418
\href{https://journals.aps.org/prd/abstract/10.1103/PhysRevD.21.418}{doi:10.1103/PhysRevD.21.418}
%55 citations counted in INSPIRE as of 07 May 2021

%\cite{Goebel:1986na}
\bibitem{Goebel:1986na}
C.~J.~Goebel,
%``On the {Sine-Gordon} S Matrix,''
Prog. Theor. Phys. Suppl. \textbf{86} (1986), 261-273
\href{https://academic.oup.com/ptps/article/doi/10.1143/PTPS.86.261/1885208}{doi:10.1143/PTPS.86.261}
%24 citations counted in INSPIRE as of 16 Apr 2021

%\cite{Arefeva:1974bk}
\bibitem{Arefeva:1974bk}
I.~Arefeva and V.~Korepin,
``Scattering in two-dimensional model with Lagrangian (1/gamma) ((d(mu)u)**2/2 + m**2 cos(u-1)),''
Pisma Zh. Eksp. Teor. Fiz. \textbf{20} (1974), 680
%37 citations counted in INSPIRE as of 18 Apr 2021


%\cite{Kalousios:2009ey}
\bibitem{Kalousios:2009ey}
C.~Kalousios, C.~Vergu and A.~Volovich,
``Factorized Tree-level Scattering in AdS(4) x CP**3,''
JHEP \textbf{09} (2009), 049
\href{https://iopscience.iop.org/article/10.1088/1126-6708/2009/09/049}{doi:10.1088/1126-6708/2009/09/049}
\href{https://arxiv.org/abs/0905.4702}{[arXiv:0905.4702 [hep-th]]}.
%20 citations counted in INSPIRE as of 19 Apr 2021

%\cite{Wulff:2017hzy}
\bibitem{Wulff:2017hzy}
L.~Wulff,
``Integrability of the superstring in AdS$_3 \times$ S$^2 \times$ S$^2 \times$ T$^3$,''
J. Phys. A \textbf{50} (2017) no.23, 23LT01
\href{https://doi.org/10.1088/1751-8121/aa70b5}{doi:10.1088/1751-8121/aa70b5}
\href{https://arxiv.org/abs/1702.08788}{[arXiv:1702.08788 [hep-th]]}.
%13 citations counted in INSPIRE as of 19 Apr 2021

%\cite{Wulff:2017vhv}
\bibitem{Wulff:2017vhv}
L.~Wulff,
``Classifying integrable symmetric space strings via factorized scattering,''
JHEP \textbf{02} (2018), 106
\href{https://link.springer.com/article/10.1007/JHEP02(2018)106}{doi:10.1007/JHEP02(2018)106}
\href{https://arxiv.org/abs/1711.00296v2}{[arXiv:1711.00296 [hep-th]]}.
%10 citations counted in INSPIRE as of 19 Apr 2021


%\cite{Khastgir:2003au}
\bibitem{Khastgir:2003au}
  S.~P.~Khastgir,
  ``Affine Toda field theory from tree unitarity,''
  Eur.\ Phys.\ J.\ C {\bf 33} (2004) 137
  \href{https://link.springer.com/article/10.1140\%2Fepjc\%2Fs2003-01523-7}{doi:10.1140/epjc/s2003-01523-7}
  \href{https://arxiv.org/abs/hep-th/0308032}{[hep-th/0308032]}.
  %%CITATION = doi:10.1140/epjc/s2003-01523-7;%%
  %1 citations counted in INSPIRE as of 15 May 2019


%\cite{Gabai:2018tmm}
\bibitem{Gabai:2018tmm}
  B.~Gabai, D.~Mazáč, A.~Shieber, P.~Vieira and Y.~Zhou,
  ``No Particle Production in Two Dimensions: Recursion Relations and Multi-Regge Limit,''
  JHEP {\bf 1902} (2019) 094
  \href{https://link.springer.com/article/10.1007\%2FJHEP02\%282019\%29094}{doi:10.1007/JHEP02(2019)094}
   \href{https://arxiv.org/abs/1803.03578v1}{[arXiv:1803.03578 [hep-th]]}.
  %%CITATION = doi:10.1007/JHEP02(2019)094;%%
  %6 citations counted in INSPIRE as of 01 May 2019


%\cite{Bercini:2018ysh}
\bibitem{Bercini:2018ysh}
  C.~Bercini and D.~Trancanelli,
  ``Supersymmetric integrable theories without particle production,''
  Phys.\ Rev.\ D {\bf 97} (2018) no.10,  105013
  \href{https://journals.aps.org/prd/abstract/10.1103/PhysRevD.97.105013}{doi:10.1103/PhysRevD.97.105013}
  \href{https://arxiv.org/abs/1803.03612v2}{[arXiv:1803.03612 [hep-th]]}.
  %%CITATION = doi:10.1103/PhysRevD.97.105013;%%
  %2 citations counted in INSPIRE as of 01 May 2019

%\cite{Braden:1990wx}
\bibitem{Braden:1990wx}
  H.~W.~Braden, E.~Corrigan, P.~E.~Dorey and R.~Sasaki,
  ``Multiple poles and other features of affine Toda field theory,''
  Nucl.\ Phys.\ B {\bf 356} (1991) 469.
  \href{https://www.sciencedirect.com/science/article/pii/055032139190317Q?via\%3Dihub}{doi:10.1016/0550-3213(91)90317-Q}
  %%CITATION = doi:10.1016/0550-3213(91)90317-Q;%%
  %82 citations counted in INSPIRE as of 01 May 2019



%\cite{Dorey:1990xa}
\bibitem{a24}
  P.~Dorey,
  ``Root systems and purely elastic S matrices,''
  Nucl.\ Phys.\ B {\bf 358} (1991) 654.
  \href{https://www.sciencedirect.com/science/article/pii/055032139190428Z?via\%3Dihub}{doi:10.1016/0550-3213(91)90428-Z}.
  %%CITATION = doi:10.1016/0550-3213(91)90428-Z;%%
  %107 citations counted in INSPIRE as of 21 Feb 2019


%\cite{Freeman:1991xw}
\bibitem{Freeman:1991xw}
  M.~D.~Freeman,
  ``On the mass spectrum of affine Toda field theory,''
  Phys.\ Lett.\ B {\bf 261} (1991) 57.
  \href{https://www.sciencedirect.com/science/article/pii/037026939191324O}{doi:10.1016/0370-2693(91)91324-O}
  %%CITATION = doi:10.1016/0370-2693(91)91324-O;%%
  %55 citations counted in INSPIRE as of 01 May 2019

%\cite{Fring:1991me}
\bibitem{Fring:1991me}
  A.~Fring, H.~C.~Liao and D.~I.~Olive,
  ``The Mass spectrum and coupling in affine Toda theories,''
  Phys.\ Lett.\ B {\bf 266} (1991) 82
  \href{https://www.sciencedirect.com/science/article/pii/037026939190747E}{doi:10.1016/0370-2693(91)90747-E}
  %%CITATION = doi:10.1016/0370-2693(91)90747-E;%%
  %67 citations counted in INSPIRE as of 01 May 2019


%\cite{Fring:1992tt}
\bibitem{Fring:1992tt}
  A.~Fring,
  ``Couplings in affine Toda field theories,''
    \href{https://arxiv.org/abs/hep-th/9212107v1}{hep-th/9212107}.
  %%CITATION = HEP-TH/9212107;%%

%\cite{Braden:1991vz}
\bibitem{Braden:1991vz}
  H.~W.~Braden and R.~Sasaki,
  ``Affine Toda perturbation theory,''
  Nucl.\ Phys.\ B {\bf 379} (1992) 377.
  \href{https://www.sciencedirect.com/science/article/pii/0550321392906017?via\%3Dihub}{doi:10.1016/0550-3213(92)90601-7}
  %%CITATION = doi:10.1016/0550-3213(92)90601-7;%%
  %18 citations counted in INSPIRE as of 01 May 2019

%\cite{Complex_Functions}
\bibitem{Complex_Functions}
Gareth. A. Jones, David Singerman, ``Complex Functions``  (Cambridge  University  Press,  1987), Chapter 4.

%\cite{Mittag_Leffler_Complex}
\bibitem{Mittag_Leffler_Complex}
G. Mittag-Leffler,  An Introduction to the Theory of Elliptic Functions. Annals of Mathematics, 24(4), second series (1923), 271. \href{https://www.jstor.org/stable/1967677?seq=1#metadata_info_tab_contents}{doi:10.2307/1967677}



%\cite{Mikhailov:1980my}
\bibitem{a1}
  A.~V.~Mikhailov, M.~A.~Olshanetsky and A.~M.~Perelomov,
  ``Two-Dimensional Generalized Toda Lattice,''
  Commun.\ Math.\ Phys.\  {\bf 79} (1981) 473.
  \href{https://link.springer.com/article/10.1007\%2FBF01209308}{doi:10.1007/BF01209308}.
  %%CITATION = doi:10.1007/BF01209308;%%
  %208 citations counted in INSPIRE as of 21 Feb 2019


%\cite{Olive:1984mb}
\bibitem{a2}
  D.~I.~Olive and N.~Turok,
  ``Local Conserved Densities and Zero Curvature Conditions for Toda Lattice Field Theories,''
  Nucl.\ Phys.\ B {\bf 257} (1985) 277.
  \href{https://www.sciencedirect.com/science/article/pii/0550321385903475?via\%3Dihub}{doi:10.1016/0550-3213(85)90347-5}.
  %%CITATION = doi:10.1016/0550-3213(85)90347-5;%%
  %160 citations counted in INSPIRE as of 21 Feb 2019




%Exact S matrix found for simply-laced b7

%\cite{Arinshtein:1979pb}
\bibitem{a16}
  A.~E.~Arinshtein, V.~A.~Fateev and A.~B.~Zamolodchikov,
  ``Quantum s Matrix of the (1+1)-Dimensional Todd Chain,''
  Phys.\ Lett.\  {\bf 87B} (1979) 389.
  \href{https://www.sciencedirect.com/science/article/abs/pii/0370269379905616?via\%3Dihub}{doi:10.1016/0370-2693(79)90561-6}.
  %%CITATION = doi:10.1016/0370-2693(79)90561-6;%%
  %205 citations counted in INSPIRE as of 21 Feb 2019
  
  %\cite{Freund:1989jq}
\bibitem{a18}
  P.~G.~O.~Freund, T.~R.~Klassen and E.~Melzer,
  ``S Matrices for Perturbations of Certain Conformal Field Theories,''
  Phys.\ Lett.\ B {\bf 229} (1989) 243.
  \href{https://www.sciencedirect.com/science/article/abs/pii/0370269389911659?via\%3Dihub}{doi:10.1016/0370-2693(89)91165-9}.
  %%CITATION = doi:10.1016/0370-2693(89)91165-9;%%
  %107 citations counted in INSPIRE as of 21 Feb 2019


%\cite{Destri:1989pg}
\bibitem{a19}
  C.~Destri and H.~J.~de Vega,
  ``The Exact S Matrix of the Affine $E(8)$ Toda Field Theory,''
  Phys.\ Lett.\ B {\bf 233} (1989) 336.
  \href{https://www.sciencedirect.com/science/article/abs/pii/0370269389913191?via\%3Dihub}{doi:10.1016/0370-2693(89)91319-1}.
  %%CITATION = doi:10.1016/0370-2693(89)91319-1;%%
  %58 citations counted in INSPIRE as of 21 Feb 2019



%\cite{Christe:1989ah}
\bibitem{a20}
  P.~Christe and G.~Mussardo,
  ``Integrable Systems Away from Criticality: The Toda Field Theory and S Matrix of the Tricritical Ising Model,''
  Nucl.\ Phys.\ B {\bf 330} (1990) 465.
  \href{https://www.sciencedirect.com/science/article/pii/055032139090119X?via\%3Dihub}{doi:10.1016/0550-3213(90)90119-X}.
  %%CITATION = doi:10.1016/0550-3213(90)90119-X;%%
  %163 citations counted in INSPIRE as of 21 Feb 2019




%\cite{Christe:1989my}
\bibitem{aa20}
  P.~Christe and G.~Mussardo,
  ``Elastic S Matrices in (1+1)-Dimensions and Toda Field Theories,''
  Int.\ J.\ Mod.\ Phys.\ A {\bf 5} (1990) 4581.
  \href{https://www.worldscientific.com/doi/abs/10.1142/S0217751X90001938}{doi:10.1142/S0217751X90001938}.
  %%CITATION = doi:10.1142/S0217751X90001938;%%
  %112 citations counted in INSPIRE as of 21 Feb 2019

%\cite{Klassen:1989ui}
\bibitem{a21}
  T.~R.~Klassen and E.~Melzer,
  ``Purely Elastic Scattering Theories and their Ultraviolet Limits,''
  Nucl.\ Phys.\ B {\bf 338} (1990) 485.
  \href{https://www.sciencedirect.com/science/article/pii/055032139090643R?via\%3Dihub}{doi:10.1016/0550-3213(90)90643-R}.
  %%CITATION = doi:10.1016/0550-3213(90)90643-R;%%
  %223 citations counted in INSPIRE as of 21 Feb 2019
  
  
%b8
%\cite{Braden:1989bg}
\bibitem{a22}
  H.~W.~Braden, E.~Corrigan, P.~E.~Dorey and R.~Sasaki,
  ``Extended Toda Field Theory and Exact S Matrices,''
  Phys.\ Lett.\ B {\bf 227} (1989) 411.
  \href{https://www.sciencedirect.com/science/article/abs/pii/0370269389909520?via\%3Dihub}{doi:10.1016/0370-2693(89)90952-0}.
  %%CITATION = doi:10.1016/0370-2693(89)90952-0;%%
  %69 citations counted in INSPIRE as of 21 Feb 2019


%\cite{Braden:1989bu}
\bibitem{a23}
  H.~W.~Braden, E.~Corrigan, P.~E.~Dorey and R.~Sasaki,
  ``Affine Toda Field Theory and Exact S Matrices,''
  Nucl.\ Phys.\ B {\bf 338} (1990) 689.
  \href{https://www.sciencedirect.com/science/article/pii/055032139090648W?via\%3Dihub}{doi:10.1016/0550-3213(90)90648-W}.
  %%CITATION = doi:10.1016/0550-3213(90)90648-W;%%
  %281 citations counted in INSPIRE as of 21 Feb 2019



%Exact S matrix found for non simply-laced b10

%\cite{Delius:1991cu}
\bibitem{a25}
  G.~W.~Delius, M.~T.~Grisaru and D.~Zanon,
  ``Exact S matrices for the non simply-laced affine Toda theories a(2)(2n-1),''
  Phys.\ Lett.\ B {\bf 277} (1992) 414
  \href{https://www.sciencedirect.com/science/article/abs/pii/037026939291804I?via\%3Dihub}{doi:10.1016/0370-2693(92)91804-I}
 \href{https://arxiv.org/abs/hep-th/9112007v1}{ [hep-th/9112007]}.
  %%CITATION = doi:10.1016/0370-2693(92)91804-I;%%
  %17 citations counted in INSPIRE as of 21 Feb 2019


%\cite{Delius:1991kt}
\bibitem{a26}
  G.~W.~Delius, M.~T.~Grisaru and D.~Zanon,
  ``Exact S matrices for non simply-laced affine Toda theories,''
  Nucl.\ Phys.\ B {\bf 382} (1992) 365
  \href{https://www.sciencedirect.com/science/article/pii/055032139290190M?via\%3Dihub}{doi:10.1016/0550-3213(92)90190-M}
  \href{https://arxiv.org/abs/hep-th/9201067v1}{[hep-th/9201067]}.
  %%CITATION = doi:10.1016/0550-3213(92)90190-M;%%
  %94 citations counted in INSPIRE as of 21 Feb 2019


%\cite{Corrigan:1993xh}
\bibitem{Corrigan:1993xh}
E.~Corrigan, P.~E.~Dorey and R.~Sasaki,
``On a generalized bootstrap principle,''
Nucl. Phys. B \textbf{408} (1993), 579
\href{https://www.sciencedirect.com/science/article/pii/055032139390381X?via\%3Dihub}{doi:10.1016/0550-3213(93)90381-X}
\href{https://arxiv.org/abs/hep-th/9304065}{[arXiv:hep-th/9304065 [hep-th]]}.
%65 citations counted in INSPIRE as of 13 Jul 2020

%\cite{Dorey:1993np}
\bibitem{Dorey:1993np}
P.~Dorey,
``A Remark on the coupling dependence in affine Toda field theories,''
Phys. Lett. B \textbf{312} (1993), 291
\href{https://www.sciencedirect.com/science/article/pii/037026939391083Y?via\%3Dihub}{doi:10.1016/0370-2693(93)91083-Y}
\href{https://arxiv.org/abs/hep-th/9304149}{[arXiv:hep-th/9304149 [hep-th]]}.
%25 citations counted in INSPIRE as of 13 Jul 2020

%\cite{Oota:1997un}
\bibitem{Oota:1997un}
T.~Oota,
``q deformed Coxeter element in non simply-laced affine Toda field theories,''
Nucl. Phys. B \textbf{504} (1997), 738-752
\href{https://www.sciencedirect.com/science/article/pii/S0550321397005555?via\%3Dihub}{doi:10.1016/S0550-3213(97)00555-5}
\href{https://arxiv.org/abs/hep-th/9706054v1}{[hep-th/9706054]}.
%22 citations counted in INSPIRE as of 03 Jul 2020




%\cite{Dorey:1991zp}
\bibitem{aa24}
  P.~Dorey,
  ``Root systems and purely elastic S matrices. 2.,''
  Nucl.\ Phys.\ B {\bf 374} (1992) 741
  \href{https://www.sciencedirect.com/science/article/pii/0550321392904073?via\%3Dihub}{doi:10.1016/0550-3213(92)90407-3}
  \href{https://arxiv.org/abs/hep-th/9110058}{[hep-th/9110058]}.
  %%CITATION = doi:10.1016/0550-3213(92)90407-3;%%
  %68 citations counted in INSPIRE as of 21 Feb 2019





%\cite{Fring:1991gh}
\bibitem{Fring:1991gh}
A.~Fring and D.~I.~Olive,
``The Fusing rule and the scattering matrix of affine Toda theory,''
Nucl. Phys. B \textbf{379} (1992), 429-447
\href{https://www.sciencedirect.com/science/article/pii/0550321392906028?via\%3Dihub}{doi:10.1016/0550-3213(92)90602-8}
%47 citations counted in INSPIRE as of 28 Jun 2020

%\cite{Braden:1990qa}
\bibitem{Braden:1990qa}
  H.~W.~Braden and R.~Sasaki,
  ``The S matrix coupling dependence for a, d and e affine toda field theory,''
  Phys.\ Lett.\ B {\bf 255} (1991) 343.
  \href{https://www.sciencedirect.com/science/article/pii/037026939190777N?via\%3Dihub}{doi:10.1016/0370-2693(91)90777-N}
  %%CITATION = doi:10.1016/0370-2693(91)90777-N;%%
  %41 citations counted in INSPIRE as of 01 May 2019
  
%\cite{Sasaki:1992sk}
\bibitem{Sasaki:1992sk}
R.~Sasaki and F.~P.~Zen,
``The affine Toda S matrices versus perturbation theory,''
Int. J. Mod. Phys. A \textbf{8} (1993), 115-134
\href{https://www.worldscientific.com/doi/abs/10.1142/S0217751X93000059}{doi:10.1142/S0217751X93000059}
%16 citations counted in INSPIRE as of 27 Oct 2021
  
%\cite{Braden:1992gh}
\bibitem{Braden:1992gh}
H.~W.~Braden, H.~S.~Cho, J.~D.~Kim, I.~G.~Koh and R.~Sasaki,
``Singularity analysis in A(n) affine Toda theories,''
Prog. Theor. Phys. \textbf{88} (1992), 1205-1212
\href{https://academic.oup.com/ptp/article/88/6/1205/1855511}{doi:10.1143/PTP.88.1205}
\href{https://arxiv.org/pdf/hep-th/9207025.pdf}{[arXiv:hep-th/9207025]}.
%6 citations counted in INSPIRE as of 27 Oct 2021
  

%\cite{Koubek:1991jj}
\bibitem{Koubek:1991jj}
A.~Koubek, G.~Mussardo and R.~Tateo,
``Bootstrap trees and consistent S matrices,''
Int. J. Mod. Phys. A \textbf{7} (1992), 3435-3446
\href{https://www.worldscientific.com/doi/abs/10.1142/S0217751X92001526}{doi:10.1142/S0217751X92001526}
%3 citations counted in INSPIRE as of 10 Jul 2021



%\cite{Delius:1990ij}
\bibitem{Delius:1990ij}
  G.~W.~Delius, M.~T.~Grisaru, S.~Penati and D.~Zanon,
  ``The Exact S matrices of affine Toda theories based on Lie superalgebras,''
  Phys.\ Lett.\ B {\bf 256} (1991) 164.
  \href{https://www.sciencedirect.com/science/article/abs/pii/037026939190668G?via\%3Dihub}{doi:10.1016/0370-2693(91)90668-G}.
  %%CITATION = doi:10.1016/0370-2693(91)90668-G;%%
  %32 citations counted in INSPIRE as of 21 Feb 2019
  
%\cite{Delius:1991sv}
\bibitem{Delius:1991sv}
  G.~W.~Delius, M.~T.~Grisaru, S.~Penati and D.~Zanon,
  ``Exact S matrix and perturbative calculations in affine Toda theories based on Lie superalgebras,''
  Nucl.\ Phys.\ B {\bf 359} (1991) 125.
  \href{https://www.sciencedirect.com/science/article/pii/0550321391902959?via\%3Dihub}{doi:10.1016/0550-3213(91)90295-9}.
  %%CITATION = doi:10.1016/0550-3213(91)90295-9;%%
  %35 citations counted in INSPIRE as of 21 Feb 2019   


%BOUNDARIES AND DEFECTS


%\cite{Bowcock:1995vp}
\bibitem{b3}
  P.~Bowcock, E.~Corrigan, P.~E.~Dorey and R.~H.~Rietdijk,
  ``Classically integrable boundary conditions for affine Toda field theories,''
  Nucl.\ Phys.\ B {\bf 445} (1995) 469
  \href{https://www.sciencedirect.com/science/article/pii/055032139500153J?via\%3Dihub}{doi:10.1016/0550-3213(95)00153-J}
 \href{https://arxiv.org/abs/hep-th/9501098}{[hep-th/9501098]}.
  %%CITATION = doi:10.1016/0550-3213(95)00153-J;%%
  %109 citations counted in INSPIRE as of 22 Feb 2019





%\cite{Penati:1995xp}
\bibitem{b6}
  S.~Penati, A.~Refolli and D.~Zanon,
  ``Classical versus quantum symmetries for Toda theories with a nontrivial boundary perturbation,''
  Nucl.\ Phys.\ B {\bf 470} (1996) 396
  \href{https://www.sciencedirect.com/science/article/pii/0550321396001630?via\%3Dihub}{doi:10.1016/0550-3213(96)00163-0}
  \href{https://arxiv.org/abs/hep-th/9512174v1}{[hep-th/9512174]}.
  %%CITATION = doi:10.1016/0550-3213(96)00163-0;%%
  %8 citations counted in INSPIRE as of 22 Feb 2019



%\cite{Bowcock:2004my}
\bibitem{d1}
  P.~Bowcock, E.~Corrigan and C.~Zambon,
  ``Affine Toda field theories with defects,''
  JHEP {\bf 0401} (2004) 056
  \href{https://iopscience.iop.org/article/10.1088/1126-6708/2004/01/056/meta}{doi:10.1088/1126-6708/2004/01/056}
  \href{https://arxiv.org/abs/hep-th/0401020}{[hep-th/0401020]}.
  %%CITATION = doi:10.1088/1126-6708/2004/01/056;%%
  %78 citations counted in INSPIRE as of 22 Feb 2019
  
  
 %\cite{Corrigan:2009vm}
\bibitem{d3}
  E.~Corrigan and C.~Zambon,
  ``A New class of integrable defects,''
  J.\ Phys.\ A {\bf 42} (2009) 475203
  \href{https://iopscience.iop.org/article/10.1088/1751-8113/42/47/475203/meta}{doi:10.1088/1751-8113/42/47/475203}
  \href{https://arxiv.org/abs/0908.3126v1}{[arXiv:0908.3126 [hep-th]]}.
  %%CITATION = doi:10.1088/1751-8113/42/47/475203;%%
  %45 citations counted in INSPIRE as of 22 Feb 2019

\bibitem{d4}
%\cite{Bristow:2016zrn}
% \bibitem{Bristow:2016zrn}
  R.~Bristow and P.~Bowcock,
  ``Momentum conserving defects in affine Toda field theories,''
  JHEP {\bf 1705} (2017) 153
   \href{https://link.springer.com/article/10.1007\%2FJHEP05\%282017\%29153}{doi:10.1007/JHEP05(2017)153}
  \href{https://arxiv.org/abs/1612.03002}{[arXiv:1612.03002 [hep-th]]}.
  %%CITATION = doi:10.1007/JHEP05(2017)153;%%
  %6 citations counted in INSPIRE as of 16 Feb 2019


%\cite{Penati:2019kep}
\bibitem{d7}
  S.~Penati and D.~Polvara,
  ``Quantum anomalies in $A^{(1)}_r$ Toda theories with defects,''
  JHEP {\bf 1906} (2019) 062
  \href{https://link.springer.com/article/10.1007\%2FJHEP06\%282019\%29062}{doi:10.1007/JHEP06(2019)062}
  \href{https://arxiv.org/abs/1902.10690}{[arXiv:1902.10690 [hep-th]]}.
  %%CITATION = doi:10.1007/JHEP06(2019)062;%%

%\cite{Steinberg_Paper}
\bibitem{Steinberg_Paper}
R.~Steinberg, ``Finite reflection groups,'' Transactions of the American Mathematical Society {\bf 91} (1959) 493.

%\cite{B. Kostant1}
\bibitem{B. Kostant1}
  B.~Kostant, ``The Principal Three-Dimensional Subgroup and the Betti Numbers of a Complex Simple Lie Group,``
  Am.\ J.\ Math. {\bf 81} (1959) 973
\href{https://www.jstor.org/stable/2372999?seq=1#metadata_info_tab_contents}{https://www.jstor.org/stable/2372999}.

%\cite{Carter_book}
\bibitem{Carter_book}
R.~Carter, ``Simple groups of Lie type'' (Wiley, New York, 1972)

%\cite{Corrigan:1994nd}
\bibitem{Corrigan:1994nd}
E.~Corrigan,
``Recent developments in affine Toda quantum field theory,''
  \href{https://arxiv.org/abs/hep-th/9412213}{[arXiv:hep-th/9412213 [hep-th]]}.
%46 citations counted in INSPIRE as of 23 Jun 2020

%\cite{Dorey:1992gr}
\bibitem{Dorey:1992gr}
P.~Dorey,
``Hidden geometrical structures in integrable models,''
Proceedings of the conference `Integrable Quantum Field Theories,' Como, 1992
\href{https://arxiv.org/abs/hep-th/9212143v2}{[arXiv:hep-th/9212143 [hep-th]]}.
%8 citations counted in INSPIRE as of 09 Feb 2021

%\cite{Olive:1982ye}
\bibitem{Olive:1982ye}
D.~I.~Olive and N.~Turok,
``The Symmetries of Dynkin Diagrams and the Reduction of Toda Field Equations,''
Nucl. Phys. B \textbf{215} (1983), 470-494
\href{https://www.sciencedirect.com/science/article/pii/0550321383902560?via=ihub}{doi:10.1016/0550-3213(83)90256-0}
%116 citations counted in INSPIRE as of 09 Feb 2021




\end{thebibliography}
\end{document}